\documentclass[12pt,english]{article}
\pdfoutput=1 
\usepackage[T1]{fontenc}
\usepackage{times}
\usepackage{lmodern}
\usepackage{threeparttable}
\usepackage[latin9]{inputenc}
\usepackage{titling}
\usepackage{geometry}
\usepackage[english]{babel}
\usepackage{bbm}
\usepackage{iftex}
\usepackage{amsmath}
\usepackage{booktabs}
\usepackage{mathtools}
\usepackage{multirow}
\usepackage{amsfonts}
\usepackage{colortbl}
\usepackage{longtable}
\usepackage{booktabs}
\usepackage{comment}
\usepackage{rotating}
\usepackage{multirow}
\usepackage{textcomp}
\usepackage{subcaption}
\usepackage{float}
\usepackage[dvipsnames]{xcolor}
\definecolor{mygray}{gray}{0.6}
\definecolor{mediumpersianblue}{rgb}{0.0, 0.4, 0.65}
\definecolor{mossgreen}{rgb}{0.68, 0.87, 0.68}
\usepackage{setspace}
\usepackage{graphicx}
\usepackage{subcaption}
\usepackage{upgreek}
\usepackage{array}
\usepackage{grffile}
\usepackage{pbox}
\usepackage{hhline}
\usepackage{etoolbox}
\usepackage[unicode=true]
{hyperref}
\hypersetup{colorlinks=true,citecolor=mediumpersianblue,linkcolor=mediumpersianblue,urlcolor  = mediumpersianblue,linktocpage=true}
\usepackage{apacite}
\makeatletter
\usepackage{babel}
\usepackage{cleveref}
\def\arrvline{\hfil\kern\arraycolsep\vline\kern-\arraycolsep\hfilneg}

\usepackage{chngcntr}
\counterwithin{figure}{section}
\counterwithin{table}{section}
\numberwithin{equation}{section}

\usepackage{tikz}
\usetikzlibrary{shapes,arrows,arrows.meta}
\usetikzlibrary{positioning}
\makeatother

\usetikzlibrary{arrows.meta,angles,quotes}
\colorlet{linecol}{black!75}
\usepackage{forest}

\tikzset{
	my rounded corners/.append style={rounded corners=2pt},
}

\title{Global temperature projections from a statistical energy balance model using multiple sources of historical data}
\author{Mikkel Bennedsen\thanks{Department of Economics and Business Economics and CREATES, Aarhus University, Fuglesangs All\'e 4, 8210
		Aarhus V, Denmark. E-mail: \href{mbennedsen@econ.au.dk}{mbennedsen@econ.au.dk}}, Eric Hillebrand\thanks{Department of Economics and Business Economics and CREATES, Aarhus University, Fuglesangs All\'e 4, 8210
		Aarhus V, Denmark. E-mail: \href{ehillebrand@econ.au.dk}{ehillebrand@econ.au.dk}}, and Jingying Zhou Lykke\thanks{Department of Economics and Business Economics and CREATES, Aarhus University, Fuglesangs All\'e 4, 8210
		Aarhus V, Denmark. E-mail: \href{jzlykke@econ.au.dk}{jzlykke@econ.au.dk}} }
\begin{document}
	\maketitle
	
	\begin{abstract}
		\noindent This paper estimates the two-component energy balance model as a linear state space system (EBM-SS model) using historical data. It is a joint model for the temperature in the mixed layer, the temperature in the deep ocean layer, and radiative forcing. The EBM-SS model allows for the modeling of non-stationarity in forcing, the incorporation of multiple data sources for the latent processes, and the handling of missing observations. We estimate the EBM-SS model using observational datasets at the global level for the period 1955 -- 2020 by maximum likelihood. We show in the empirical estimation and in simulations that using multiple data sources for the latent processes reduces parameter estimation uncertainty. When fitting the EBM-SS model to eight observational global mean surface temperature (GMST) anomaly series, the physical parameter estimates and the GMST projection under Representative Concentration Pathway (RCP) scenarios are comparable to those from Coupled Model Intercomparison Project 5 (CMIP5) models and the climate emulator Model for the Assessment of Greenhouse Gas Induced Climate Change (MAGICC) 7.5. This provides evidence that utilizing a simple climate model and historical records alone can produce meaningful GMST projections. 
		\end{abstract}
	\section{Introduction}
	In this paper, we propose a state space representation (EBM-SS model) of the two-component energy balance model (EBM) (or two-layer EBM), which is initially introduced in \citeA{paltridge1981thermodynamic} and subsequently extended in \citeA{gregory2000vertical} and \citeA{held2010probing}. Like other versions of EBMs \shortcite<e.g.,>{north1981energy}, the two-component EBM is a simplified mathematical representation of the complicated dynamics underlying temperature changes as energy imbalance between the incoming solar radiation and the outgoing terrestrial radiation for the earth system. It extends the zero-dimensional EBM \shortcite<e.g.,>{budyko1969effect,sellers1969global,north1981energy,imkeller2001energy} by accounting for the vertical resolution of the Earth system and two distinct time scales of the global temperature response to external perturbations \shortcite{hasselmann1993cold,held2010probing,geoffroy2013transient}.  \\
	\indent The EBM-SS model enables statistical inference to evaluate parameter estimation uncertainties. Climate modeling is inevitably accompanied with parameter uncertainty \shortcite<e.g.,>{winsberg2012values,reyer2016integrating,gillingham2018modeling}. Accounting for parameter uncertainty plays a crucial role in obtaining reliable evaluations of climate change and conducting robust projection.\\
	\indent Other approaches to quantify parameter uncertainty for the two-component EBM include sensitivity analysis \shortcite{soldatenko2019climate,colman2020understanding}, Monte Carlo simulation \shortcite{gillingham2018modeling,smith2018fair,jimenez2019emergent}, and Bayesian estimation \shortcite{jonko2018towards,nijsse2020emergent}. The main drawback of sensitivity analysis is its disconnection with any measure of probability, while Monte Carlo simulation assumes the input parameter estimates as the true values.  \cite<e.g.,>{cox1981methods}. The performance of Bayesian estimation depends on the prior distributions \shortcite<e.g.,>{kim2020comparison}. Alternatively, state space methods obtain maximum likelihood estimators based on the frequentist principle and easily quantify parameter uncertainty using asymptotic properties \cite{durbin2012time}. \\
	\indent The EBM-SS model reduces estimation uncertainty by using multiple data sources. There are different datasets available for GMST. All of the GMST anomalies series from separate research groups can be regarded as different measurements for the same variable of interest -- the temperature in the mixed layer in the two-component EBM. As we will show, employing different data sources reduces information loss and improves estimation accuracy.\\ 
	\indent The EBM-SS model provides alternatives and extensions to two current contributions by \shortciteA{pretis2020econometric} and \shortciteA{cummins2020optimal}, who obtain parameter values of the two-component EBM using maximum likelihood. \citeA{pretis2020econometric} shows the mathematical equivalence between the two-component EBM and a cointegrated VAR model (EBM-CVAR). Instead of including the temperature in the deep ocean layer, he includes ocean heat content (OHC) in his model. In his discretization, two of the parameters of the two-component EBM,  heat capacity in the mixed layer and the coefficient for the heat transfer, are not recovered in the output EBM-CVAR model. In this paper, we maintain the original parametrization by modeling the temperature in the deep layer but also incorporate OHC as an additional measurement for it, which helps constrain the parameters. We maintain a one-to-one mapping relationship between the two-component EBM and our state space model so that all of the physical parameters can be estimated and interpreted accordingly.  \\
	\indent \shortciteA{cummins2020optimal} present a state space representation of the $k$-layer EBM and report parameter estimates for the cases where $k=2$ and $k=3$. Our paper only considers the case $k=2$ and differs from \shortciteA{cummins2020optimal} in several ways: (1) we employ observational datasets instead of the abrupt $4\times$CO$_2$ experiment data from CMIP5; (2) we model radiative forcing as a non-stationary process instead of a stationary red noise process; (3) we use instrumental records of ocean temperature, OHC, and effective radiative forcing instead of top-of-the-atmosphere net downward radiative flux as measurements in the state space model; and (4) we incorporate multiple data sources for the latent states. \\
\indent {In this paper, using historical datasets, we obtain estimates for the physical parameter in the two-component EBM that are comparable to the estimates in \citeA{cummins2020optimal} that are obtained from CMIP5 model outputs. Meanwhile, the GMST projection results under RCP 2.6, RCP 4.5, and RCP 6.0 scenarios have a high degree of agreement with the outputs from the climate emulator MAGICC 7.5 and CMIP5 models. These results require two ingredients. One is the inclusion of multiple historical records as different data sources into the EBM-SS model. The other is the inclusion of both ocean temperature and OHC datasets. Our results indicate that utilizing a simple climate model with historical records alone can produce meaningful physical parameter estimates and GMST projections.}\\
	\indent The remainder of the paper is organized as follows: Section \ref{sec2} presents the two-component EBM. Section \ref{sec3} describes the method and technical details on mapping the two-component EBM into a state space representation.  Sections \ref{secsim} presents simulation results on the performance of the EBM-SS model. Sections \ref{sec4} and \ref{secEst} introduce the datasets and their use as measurements in the empirical study. Section \ref{scenario} contains an application of the EBM-SS model to GMST projections using RCP scenarios. Section \ref{sec7} concludes. 
	\section{Two-component energy balance models}
	\label{sec2}
	\indent The two-component EBM \cite<e.g.,>{gregory2000vertical,held2010probing} divides the earth system into two thermal reservoirs  (also referred to as ``layer'' , ``box'' , or ``component'') that are characterized by different heat capacities to measure thermal inertia. Each of the reservoirs contains components of global climate responses with both fast and slow time scales. \\
	\indent The first layer is usually called the ``mixed layer'' and consists of the atmosphere, the land surface, and the upper ocean layer. The second layer is called the ``deep ocean layer''.  The depth of the upper ocean layer varies seasonally and geographically. \citeA[Chapter 7]{hartmann2015global} argues that the global mean depth of the ocean in the mixed layer is 70 m. \citeA{gregory2000vertical} considers the upper ocean layer as the part that shows consistent temperature variations with the surface temperature. He chooses 150 m as the depth for the upper layer based on the temporal correlation between the heating rates of different ocean layers and that of the surface. In addition, he defines 2,400 m as the lower bound for the deep layer and describes the part beyond this level as an ``isolated basin''. 
		\begin{figure}[ht]
		\tikzstyle{block} = [rectangle, draw,
		text width=7em, text centered, rounded corners, minimum height=3em,node distance=2cm]
		\caption{\footnotesize Dynamics of the two-component EBM}
		\begin{tikzpicture}[node distance = 3cm, auto]
			\footnotesize
			\node [block,text width=4.5cm] (expert) {\textbf{Incoming radiation} \\ Radiative forcing $F$};
			\node [block,below of=expert,text width=5cm] (init) {\textbf{Mixed Layer} \\(Net radiative flux: $F-\lambda T_m$)};
			\node [block, right of=expert] (system) [right=0.1cm of expert, text width=3.5cm]  {\textbf{Outgoing radiation} \\$\lambda T_m$};
			\node [block, left of=init, node distance=3cm] (update)[left=0.1cm of init, text width=3.5cm] {\textbf{Heat transport} $\mathcal{H}$\\ $\gamma\left(T_m-T_d\right)$};
			\node [block, below=0 cm of init,text width=5cm,minimum height=6em] (decide) {\textbf{Deep ocean layer}};
			\draw [-{Stealth[scale=1.2]}] (update) |-  (decide);
			\draw [-{Stealth[scale=1.2]}] (init) --(update);
			\draw [-{Stealth[scale=1.2]}](expert) -- (init);
			\draw [-{Stealth[scale=1.2]}] (init)-| (system);
		\end{tikzpicture}
		\label{flowchart}
	\end{figure}
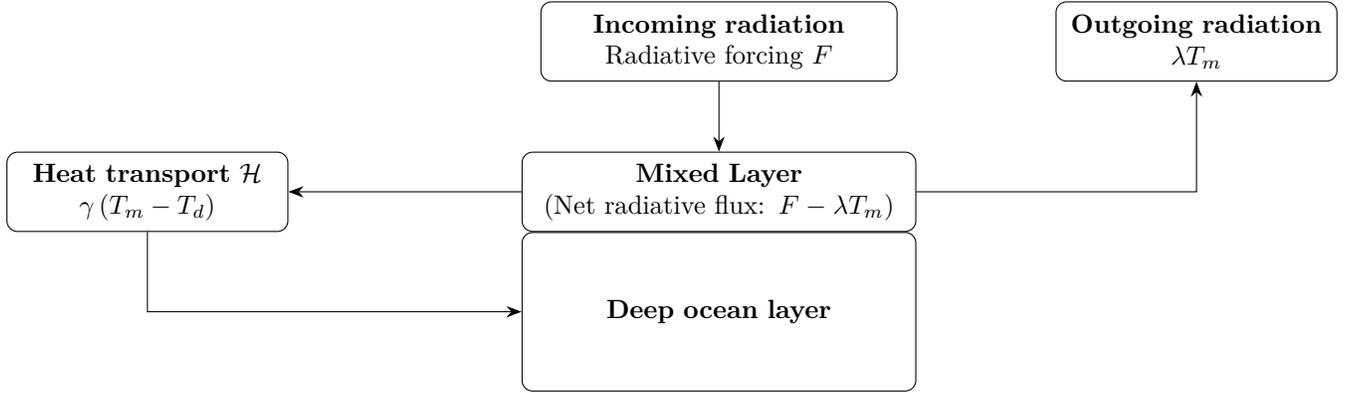 \\
	\indent Figure \ref{flowchart}  gives a graphic illustration of the dynamics underlying the two-component EBM. The change in the heat content of the mixed layer is driven by two sources in opposite directions: external and internal. The external source comes from the net heat flux, which is characterized as $F-\lambda T_{m}$, where the incoming heat radiation is represented by the effective radiative forcing $F$ (measured in $\mathrm{Wm}^{-2} $), and the outgoing longwave radiation (OLR)  is modeled as a linear function of the temperature in the mixed layer $T_m$ with the slope $\lambda$ (known as the climate feedback parameter, $\mathrm{Wm}^{-2} \mathrm{K}^{-1}$). The practice of using $\lambda T_m$ to approximate OLR follows a linear approximation widely used in physical science. The linearity between $T$ and OLR is validated by satellite measurements and can be explained in physics by the offsetting of two non-linear processes \cite{koll2018}. The internal source involves a downward heat transport $\mathcal{H}$ (measured in $ \mathrm{Wm}^{-2} $) from the mixed layer to the deep ocean layer, which changes proportionally to the temperature difference $T_m-T_d$ between these two layers with a coefficient $\gamma$ ($\mathrm{Wm}^{-2} \mathrm{K}^{-1}$).  This term is deemed a reasonable approximation of all of the small perturbations constituting the heat transfer \cite{gregory2000vertical}. The heat exchange term, $\mathcal{H}=\gamma\left(T_m-T_d\right)$, poses the only source of energy for the deep ocean layer. 
Summarizing the dynamics described in Figure \ref{flowchart} into a differential equation system, the two-component EBM is specified as:
	\begin{equation}
		\begin{aligned}
		C_{m} \frac{d T_{m}}{d t} &=F-\lambda T_{m}-\gamma\left(T_{m}-T_{d}\right), \\
		C_{d} \frac{d T_{d}}{d t} &=\gamma\left(T_{m}-T_{d}\right),
		\label{2comEBM}
	\end{aligned}
	\end{equation}
where $C_m$ and $C_d$ (measured in W year m $^{-2} \mathrm{~K}^{-1}$) denote heat capacities of the mixed layer and of the deep ocean layer, respectively. It holds that $C_m<C_d$, indicating that the deep ocean requires a greater amount of energy than the upper layer for a unit change in the temperature. The terms $C_{m} \frac{d T_{m}}{d t} $ and $C_{d} \frac{d T_{d}}{d t}$ describe the rates at which the corresponding temperatures change. 
 In this framework, the radiative forcing $F$ is generated exogenously to the system (e.g., by anthropogenic factors). There are four physical parameters to be estimated: $\lambda$, $\gamma$, $C_m$, and $C_d$. In the next sections, we map the two-component EBM into a state space model and estimate the values of these physical parameters using maximum likelihood methods. 
 	\subsection{Including ocean heat content in the system}
Ocean heat content  (OHC) (measured in J $\mathrm{m}^{-2}$) measures the amount of heat stored in the ocean. {Mathematically, ocean heat content $O$ between ocean depths $h_1$ and $h_2$ is calculated as:
 	\begin{equation}
 		O=	\rho C \int_{h 2}^{h 1} T(x) d x, 
 		\label{OHC}
 	\end{equation}
 	where $\rho$, $C$, and $T(\cdot)$ denote the seawater density,  heat capacity, and the temperature at a specific depth, respectively \cite{dijkstra2008dynamical}}. We denote $\rho C$ as $C_d$ and $\int_{h 2}^{h 1} T(x) d x$ as $T_d$, which represents the integrated average ocean temperature between $h_1$ and $h_2$.  Then, Equation \eqref{OHC} is rewritten as: 
 	\begin{equation}
 	O=	C_dT_d. 
 	\label{T_d}
 \end{equation}
 This relationship  implies the heat content expression $C_d \frac{d T_d}{d t}=\frac{d O}{d t}$ for the term $C_d \frac{d T_d}{d t}$ in the two-component EBM \eqref{2comEBM}. Relating the temperature to heat content in this way is a common practice in energy balance models, and it is motivated by empirical evidence \cite{schwartz2007heat}. \\
 \indent In our state space model of the two-component EBM introduced later, we include OHC in the measurement equation using the linear relationship  $O=C_dT_d$. In our specification, OHC plays the role of a second measurement for the latent state $T_d$, in addition to the direct empirical observations of $T_d$ as the first measurement. As shown in Equation \eqref{OHC}, OHC is a function of the deep ocean temperature, and in practice, the observational OHC series is compiled using the records of ocean temperature data plus salinity information \cite{levitus2012world}. Hence, our choice to include OHC as an additional measurement for $T_d$ is grounded on both the theory and the empirical data construction process. In Section	\ref{secEst}, we will show that using information of both ocean temperature and OHC helps constrain the parameter estimates to be more realistic compared with using ocean temperature alone. 
	\section{Mapping the two-component EBM into a state space model }
	\label{sec3}
	The objective of this section is a discrete-time state space representation of the two-component EBM that enables estimation using empirical data. Particularly, we focus on the multivariate linear Gaussian state space model. For details on the estimation of these models, see \citeA{durbin2012time}. State space models distinguish unobserved states from observations, where the observations are employed to infer the states. They also allow for using multiple observation series as measures of the latent state of interest. 
	\subsection{Decomposition of radiative forcing}
	\label{decomF}
\begin{figure}[h!]
		\centering
		\caption{\footnotesize Components of radiative forcing (1850 -- 2018, $\mathrm{W} \mathrm{m}^{-2}$) from \protect\citeA{ha06510a} }
		\begin{subfigure}{0.49\textwidth}
			\subcaption{\footnotesize forcing from different agents}
			\includegraphics[width=\linewidth]{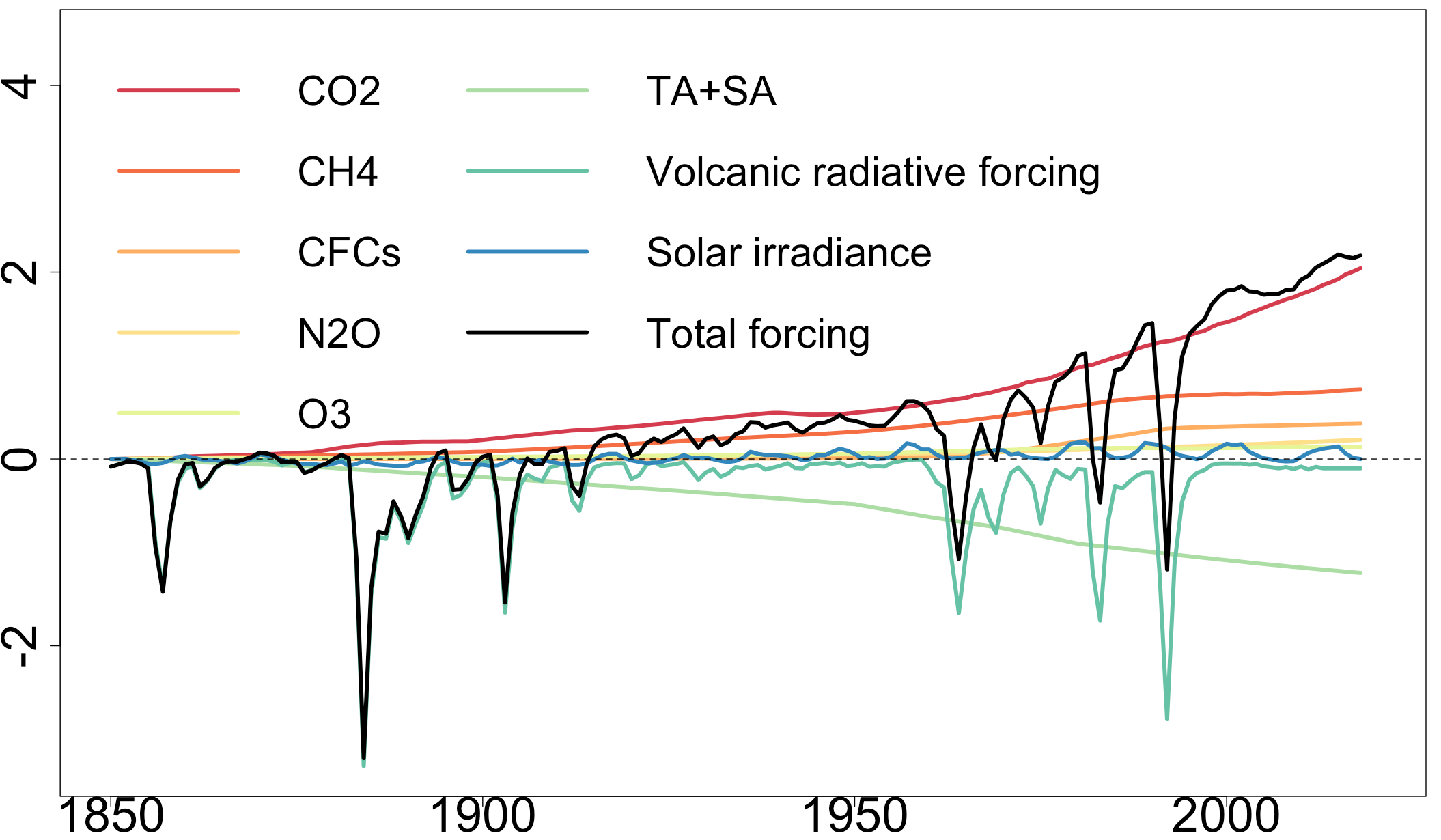}
			\label{9com}
		\end{subfigure}\hfill
		\begin{subfigure}{0.49\textwidth}
			\subcaption{\footnotesize natural and anthropogenic forcing}
			\includegraphics[width=\linewidth]{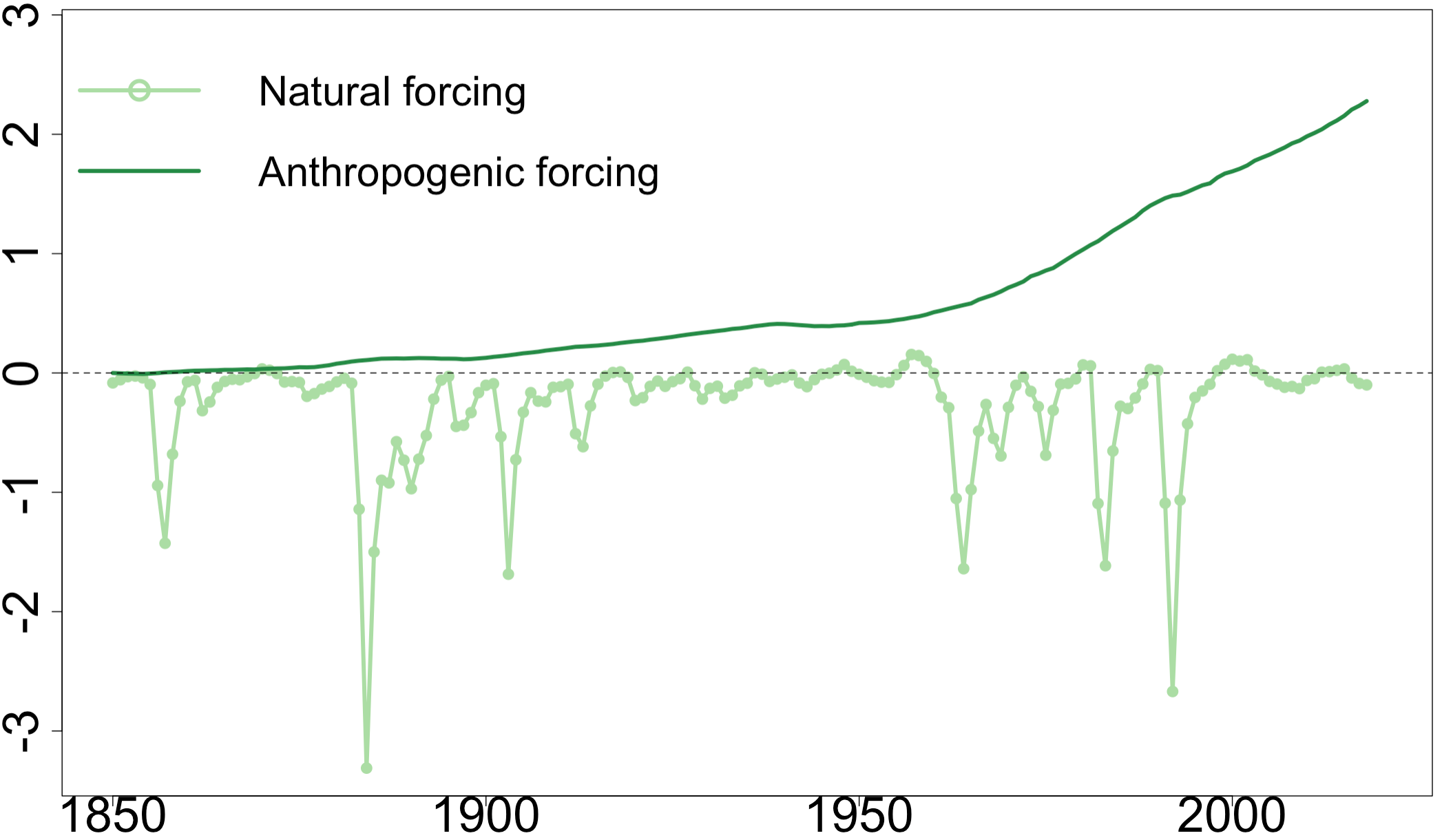}
			\label{NAforcing}
		\end{subfigure}
		\label{hansen}
	\end{figure}

	We decompose the state of radiative forcing, $F$, into two components by source and treat them separately. Total radiative forcing is disaggregated into natural forcing and anthropogenic forcing, where the latter can be further decomposed by the forcing agents.  Figure \ref{treeForcing} shows the decomposition employed in the fifth IPCC assessment report \citeA[Chapter 8]{ip05000f}. As seen in Figure \ref{treeForcing}, natural forcing is mainly driven by two contributors: solar irradiance and volcanic forcing. Anthropogenic forcing is subject to human influences and consists of forcing from greenhouse gases, land surface changes, and human-made aerosols. 
	\begin{figure}[ht!]
		\centering
		\caption{\footnotesize Components of radiative forcing}
		\scriptsize
		\begin{forest}
			for tree={
				line width=1pt,
				if={level()<2}{
					my rounded corners,
					draw=linecol,
				}{},
				edge={color=linecol, >={Triangle[]}, ->},
				if level=0{%
					l sep+=0.3cm,
					align=center,
					parent anchor=south,
					tikz={
						\path (!1.child anchor) coordinate (A) -- () coordinate (B) -- (!l.child anchor) coordinate (C) ;
					},
				}{%
					if level=1{%
						parent anchor=south west,
						child anchor=north,
						tier=parting ways,
						align=center,
						font=\bfseries,
						for descendants={
							child anchor=west,
							parent anchor=west,
							anchor=west,
							align=left,
						},
					}{
						if level=2{
							shape=coordinate,
							no edge,
							grow'=0,
							calign with current edge,
							xshift=10pt,
							for descendants={
								parent anchor=south west,
								l sep+=-40pt
							},
							for children={
								edge path={
									\noexpand\path[\forestoption{edge}] (!to tier=parting ways.parent anchor) |- (.child anchor)\forestoption{edge label};
								},
								font=\bfseries,
								for descendants={
									no edge,
								},
							},
						}{},
					},
				}%
			},
			[\textbf{Total radiative forcing}
			[Natural forcing
			[
			[Solar irradiance
			]
			[Volcanic radiative forcing 
			]
			]
			]
			[Anthropogenic forcing
			[
			[Greenhouse gases:
			[{Well-mixed greenhouse gasses}\\ Ozone ($\mathbf{O}_3$)(tropospheric and stratospheric)\\ Stratospheric water vapour]
			]
			[Land surface changes:
			[Land use changes\\snow albedo changes]
			]
			[Human-made aerosols:
			[Reflective aerosols\\Aerosol Indirect Effect \\Black carbon in snow and ice]
			]
			]
			]
			]
			]
		\end{forest}
		\label{treeForcing}
	\end{figure} \\
	\normalsize
	\indent We obtain the latest version of the effective forcing dataset from \shortciteA{ha06510a} as the measurement for $F_t$ and show the data in Figure \ref{hansen}. It provides information on forcing from different greenhouse gases and summarizes the forcing due to the land surface changes and human-made aerosols into a category ``TA+SA'' (tropospheric aerosols and surface albedo forcings combined). \\
	\indent   As seen in Figure \ref{9com}, the observation-based natural forcing $Y_{N,t}$ is dominated by the volcanic forcing, which appears as negative spikes due to temporary cooling periods lasting approximately three years after major volcanic eruptions, while solar irradiance varies around zero over time and exhibits cyclical behavior. Anthropogenic forcing is attributed to human activities, of which the dominant contributors are the well-mixed greenhouse gases. Figure \ref{NAforcing} shows the two time series for natural forcing and anthropogenic forcing we use in this paper. Natural forcing exhibits large negative spikes and remains otherwise close to zero, while anthropogenic forcing is upward trending. \\
	\indent As natural forcing and anthropogenic forcing have distinct time series characteristics, we treat them separately. In our state space model, we let $F_t=A_t + N_t$, where $A_t$ is anthropogenic forcing and $N_t$ is natural forcing. We model $A_t$ but we treat $N_t$ as an exogenous regressor and use historical data for it. Considering the small magnitude of solar irradiance, we do not introduce extra structure to model its cyclical feature explicitly. Anthropogenic greenhouse gas emissions following industrialization increase anthropogenic forcing (denoted as $Y_{A,t}$) and render it non-stationary \shortcite{kaufmann2013does,chang2020evaluating}. In the two-component EBM system, the increasing anthropogenic forcing raises temperatures at the surface and in the ocean, thus, temperatures become non-stationary, too. Anthropogenic forcing is external to the system of surface temperature and ocean heat uptake, and, hence, it can be regarded as the major source of non-stationarity in the system. Non-stationarity poses statistical challenges such as spurious regression \cite<e.g.,>{granger1974spurious}. State space methods can be used to specify systems of non-stationary variates while retaining valid statistical inference \cite{caines2018linear}. 
	\\
	\begin{figure}[h!]
		\centering
		\caption{\footnotesize First and second-order differences of the time series of anthropogenic forcing $Y_{A, t}$ $\left( \mathrm{Wm}^{-2}\right)$ during 1850 -- 2018. The two vertical dashed lines mark the year 1958 and 1978 when the measurement scheme method changes. The gray area is the time horizon 1955 -- 2020 for the empirical study in this paper.}
		\begin{subfigure}{0.5\textwidth}
			\subcaption{\footnotesize first-order difference: $\Delta Y_{A,t}$}
			\includegraphics[width=\linewidth]{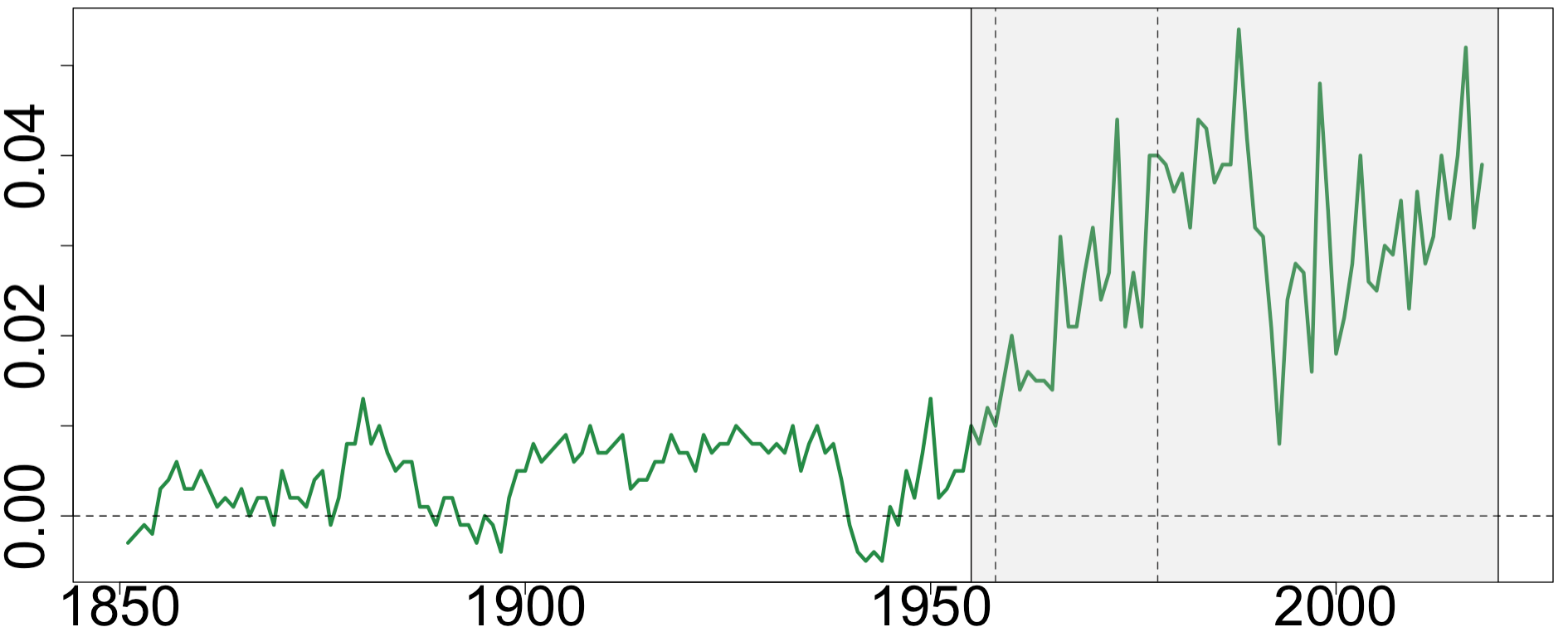}
			\label{1st}
		\end{subfigure}\hfill
		\begin{subfigure}{0.5\textwidth}
			\subcaption{\footnotesize second-order difference: $\Delta^2 Y_{A,t}$}
			\includegraphics[width=\linewidth]{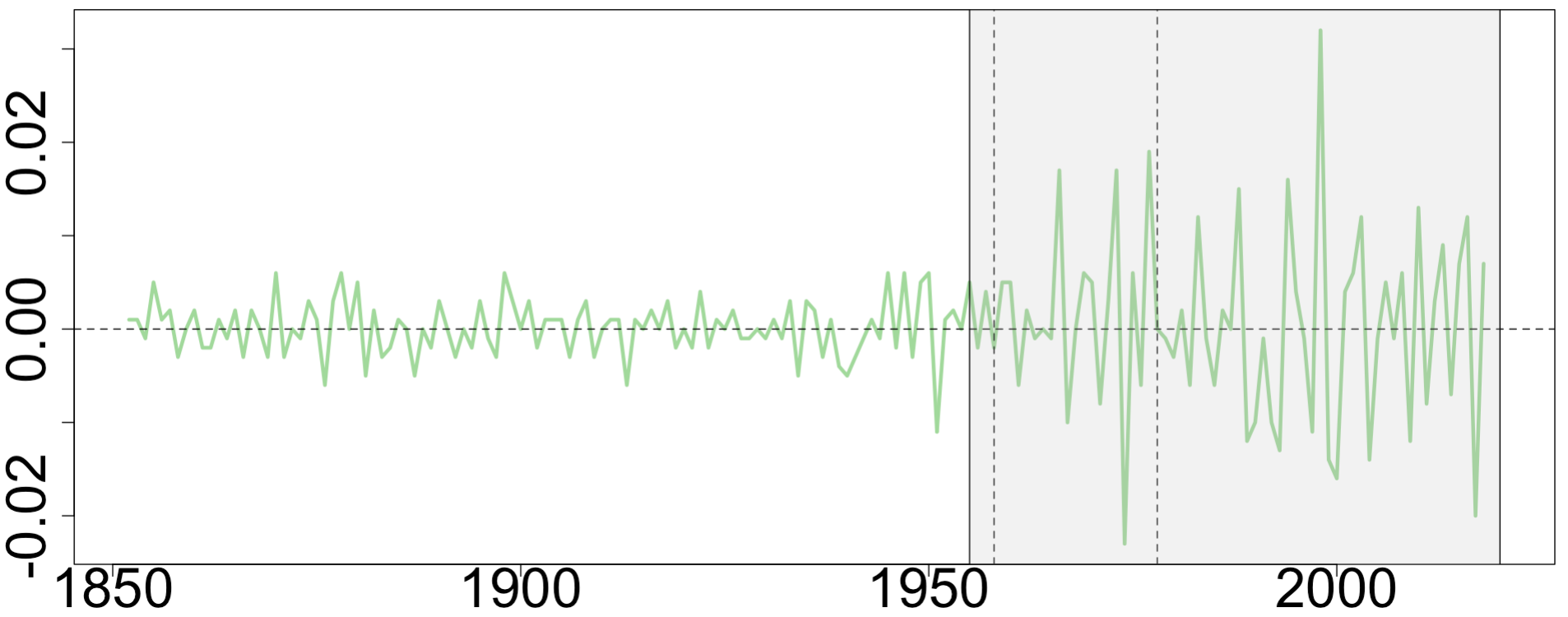}
			\label{2nd}
		\end{subfigure}
	\end{figure} 
	\indent As shown in Figure \ref{1st}, the first-order difference of anthropogenic forcing, $\Delta Y_{A,t}$, appears non-stationary with both trends and shifts of trends. The second-order difference in Figure \ref{2nd} exhibits an abrupt increase in variance in the 1950s. The shifts in the trends and the variance are mainly the result of changes in measurement methods. For example, greenhouse gases are initially measured using ice core records and later measured directly from the atmosphere \cite<e.g.,>{raynaud1993ice}. This switch applies since the year 1958 for $\text{CO}_2$ and 1978 for $\text{CH}_4$ and  $\text{N}_2$O. These structural shifts may indicate the existence of wide-sense non-stationarity in anthropogenic forcing, which invalidates the method of obtaining a stationary process by taking differences \cite{castle2019modelling,castle2020climate}. An in-depth analysis of wide-sense non-stationarity is beyond the scope of this paper, but we investigate the integration order of anthropogenic forcing. Unit root test results are reported in Table \ref{ADF} in Appendix \ref{unitroot}. As shown in Table \ref{ADF}, the components of forcing exhibit different integration orders. Total radiative forcing, $Y_{F,t}$, is an $I(1)$ process, i.e., it becomes stationary after taking first-order difference. This observation is consistent with the statement in \citeA{pretis2020econometric} that the individual forcing components integrate to an $I(1)$ total forcing upon summation. Table \ref{ADF} also implies that total anthropogenic forcing, $Y_{A,t}$, is an $I(1)$ process when no lag or only the first-order lag is included in the unit root test equation, but it is an $I(2)$ process if a higher-order lag is included. \\
	\indent Guided by the unit root tests and visual inspection of $Y_{A}$ and its differences, we represent $A_t$ using a local linear trend model \cite[Chapter 3]{durbin2012time}, where $A_t$ is a random walk process with a stochastic trend $\beta_t$: 
	\begin{equation}
		\begin{aligned}
			A_{t} =& \beta_t + 	A_{t-1} + \eta_{A,t}, \\
			\beta_t =& \beta_{t-1} + \eta_{\beta,t}, 
		\end{aligned}
	\label{locall}
	\end{equation}
	where $ \eta_{A,t}$ and $\eta_{\beta,t}$ are independent Gaussian white noise processes with variances $\sigma^2_{\eta_{A}}$ and $\sigma_{\eta_{\beta}}^2$, respectively. The second-order difference $\Delta^2 A_{t}$ is thus a linear function of two white noise processes and hence stationary:\footnote{If we set $\sigma_{\eta_A}^2=0$, Equation \eqref{ddA} is reduced to $ \Delta^2 A_t = \eta_{\beta,t}$, which is called integrated random walk model \shortcite{young1991recursive,durbin2012time}.}
	\begin{equation}
		\begin{aligned}
			\Delta^2 A_t  =\eta_{\beta,t}+\eta_{A,t}-\eta_{A, t-1}. 
		\end{aligned}
		\label{ddA}
	\end{equation}
	As a result, $A_t$ is modeled as an $I(2)$ process. 
	Figure \ref{2nd} shows a variance increase and hence suggests the existence of heteroskedasticity in $\Delta^2 A_t$. One method to accommodate such a variance shift is to impose a multiplicative constant on the variance of the measurement error of forcing $\sigma_{\varepsilon, Y_{F}}^{2}$ from the year where the measurement scheme shifts onwards, and this constant can be estimated using maximum likelihood. Inclusion of such a constant yielded insignificant estimates on the sample period 1955 -- 2020, and hence we omit it from the model. 
	\subsection{EBM-State Space (EBM-SS) model -- state equation}
	\label{secStateEqn}
	\indent The state equation in a state space model describes the dynamics of the latent state variables as a first-order autoregressive process. For our EBM-SS model, we define the following five variables as unobservable latent states: the temperature in the mixed layer $T_m$, the temperature in the deep ocean layer $T_d$, natural forcing $N$, anthropogenic forcing $A$, and the stochastic trend $\beta$. \\
\indent According to the dynamics in the two-component EBM (differential equation system \eqref{2comEBM}) and the decomposition of $F$ in Section \ref{decomF}, we formulate the state equation system as the following system of linear equations: 
	\begin{equation}
		\begin{aligned}
			{T}_{m, t}=&\left(1-\frac{(\lambda+\gamma)}{C_{m}}\right) T_{m,t-1}+\frac{\gamma}{C_{m}} T_{d,t-1}+\frac{1}{C_{m}} (N_{t-1} + A_{t-1})+ \eta_{T_{m}, t}, \\
			T_{d, t}=&\frac{\gamma}{C_{d}} T_{m, t-1}+\left(1-\frac{\gamma}{C_{d}}\right) T_{d, t-1}+\eta_{T_{d}, t}, \\
			N_t =& Y_{N,t}, \\
			A_{t}=&\beta_{t}+A_{t-1}+\eta_{A,t},\\
			\beta_{t} = & 	\beta_{t-1} +\eta_{\beta, {t}},  
		\end{aligned}
		\label{stateEqn}
	\end{equation}
	where $\eta_{T_{m}, t}$, $\eta_{T_{d}, t}$, $\eta_{A, t}$, respectively, and $\eta_{\beta,t}$ are the state disturbances to the states $T_{m,t}$, $T_{d,t}$, $A_t$ and $\beta_{t}$, and they capture the deviations from the assumed linear relations implied by the two-component EBM and the local linear trend model (Equation \eqref{locall}). We assume the state disturbances to be independent and $\eta_{\cdot, t}\sim\mathcal{N}(0, 	\sigma^2_{\eta,\cdot} )$. 
		The system of equations \eqref{stateEqn} allows for modeling non-stationarity.  It establishes temperature changes as a response to perturbations in radiative forcing and represents the non-stationarity in radiative forcing using a local linear trend model, which is the only source of a stochastic trend in the system, and the stochastic trend propagates to temperatures at the surface and in the ocean through linear relationships. The equation $N_t=Y_{N,t}$ captures that we use observational data $Y_{N,t}$ on natural forcing for the latent process $N_t$, i.e., we treat it as exogenous. 
	\subsection{EBM-SS model -- measurement equation}
\label{secMeasEqn}
The measurement equation in a state space model connects the observational data vector to the latent state vector linearly. We consider the  following measurement equations: 
\begin{equation}
	\begin{aligned}
		Y_{T_{m}, t}=& T_{m,t}+ \varepsilon_{T_{m}, t}, \\
		Y_{T_{d}, t}=& T_{d,t}+ \varepsilon_{T_{d}, t}, \\
		Y_{O, t}=&C_d T_{d,t} + \varepsilon_{O, t},\\
		Y_{F, t}=& N_t + A_t +  \varepsilon_{F, t},
	\end{aligned}
	\label{MeasEqn_explicit}
\end{equation}
where $Y_{*,t}$ denotes the measurement for the latent process $*$. Following the assumption $O=C_dT_d$, the third equation in \eqref{MeasEqn_explicit} expresses the OHC series, $Y_{O,t}$, as an alternative measurement for $T_d$. Note that the data for $Y_{O,t}$ and $Y_{T_d,t}$ are retrived from the same institution and cover the same ocean depth. We assume that $\varepsilon_{\cdot, t}\sim \mathcal{N}(0,\sigma^2_{Y_\cdot})$, and the correlation $\operatorname{Corr}\left(\varepsilon_{T_d,t},\varepsilon_{O,t}\right)=\rho$, where $\rho$ can be estimated using maximum likelihood like other parameters. Allowing for correlation between the measurement errors of ocean temperature and OHC accounts for the highly correlated data compilation processes for these two series.  
	\subsubsection{Incorporation of multiple measurements }
\label{secMultiMeas}
\citeA{cummins2020optimal} fit a two-box EBM to datasets from 16 Earth System Models (ESMs) in CMIP5 separately  and further consider a joint data series that is the average of these 16 datasets. The parameter estimates vary across different datasets due to the heterogeneity of the ESMs. We employ an alternative strategy. We include different data sources simultaneously in the measurement equation as multiple measurements for the latent states. This approach only produces one set of parameter estimates regardless of the number of measurements we include. We focus on using multiple data sources of GMST, ocean temperature, and OHC.\footnote{We use one data source for radiative forcing. This is because we could not find an alternative observation-based dataset that provides complete information on the total radiative forcing and its components as the one by \citeA{ha06510a}.} The two latent processes $T_{m,t}$ and $T_{d,t}$ are linearly linked to multiple observational measurements. For example, the $K$ ($K\in \mathbb{N}$) GMST anomalies $Y_{T_{m}, t}^{1}$, ..., $Y_{T_{m}, t}^{K}$ share the same driver -- the latent process $T_{m,t}$. They are distinguished from each other by separate measurement errors $\varepsilon_{T_m,t}^1$, ..., $\varepsilon_{T_m,t}^K$. For ocean temperature and OHC, we include series from the same institution in pairs, as they are correlated measurements for $T_d$. The measurement equations that include $J$ and $K$ ($J\in \mathbb{N}$) data sources for GMST and ocean data are formulated as: 
\begin{equation}
	\begin{aligned}
		Y_{T_{m}, t}^1&= T_{m,t}+ \varepsilon_{T_{m}, t}^1, \\
		& \vdotswithin{ = }\notag \\
		Y_{T_{m}, t}^K& = T_{m,t}+ \varepsilon_{T_{m}, t}^K, \\
		Y_{T_{d}, t}^1 &= T_{d,t}+ \varepsilon_{T_{d}, t}^1, \enspace 	Y_{O, t}^1  = C_d T_{d,t} + \varepsilon_{O, t}^1, \\
	& \vdotswithin{ = } \enspace \qquad  \qquad \qquad \enspace \enspace \vdotswithin{ = }\notag \\
	Y_{T_{d}, t}^J & = T_{d,t}+ \varepsilon_{T_{d}, t}^J,\enspace Y_{O, t}^J  = C_dT_{d,t}+ \varepsilon_{O, t}^J,  \\
		Y_{F, t} & = N_t + A_t +  \varepsilon_{F, t}.
	\end{aligned}
	\label{MeasEqn_Multi}
\end{equation}
Similar to the one-data-source case, $\varepsilon_{T_m, t}^k \sim \mathcal{N}\left(0, \sigma_{\varepsilon,Y_{T_m}^k}^{2}\right)$, $\varepsilon_{T_d, t}^j \sim \mathcal{N}\left(0, \sigma_{\varepsilon,Y_{T_d}^j}^{2}\right)$, $\varepsilon_{O, t}^j \sim \mathcal{N}\left(0, \sigma_{\varepsilon,Y_{O}^j}^{2}\right)$, $\varepsilon_{F, t} \sim \mathcal{N}\left(0, \sigma_{\varepsilon,Y_{F}}^{2}\right)$, and $\operatorname{Corr}\left(\varepsilon_{T_d,t}^j,\varepsilon_{O,t}^j\right)=\rho_j$. 
\subsection{EBM-SS model -- matrix form }
		\label{EBM-SS}
\begin{figure}[h!]
	\centering
	\caption{\footnotesize Diagram of the state space model with multiple data sources, i.e.,  $K$ GMST anomalies, $J$ pairs of ocean temperature and OHC anomalies, and one forcing as measurements. }
	\begin{tikzpicture}
		\tikzstyle{rectangle_style}=[rectangle, draw]
		\tikzstyle{dividedrectangle_style}=[draw, rectangle split, rectangle split parts=2, rotate = 90, minimum height = 15mm, minimum width = 10mm]

		\node at (7, -0.5) [circle] (Temp) {$T_{m,t}$};
		\node at (7, -2.2) [circle] (O) {$T_{d,t}$};
		\node at (7, -3.7) [circle] (F) {$F_{t}$};
		
		\node at (9, -0.25) [rectangle_style] (Z1) {$\mathbf{Z}$};
		\node at (9, -2.2) [rectangle_style] (Z2) {$\mathbf{Z}$};
		\node at (9, -4) [rectangle_style] (Z3) {$\mathbf{Z}$};
		
		\node at (5, -0.1) [circle] (mM) {${\boldsymbol{\upeta}}_{t}$};
		\node at (1.2, -1.5) [circle] (mX1) {$\mathbf{X}_{t-1}$};
		\node at (3.1, -3.5) [rectangle_style] (mT) {$\mathbf{T}$};
		
		\node at (5, -1.5) [circle] (mX) {$\mathbf{X}_t$};
		
		\foreach \x in {1,...,3};
		
		\draw node at (12.5, 0.5) [circle] (M1) {$Y_{T_{m}, t}^{1}$};
		\draw node at (12.5, -1) [circle] (M2) {$Y_{T_{m}, t}^{K}$};
		\draw node at (12.5, -1.8) [circle] (M3) {$\left(Y_{T_d, t}^{1}, Y_{O, t}^{1}\right)$};
		\draw node at (12.5, -3.3) [circle] (M4) {$\left(Y_{T_d, t}^{J}, Y_{O, t}^{J}\right)$};
		\draw node at (12.5, -4.2) [circle] (M5) {$Y_{F, t}$};
		\foreach \x in {1,...,3}
		\fill (12.5, -2.3 - \x*0.15) circle (1pt);
		\foreach \x in {1,...,3}
		\fill (12.5, 0 - \x*0.15) circle (1pt);

		\path[-] (M1) edge node[pos=.5, above] {\scriptsize$\varepsilon_{T_{m}, t}^1$} (Z1);
		\path[-] (M2) edge node[pos=.5, below] {\scriptsize$\varepsilon_{T_{m}, t}^K$} (Z1);
		\path[-] (M3) edge node[pos=.5, above] {\scriptsize$\left(\varepsilon_{T_{d}, t}^1,\varepsilon_{O, t}^1\right)$} (Z2);
		\path[-] (M4) edge node[pos=.5, below] {\scriptsize$\left(\varepsilon_{T_{d}, t}^J,\varepsilon_{O, t}^J\right)$} (Z2);
		\path[-] (M5) edge node[pos=.5, above] {\scriptsize$\varepsilon_{F, t}$} (Z3);

		\path[-] (mX) edge node[] {} (Temp);
		\path[-] (mX) edge node[] {} (O);
		\path[-] (mX) edge node[] {} (F);
		\path[<-] (Z1) edge node[above, midway] {$$} (Temp);
		\path[<-] (Z2) edge node[above, midway] {$$} (O);
		\path[<-] (Z3) edge node[above, midway] {$$} (F);
		
		\path[<-] (mX) edge node[] {} (mM);
		\draw [->] (mX1) |-  (mT);
		\draw [->] (mT) -|  (mX);
		
	\end{tikzpicture}
	\label{diagram}
\end{figure}
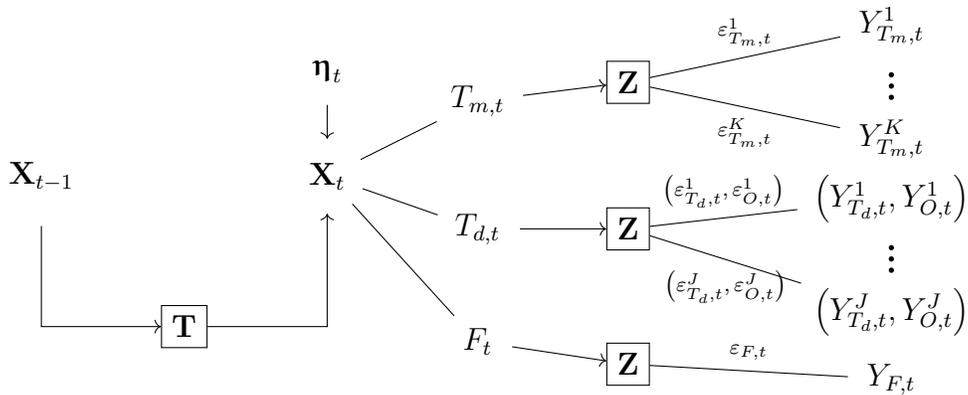

		In this section, we integrate Section \ref{secStateEqn} and Section \ref{secMeasEqn} and present the matrix form of the EBM-SS model. We only discuss the specification with multiple data sources, which nests the one-data-source case. We denote $\mathbf{X}_t = \begin{pmatrix}
			T_{m,t} & T_{d,t} & N_t & A_t & \beta_t & 1
		\end{pmatrix}^{\top}$ as the state vector and $\mathbf{Y}_t=\begin{pmatrix}
		Y_{T_m,t}^1 & \cdots &Y_{T_m,t}^K & Y_{T_d,t}^1 & \cdots &  Y_{T_d,t}^J & Y_{O,t}^1 & \cdots & Y_{O,t}^J  & Y_{F,t}
	\end{pmatrix}^\top$ as the observational vector. The processes in $X_t$ are unobserved, with the exceptions of natural forcing $N_t$ and the constant state ``1''. The latter is a technical way to equate the state $N_t$ with observations $Y_{N,t}$ without errors, or in other words, to treat them as an exogenous regressor, as seen in Equation \eqref{StatEqn}. The observational vector contains data on the elements of the state vector and on OHC ($Y_{O,t}^1$, ..., $Y_{O,t}^J$). 
		We write the state disturbances into a vector  $\boldsymbol{\upeta}_t=\left(\begin{array}{lllllll}
			\eta_{T_{m,t}} & \eta_{T_{d,t}} & 0 & \eta_{A,t} & \eta_{\beta, t} & 0
		\end{array}\right)^{\top}$ and the mearsurement errors into a vector $\boldsymbol{\upvarepsilon}_t=\begin{pmatrix}
		\varepsilon_{T_m,t}^1 & \cdots &\varepsilon_{T_m,t}^K & \varepsilon_{T_d,t}^1 & \cdots &  \varepsilon_{T_d,t}^J & \varepsilon_{O,t}^1 & \cdots & \varepsilon_{O,t}^J  & \varepsilon_{F,t}
	\end{pmatrix}^\top$, where the two 0's in $\boldsymbol{\upeta}_t$ are due to the absence of state disturbances for the natural forcing state and the constant state. 
The EBM-SS model is written as a standard discrete-time state space form, as defined in, e.g., \citeA{durbin2012time}: 
\begin{equation}
	\begin{aligned}
		\mathbf{X}_{t+1}&= \mathbf{T} \mathbf{X}_{t}+\boldsymbol{\upeta}_{t},   \enspace & \boldsymbol{\upeta}_{t} \sim \mathcal{N}(0, \mathbf{Q}), \\
		\mathbf{Y}_{t}&= \mathbf{Z} \mathbf{X}_{t}+\boldsymbol{\upvarepsilon}_{t}, \enspace & \boldsymbol{\upvarepsilon}_{t} \sim \mathcal{N}(0, \mathbf{H}).
	\end{aligned}
	\label{EM}
\end{equation}
The matrices $\mathbf{T}$, $\mathbf{Q}$, $\mathbf{H}$, and $\mathbf{Z}$ are time-invariant. The state equation $\mathbf{X}_{t+1}= \mathbf{T} \mathbf{X}_{t}+\boldsymbol{\upeta}_{t}$ is written explicitly as: 
	\begin{equation}
	\footnotesize
	\left(\begin{array}{c}T_{m, t+1} \\ T_{d,t+1} \\N_{t+1} \\A_{t+1} \\  \beta_{t+1} \\ 1\end{array}\right)=
	\left(\begin{array}{cccccccccc}\frac{-(\lambda+\gamma)}{C_{m}}+1 & \frac{\gamma}{C_{m}} & \frac{1}{C_{m}} &  \frac{1}{C_{m}} & 0 & 0 \\ \frac{\gamma}{C_{d}} & -\frac{\gamma}{C_{d}}+1 & 0 & 0 & 0 & 0  \\ 0 & 0 & 0 & 0 & 0 & Y_{N,{t+1}} \\ 0 & 0 & 0 & 1 & 1 & 0  \\ 0 & 0 & 0 & 0 & 1 & 0 \\ 0 & 0 & 0 & 0 & 0 & 1 \end{array}\right)	\left(\begin{array}{c}T_{m, t} \\ T_{d,t}  \\ N_{t}\\A_{t} \\ \beta_{t} \\ 1\end{array}\right)+\left(\begin{array}{c}\eta_{T_{m, t}} \\ \eta_{T_d, t} \\ 0 \\ \eta_{A, t}  \\ \eta_{\beta,t} \\ 0\end{array}\right),
	\label{StatEqn}
\end{equation}
and $\boldsymbol{\upeta}_t \sim \mathcal{N}(0, {\mathbf{Q}})$, where
		\begin{equation}
			\footnotesize
			{\mathbf{Q}}= \begin{pmatrix}
				\sigma^2_{\eta,T_m} & 0 & 0 & 0 & 0 & 0 \\0 & \sigma^2_{\eta,T_d}  & 0 & 0 & 0 & 0\\ 	
				0 & 0 & 0 & 0 & 0 & 0 \\
				0 & 0 & 0  &  \sigma^2_{\eta,A} & 0  & 0\\
				0 & 0 &0 & 0 & \sigma^2_{\eta,\beta} & 0 \\
					0 & 0 & 0 & 0 & 0 & 0
			\end{pmatrix}.
			\label{Q}
		\end{equation} 
	The measurement equation $\mathbf{Y}_{t}= \mathbf{Z} \mathbf{X}_{t}+\boldsymbol{\upvarepsilon}_{t}$ with multiple data sources is written explicitly as: 
		\begin{equation}
		\footnotesize
		\left(\begin{array}{c}Y^1_{T_{m},t}  \\  \vdots \\  Y^K_{T_{m},t} \\ Y_{T_d,t}^1  \\ \vdots \\  Y_{T_d,t}^J \\ Y_{O,t}^1  \\ \vdots \\  Y_{O,t}^J \\ Y_{F,t}  \end{array}\right)=\left(\begin{array}{cccccccccccccccccc}
			1 & 0 & 0 & 0 & 0  & 0 \\
			\vdots &	\vdots& 	\vdots & 	\vdots & 	\vdots  & \vdots \\
			1 & 0 & 0 & 0 & 0  & 0  \\
			0 & 1 & 0 & 0 & 0
			& 0  \\
			\vdots & \vdots &  \vdots &  \vdots &  \vdots  &\vdots  \\
			0 & 1 & 0 & 0 &0 &0 \\
			0 & C_d & 0 & 0 & 0
			& 0  \\
			\vdots & \vdots &  \vdots &  \vdots &  \vdots  &\vdots \\
			0 & C_d & 0 & 0 & 0 &0
			\\ 0 & 0 & 1 & 1 & 0  & 0 
		\end{array}\right)\left(\begin{array}{c}T_{m, t} \\ T_{d,t}  \\ N_{t}\\A_{t} \\ \beta_{t} \\ 1\end{array}\right) + 	\left(\begin{array}{c}\varepsilon_{T_{m},t}^1\\ \vdots \\\varepsilon_{T_{m},t}^K\\ \varepsilon_{T_{d},t}^1 \\ \vdots \\\varepsilon_{T_{d},t}^J\\ \varepsilon_{O,t}^1 \\ \vdots \\ \varepsilon_{O,t} ^J \\ \varepsilon_{F,t}  \end{array}\right),
		\label{local_level2}
	\end{equation}
 and $\boldsymbol{\upvarepsilon}_{t}\sim \mathcal{N}(0,\mathbf{H})$, where
		\begin{equation}
		\mathbf{H} = 	\scriptsize \begin{pmatrix}
			\sigma_{{\varepsilon}, Y_{T_m}^1}^2 & 0 & \cdots & 0 & 0  & 0 &  \cdots  &0 & 0 &  \cdots  &0 & 0 \\
			0 & 	\sigma_{{\varepsilon}, Y_{T_m}^2}^2 & \ddots & 0 & 0  & 0 &  \cdots  & 0 & 0 &  \cdots  &0 & 0 \\
			\vdots & \ddots & \ddots & \ddots  & \vdots &  \vdots  & \ddots & \vdots &  \vdots  & \ddots & \vdots  & \vdots \\
			0  & 0  & \ddots & 	\sigma_{{\varepsilon}, Y_{T_m}^{K-1}}^2 & 0  &  0	 & 	\cdots  & 	0  &   0	 & 	\cdots  & 	0  & 0   \\
			0 & 0 & \cdots & 0  & 	\sigma_{{\varepsilon}, Y_{T_m}^{K}}^2 & 0 & 	\cdots  & 0  &   0	 & 	\cdots  & 	0  & 0 \\
			0 & 0  & \cdots & 0 &0 & 	\sigma_{{\varepsilon}, Y_{T_d}^1}^2 & \ddots & 0 &   \rho \sigma_{{\varepsilon}, Y_{T_d}^1}  \sigma_{{\varepsilon}, Y_{O}^1}	 & 	\ddots  & 	0   & 0  \\
			\vdots &  \vdots &  \ddots &  \vdots & \vdots & \ddots & \ddots & \ddots &  \ddots & \ddots & \ddots & \vdots  \\
			0&  0 & \cdots & 0 &0 & 0& \ddots & 	\sigma_{{\varepsilon}, Y_{T_d}^J}^2 &  0	 & 	\ddots  & 	 \rho\sigma_{{\varepsilon}, Y_{T_d}^J} \sigma_{{\varepsilon}, Y_{O}^J}	   & 0  \\
			0 & 0  & \cdots & 0 &0 &  \rho \sigma_{{\varepsilon}, Y_{T_d}^1}  \sigma_{{\varepsilon}, Y_{O}^1} & \ddots & 0 & 	\sigma_{{\varepsilon}, Y_{O}^1}^2 &  \ddots	 & 0  & 0  \\
			\vdots &  \vdots &  \ddots &  \vdots & \vdots & \ddots & \ddots & \ddots & \ddots & \ddots & \ddots & \vdots  \\
			0&  0 & \cdots & 0 &0 & 0& \ddots &  \rho \sigma_{{\varepsilon}, Y_{T_d}^J}  \sigma_{{\varepsilon}, Y_{O}^J} & 0 & \ddots & 	\sigma_{{\varepsilon}, Y_{O}^J}^2 & 0 
			\\  0 & 0  & \cdots &0 & 0 & 0 &\cdots  & 0 & 0 & \cdots & 0 & \sigma_{{\varepsilon}, Y_{F}}^2 
		\end{pmatrix}.
	\label{matrixH}
	\end{equation} 
 Since the observation vector $\mathbf{Y}_t$ contains data measurements of the latent states, the measurement equation matrix $\mathbf{Z}$ contains 1's on the diagonal with the exception of the equation for OHC, where the heat capacity parameter $C_d$ enters. When $K=J=1$, the model collapses to the specification with one data source for the latent states. The EBM-SS model with multiple data sources is schematically depicted in Figure \ref{diagram}. 
	\section{Estimation and simulation of the EBM-SS}
\label{secsim}
\indent The parameters to be estimated in EBM-SS model include the four physical parameters $\lambda$, $\gamma$, $C_m$, and $C_d$; the parameters in the variance-covariance matrices of the state disturbances, ${\mathbf{Q}}$, and of the measurement errors, $\mathbf{H}$; and constants $\boldsymbol{\upmu}_{T_d}$ for the observations of $T_d$. These parameters are collected in the parameter vector $\boldsymbol{\uptheta}$. 
The estimate of  $\boldsymbol{\uptheta}$ is obtained by maximum likelihood, where the log-likelihood function is evaluated using the Kalman filter, see \citeA{durbin2012time}. We implement the EBM-SS model using the R package KFAS \cite{helske2017kfas} and optimize the log-likelihood value using a combination of ``solnp''\footnote{The ``solnp'' optimizer implements the general nonlinear augmented Lagrange multiplier method.} \cite{ye1988interior} and the Nelder-Mead simplex method \cite{nelder1965simplex}. 
 For initialization, we adopt the ``Big K'' technique (illustrated in \citeA{durbin2012time}) to approximate the diffuse initialization, and we set the initial variances of the non-stationary states as $10^6$.  \\
\indent We conduct a Monte Carlo simulation to explore the small-sample properties of the EBM-SS model. We choose a sample size of 66, the length of the historical dataset, and perform 1,000 simulation replications. We consider the EBM-SS base model, where there is one data source for each latent state, and the EBM-SS full model, where there are eight GMST series, two pairs of ocean temperature and OHC series, and one radiative forcing series, all of which are simulated in this exercise. We also allows for correlations between each pair of ocean temperature and OHC. For the EBM-SS full model, we employ the empirical estimates reported in panel A in Table \ref{empirical} as the data-generating parameters to simulate the data. First, we simulate the data for the EBM-SS full model, and we chose the first GMST, first ocean temperature, first OHC, and the radiative forcing series from this set of data as the simulated data for the EBM-SS base model. Hence, these two sets of simulated data share the same source of randomness. \\
\indent As shown in Equation \eqref{ddA}, the anthropogenic forcing state $A_t$ is of integration order 2, as it cumulates a stochastic trend $\beta_{t}$. Simulating the anthropogenic forcing series unrestrictedly inevitably generates trajectories with downward trends, which contradicts the pronounced upward trends observed in the historical series (Figure \ref{NAforcing}). To obtain simulated paths that are comparable with the observational records, we apply rejection sampling \shortcite{wells2004generalized} and only retain the $i$th simulated trajectory $\left\{Y^{i, \text{sim}}_{A,t}\right\}_{t=1}^T$ , if it satisfies that:
\begin{equation}
Y^{i, \text{sim}}_{A,\frac{T}{2}}\geq0.75Y_{A, {\frac{T}{2}}}\enspace \text{and} \enspace Y^{i, \text{sim}}_{A,T}\geq0.75Y_{A,{T}},
\end{equation} 
where $Y_{A, {\frac{T}{2}}}$ and $Y_{A, T}$ are the mid-point and the endpoint in the historical anthropogenic forcing series from \citeA{ha06510a}. This ensures that the simulated trajectories are consistent with the historical series in trending upward. \\
\indent We calculate equilibrium climate sensitivity (ECS) using the relationship as in, e.g., the IPCC Sixth Assessment Report (AR6) \cite[Chapter 7]{ipcc2021c7} : $\text{ECS}=\frac{F_{2 \times \mathrm{CO}_{2}}}{\lambda} $, where  $F_{2 \times \mathrm{CO}_{2}}$ is the radiative forcing in response to a doubling of the $\mathrm{{CO}}_{2}$ concentrations in the atmosphere. We use the updated best estimate of $F_{2 \times \mathrm{CO}_{2}} \approx 3.93$ $\left(\pm0.47,\enspace5\%-95\%\enspace \text{CI}\right) \mathrm{~W} \mathrm{~m}^{-2}$ from the IPCC AR6 \cite[Chapter 7]{ipcc2021c7}. \\
\begin{table}[ht!]
	\centering
	\caption{\footnotesize Data-generating parameter (DGP) values, estimation biases, standard deviations, and root mean squared errors (RMSEs), and mean absolute errors (MAEs) of the Monte Carlo simulation for the EBM-SS base model and EBM-SS full model. Here, $\sigma_{\varepsilon,{Y_{T_m}^k}}^2$, $\sigma_{\varepsilon,{Y_{T_d}^j}}^2$, and $\sigma_{\varepsilon,{Y_{O}^j}}^2$ denote the variance of measurement error of the $k$th GMST series, of the $j$th ocean temperature series, and of the $j$th OHC series, respectively.}
	\setlength{\tabcolsep}{1pt} 
	\setstretch{1.2}
	\begingroup\scriptsize
	\begin{tabular}{l|cccccccccccccc}
		\hline \hline 
		\multicolumn{10}{c}{\footnotesize \textbf{EBM-SS base model}}\\
		\hline
		&  \multicolumn{5}{c}{\footnotesize \textbf{physical parameters}} &  \multicolumn{4}{c}{\footnotesize \textbf{variances of state disturbance}}& \\
		\cmidrule(lr){2-6}\cmidrule(lr){7-10}
		& $\lambda$ & $\gamma$ & $C_m$ & $C_d$ & $\text{ECS}$&$\sigma_{\eta,{T_m}}^2$ & $\sigma_{\eta,{T_d}}^2$ & $\sigma_{\eta,A}^2$ & $\sigma_{\eta,\beta}^2$  \\ 
		\hline
  		{DGP value} & 1.0828 & 1.3027 & 9.6376 & 98.4886 & 3.6294 & 0.0122 & $3.66 \times 10^{-5}$ & $4.73 \times 10^{-5}$ & $9.72 \times 10^{-6}$ \\ 
  	{estimation bias} & 0.0139 & 0.0308 & 0.2696 & $-0.0279$ & 0.1993 & $-0.0012$ & $-8.59 \times 10^{-6}$ & $-1.82 \times 10^{-5}$ & $6.51 \times 10^{-6}$ \\ 
  {standard deviation} & 0.2745 & 0.2674 & 2.6176 &  0.8397 & 1.0595 & 0.00354 & $3.54 \times 10^{-5}$ & $1.72 \times 10^{-5}$ & $6.67 \times 10^{-6}$ \\ 
  	{RMSE}  & 0.2747 & 0.2691 & 2.6301 &  0.8397 & 1.0776 & 0.00375 & $3.65 \times 10^{-5}$ & $2.5 \times 10^{-5}$ & $9.32 \times 10^{-6}$ \\ 
  	{MAE} &  0.2137 & 0.2049 & 1.9236 &  0.6438 & 0.7705 & 0.00295 & $2.33 \times 10^{-5}$ & $2.03 \times 10^{-5}$ & $7.2 \times 10^{-6}$ \\
		\hline 
		& \multicolumn{8}{c}{\footnotesize \textbf{variances of measurement errors (I)}} \\
		\cmidrule(lr){2-9}
		&$\sigma_{\varepsilon,{Y_{T_m}^1}}^2$  & $\sigma_{\varepsilon,{Y_{T_m}^2}}^2$& $\sigma_{\varepsilon,{Y_{T_m}^3}}^2$ & $\sigma_{\varepsilon,{Y_{T_m}^4}}^2$ & $\sigma_{\varepsilon,{Y_{T_m}^5}}^2$& $\sigma_{\varepsilon,{Y_{T_m}^6}}^2$ & $\sigma_{\varepsilon,{Y_{T_m}^7}}^2$& $\sigma_{\varepsilon,{Y_{T_m}^8}}^2$  &  \\
		\cline{1-9}
			{DGP value} & 0.0010 &  &  &  &  &  &  & \\ 
  	{estimation bias} & 0.00057 &  &  &  &  &  &  &  \\ 
  	{standard deviation} & 0.0020 &  &  &  &  &  &  &  \\ 
  	{RMSE} & 0.0021 &  &  &  &  &  &  &  \\ 
  	{MAE} & 0.0016 &  &  &  &  &  &  & \\
		\hline 
		& \multicolumn{5}{c}{\footnotesize \textbf{variances of measurement errors (II)}} &  \multicolumn{2}{c}{\footnotesize \textbf{constant of $T_d$}} &  \multicolumn{2}{c}{\footnotesize \textbf{$\boldsymbol{\rho}_{T_d,O}$}}\\
		\cmidrule(lr){2-6} 	\cmidrule(lr){7-8} 	\cmidrule(lr){9-10}
		& $\sigma_{\varepsilon,{Y_{T_d}^1}}^2$ & $\sigma_{\varepsilon,{Y_{T_d}^2}}^2$ & $\sigma_{\varepsilon,{Y_{O}^1}}^2$ & $\sigma_{\varepsilon,{Y_{O}^2}}^2$ & $\sigma_{\varepsilon,Y_{F}}^2$  & $\mu_{T_d,1}$ &  $\mu_{T_d,2}$ & $\rho_{1}$& $\rho_{2}$ \\ 
		\hline	
	{DGP value}  	& 0.00014 &  &  1.5311 &  & $1.61 \times 10^{-11}$ & $-0.2738$ &  &  0.9092 &  \\ 
  	{estimation bias} & $8.77 \times 10^{-6}$ &  & $-1.2819 $&  & $5.29 \times 10^{-6}$ &  $-0.0004$ &  &$ -0.0059 $&  \\ 
  	{standard deviation} & $3.93 \times 10^{-5}$ &  &  0.0479 &  & $5.81 \times 10^{-6}$ &  0.0751 &  &  0.1339 &  \\ 
  	{RMSE} & $4.03 \times 10^{-5}$ &  &  1.2828 &  & $7.85 \times 10^{-6}$ &   0.0705 &  &  0.1339 &  \\ 
  	{MAE}& $3.10 \times 10^{-5}$ &  &  1.2819 &  & $5.29 \times 10^{-6}$ &  0.0586 &  &  0.0316 &  \\ 
		\hline \hline
			\multicolumn{10}{c}{\footnotesize }\\
			\multicolumn{10}{c}{\footnotesize \textbf{EBM-SS full model}}\\
		\hline
		&  \multicolumn{5}{c}{\footnotesize \textbf{physical parameters}} &  \multicolumn{4}{c}{\footnotesize \textbf{variances of state disturbance}}& \\
		\cmidrule(lr){2-6}\cmidrule(lr){7-10}
		& $\lambda$ & $\gamma$ & $C_m$ & $C_d$ & $\text{ECS}$&$\sigma_{\eta,{T_m}}^2$ & $\sigma_{\eta,{T_d}}^2$ & $\sigma_{\eta,A}^2$ & $\sigma_{\eta,\beta}^2$  \\ 
		\hline
 {DGP value} & 1.0828 &  1.3027 &  9.6376 & 98.4886 & 3.6294 & 0.0122 & $3.66 \times 10^{-5}$ & $4.73 \times 10^{-5}$ & $9.72 \times 10^{-6}$ \\ 
   {estimation bias} & 0.0309 & $-0.0205$ & $-0.1332$ &$ -0.0021$ & 0.1166 & $-0.00035$ & $-3.80 \times 10^{-6}$ & $-1.81 \times 10^{-5}$ & $6.52 \times 10^{-6}$ \\ 
   {standard deviation} & 0.2686 &  0.2695 &  2.2429 &  0.2052 & 0.9639 & 0.00211 & $2.54 \times 10^{-5}$ & $1.71 \times 10^{-5}$ & $6.65 \times 10^{-6}$ \\ 
  {RMSE} & 0.2703 &  0.2702 &  2.2458 &  0.2051 & 0.9705 & 0.00214 & $2.56 \times 10^{-5}$ & $2.49 \times 10^{-5}$ & $9.31 \times 10^{-6}$ \\ 
  {MAE} & 0.2093 &  0.1985 &  1.7106 &  0.1587 & 0.7174 & 0.00169 & $1.55 \times 10^{-5}$ & $2.02 \times 10^{-5}$ & $7.2 \times 10^{-6}$ \\ 
		\hline 
		& \multicolumn{8}{c}{\footnotesize \textbf{variances of measurement errors (I)}} \\
		\cmidrule(lr){2-9}
		&$\sigma_{\varepsilon,{Y_{T_m}^1}}^2$  & $\sigma_{\varepsilon,{Y_{T_m}^2}}^2$& $\sigma_{\varepsilon,{Y_{T_m}^3}}^2$ & $\sigma_{\varepsilon,{Y_{T_m}^4}}^2$ & $\sigma_{\varepsilon,{Y_{T_m}^5}}^2$& $\sigma_{\varepsilon,{Y_{T_m}^6}}^2$ & $\sigma_{\varepsilon,{Y_{T_m}^7}}^2$& $\sigma_{\varepsilon,{Y_{T_m}^8}}^2$  &  \\
		\cline{1-9} 	   
    {DGP value} & 0.0010 & 0.00090 & 0.00241 & 0.0118 & 0.00282 & 0.00659 & 0.00074 & 0.00026 \\ 
   {estimation bias}  & $-7.18 \times 10^{-6}$ & $1.8 \times 10^{-5}$ & $-4.65 \times 10^{-6}$ & 0.00023 & $9.68 \times 10^{-5}$ & $-6.97 \times 10^{-5}$ & $-1.97 \times 10^{-6}$ & $-2.44 \times 10^{-6}$ \\ 
   {standard deviation} & 0.00021 & 0.00018 & 0.00045 & 0.0021 & 0.00049 & 0.0012 & 0.00016 & $8.73 \times 10^{-5}$ \\ 
  {RMSE} & 0.00021 & 0.00019 & 0.00045 & 0.0021 & 0.00050 & 0.00115 & 0.00016 & $8.73 \times 10^{-5}$ \\ 
  {MAE} & 0.00017 & 0.00015 & 0.00035 & 0.00163 & 0.00040 & 0.00091 & 0.00013 & $6.97 \times 10^{-5}$ \\ 
		\hline 
		& \multicolumn{5}{c}{\footnotesize \textbf{variances of measurement errors (II)}} &  \multicolumn{2}{c}{\footnotesize \textbf{constant of $T_d$}} &  \multicolumn{2}{c}{\footnotesize \textbf{$\boldsymbol{\rho}_{T_d,O}$}}\\
		\cmidrule(lr){2-6} 	\cmidrule(lr){7-8} 	\cmidrule(lr){9-10}
		& $\sigma_{\varepsilon,{Y_{T_d}^1}}^2$ & $\sigma_{\varepsilon,{Y_{T_d}^2}}^2$ & $\sigma_{\varepsilon,{Y_{O}^1}}^2$ & $\sigma_{\varepsilon,{Y_{O}^2}}^2$ & $\sigma_{\varepsilon,Y_{F}}^2$  & $\mu_{T_d,1}$ &  $\mu_{T_d,2}$ & $\rho_{1}$& $\rho_{2}$ \\ 
		\hline
   	{DGP value} & 0.00014 & 0.00015 & 1.5311 & 1.2805 & $1.61 \times 10^{-11}$ & $-0.2738$ & $-0.2802$ & 0.9092 & 0.9943 \\ 
  	{estimation bias} & $2.41 \times 10^{-6}$ & $2.43 \times 10^{-5}$ & 0.0054 & 0.2368 & $5.29 \times 10^{-6}$ & $-0.0015$ & $-0.0021$ &  0.0004 &  0.0007 \\ 
  	{standard deviation} & $2.96 \times 10^{-5}$ & $3.49 \times 10^{-5}$ & 0.3195 & 0.3129 & $5.78 \times 10^{-6}$ &  0.6907 &  0.6908 & 0.0246 & 0.0016 \\ 
   {RMSE} & $2.97 \times 10^{-5}$ & $4.25 \times 10^{-5}$ & 0.3194 & 0.3923 & $7.84 \times 10^{-6}$ &  0.6904 &  0.6905 & 0.0246 & 0.0017 \\ 
   {MAE} & $2.28 \times 10^{-5}$ & $3.35 \times 10^{-5}$ & 0.2462 & 0.3111 & $5.29 \times 10^{-6}$ &  0.1099 &  0.1098 &  0.019 & 0.0013 \\ 
		\hline \hline
	\end{tabular}
	\endgroup
	\label{sim}
\end{table}
\begin{figure}[h!]
	\centering
	\caption{\footnotesize Simulated distributions of $\theta_i -{\theta}_i^0$ for the five physical parameters from 1,000 Monte Carlo simulations of the EBM-SS base model and the EBM-SS full model. ``d'' in the titles denotes the difference between the estimate in the simulation and the data-generating value, i.e., the bias. The red, blue, and yellow vertical lines represent the data-generating parameter values, the means, and the medians of the 1,000 estimates from the simulation, respectively.}
	\centering
	\begin{subfigure}{\textwidth}
	\subcaption{\footnotesize Base model}
	\includegraphics[width=\linewidth]{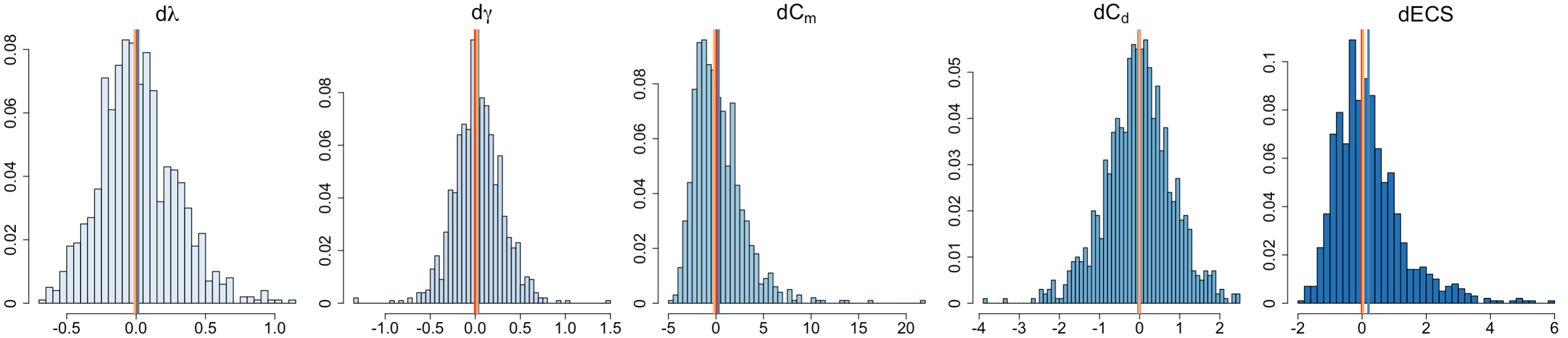}
\end{subfigure}\hfill
		\begin{subfigure}{\textwidth}
		\subcaption{\footnotesize Full model}
		\includegraphics[width=\linewidth]{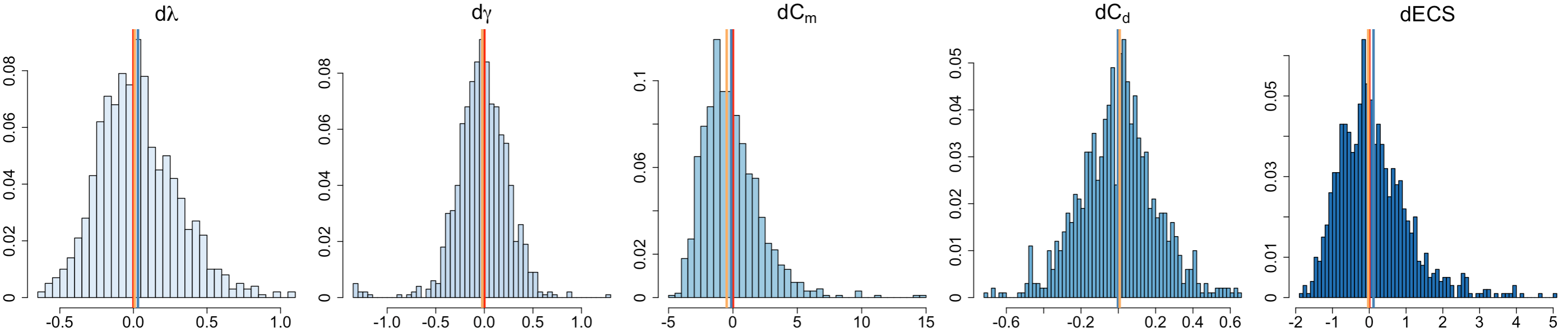}
	\end{subfigure}\hfill
	\label{MC}
\end{figure}
\indent Table \ref{sim} reports data-generating parameter values,  estimation biases, standard deviations, root mean squared errors (RMSEs), and mean absolute errors (MAEs) of the simulation exercise. Figure \ref{MC} shows the distributions of the deviations of the physical parameter estimates in the simulation relative to the true values (denoted as $d\cdot$), which are centered around zero. Among these physical parameters, $d\gamma$ and $dC_d$ exhibit the feature of normality, and $dC_d$ appears much less dispersed for the full model. Under both models, $d\lambda$ is slightly positive-skewed. Due to the inverse relationship between ECS and $\lambda$, the small magnitudes of the estimates of $\lambda$ produce large ECS estimates, thus the long right tail in $d$ECS. There are positive skewness  and a long right tail in $dC_m$ as a result of a few large outliers. The small values of the deviations measures in Table \ref{sim}, together with the distributions in Figure \ref{MC}, demonstrate good finite-sample properties of the EBM-SS model.\\
\indent Comparing the simulation results for the base model and for the full model in Table \ref{sim}, we note that estimation biases, standard deviations, RMSEs, and MAEs decrease, as we include more data series. This provides evidence that including multiple data sources helps decrease the estimation uncertainty.
\section{Data}
			\label{sec4}
						\label{measurement}
		This section presents the historical records of anomalies employed in the empirical investigation and the synchronization of these anomalies to a common baseline. For the empirical analysis in this paper, we have collected eight observational GMST datasets from separate research groups, three pairs of ocean temperature and OHC data series, and one effective radiative forcing series (summarized in Table \ref{tableData}) as the measurements for the latent processes in the state equation \eqref{stateEqn} of the EBM-SS model. 
		\begin{table}[h!]
		\caption{\footnotesize Summary of the anomaly datasets employed in the empirical analysis. The last column ``baseline'' indicates the reference period or the year upon which the anomalies are constructed.}
		\small
		\setlength{\tabcolsep}{2.7pt} 
		\renewcommand{\arraystretch}{1.3} 
		\begin{threeparttable}
			\begin{tabular}{l|ll|l|ll}
			\hline\hline
			\footnotesize \textbf{Variable}& 	\footnotesize \textbf{Acronym/Type} &\footnotesize \textbf{Institution/Authors}&\footnotesize \textbf{ Coverage}& \footnotesize \textbf{Baseline}\\
			\hline
			\footnotesize \multirow{8}{*}{\textbf{GMST Anomalies}} &	\footnotesize  \href{https://data.giss.nasa.gov/gistemp/}{GISTEMP}  &	\footnotesize NASA &	\footnotesize1880 -- 2020 & 	 	\footnotesize 1951 -- 1980\\
			&	\footnotesize \href{https://www.ncdc.noaa.gov/cag/global/time-series/globe/land_ocean/ytd/12/1880-2019}{NOAAGlobalTemp}&	\footnotesize NOAA &	\footnotesize1880 -- 2019 & 	\footnotesize  1901 -- 2000\\
			&	\footnotesize \href{https://www.metoffice.gov.uk/hadobs/hadcrut5/data/current/download.html}{HadCRUT5}&	\footnotesize Met Office Hadley Center &	\footnotesize1850 -- 2020 & 	\footnotesize  1961 -- 1990\\
			&	\footnotesize \href{http://berkeleyearth.org/data/}{BEST}&	\footnotesize Berkeley Earth &	\footnotesize1850 -- 2020 & 	\footnotesize 1951 -- 1980\\
			&	\footnotesize \href{https://www-users.york.ac.uk/~kdc3/papers/coverage2013/series.html}{CW14}&	\footnotesize Cowtan and Way(2014) &	\footnotesize1850 -- 2020 & 	\footnotesize 1961 -- 1990\\
			& 	\footnotesize \href{https://ds.data.jma.go.jp/tcc/tcc/products/gwp/temp/list/year_wld.html}{JMA} &	\footnotesize Japanese Meteorological Agency &	\footnotesize1891 -- 2020 & 	\footnotesize 1981 -- 2010 \\
			& 	\footnotesize \href{https://climate.copernicus.eu/2020-warmest-year-record-europe-globally-2020-ties-2016-warmest-year-recorded}{ERA-Interim}&	\footnotesize Copernicus &	\footnotesize1970 -- 2020 & 	\footnotesize pre-industrial $^\text{c}$  \\
			& 	\footnotesize \href{https://climate.copernicus.eu/2020-warmest-year-record-europe-globally-2020-ties-2016-warmest-year-recorded}{JRA-55}	&	\footnotesize Japanese Meteorological Agency &	\footnotesize1970 -- 2019 & 	\footnotesize pre-industrial $^\text{d}$ \\
				\hline
			\footnotesize \multirow{2}{2.8cm}{\textbf{Global Ocean Temperature Anomalies}} &  \footnotesize \href{https://www.ncei.noaa.gov/access/global-ocean-heat-content/basin_avt_data.html}{NOAA yearly, 0-700m} \footnotesize  &  \footnotesize NOAA & \footnotesize 1955 -- 2020& \footnotesize  not specified \\
			&   \footnotesize \href{http://www.ocean.iap.ac.cn/pages/dataService/dataService.html?languageType=en\&navAnchor=dataService}{IAP yearly 0-700m, }\href{http://www.ocean.iap.ac.cn/pages/dataService/dataService.html?languageType=en\&navAnchor=dataService}{0-2000m}  &  \footnotesize  Institute of Atmospheric Physics & \footnotesize 1940 -- 2020& \footnotesize 1981 -- 2010\\
			\hline
			\footnotesize \multirow{2}{2.8cm}{\textbf{Global OHC Anomalies}} 
			&  \footnotesize \href{https://www.ncei.noaa.gov/access/global-ocean-heat-content/basin_heat_data.html}{NOAA yearly, 0-700m} \footnotesize  &  \footnotesize NOAA & \footnotesize 1955 -- 2020& \footnotesize  not specified \\
			&   \footnotesize \href{http://www.ocean.iap.ac.cn/pages/dataService/dataService.html?languageType=en&navAnchor=dataService}{IAP yearly 0-700m, }\href{http://159.226.119.60/cheng/images_files/OHC2000m_annual_timeseries.txt}{0-2000m}   &  \footnotesize  Institute of Atmospheric Physics & \footnotesize 1940 -- 2020& \footnotesize 1981 -- 2010\\
			\hline 
			\footnotesize {\textbf{Forcing}}  &	\footnotesize \href{http://www.columbia.edu/~mhs119/Forcings/}{Effective Radiative Forcing}   & 	\footnotesize  Hansen et al. (2011) & 	\footnotesize 1850 -- 2018 & \footnotesize 1850\\
			\hline \hline
		\end{tabular}
	\begin{tablenotes}
		\footnotesize
		\item [c,d] The ERA 5 and JRA 55 yearly series are downloaded from the Copernicus Climate Change Service (C3S) Climate Data Store, which are processed according to \citeA{simmons2017reassessment}.  These two datasets have already been transformed with respect to the pre-industrial period. 
	\end{tablenotes}
		\end{threeparttable}
		\label{tableData}
	\end{table}\\
	\indent Our choice of the eight GMST datasets is in accordance with that in the IPCC Global Warming of $1.5^{\circ} \mathrm{C}$ report \shortcite{2019teallenchnical}, which includes: the GISS Surface Temperature Analysis (GISTEMP) \cite{GISS2021,lenssen2019improvements},  the NOAA Merged Land Ocean Global Surface Temperature Analysis (NOAAGlobalTemp) \cite{noaatemp2021}, HadCRUT5 by the Met Office Hadley Centre \cite{morice2020updated}, the Berkeley Earth Surface Temperatures Land + Ocean  (BEST) \cite{rohde2020berkeley}, the Cowtan-Way temperature series (CW2014) \cite{cowtan2014coverage}, JMA annual anomalies \cite{JMAtemp}, ERA-Interim reanalysis \cite{ERA}, and the Japanese 55-year reanalysis (JRA-55) \cite{kobayashi2015jra}. We assume that each of these datasets represents an independent assessment of the global mean temperature variations. \\
	\indent All of these eight GMST series are in the form of anomalies, i.e., they measure the departures from an average of the observations over a long period (called ``reference period'' or ``baseline'') that usually spans thirty years or longer. It is common practice to record and construct the anomalies rather than the absolute value of the observations \cite{hawkins2016connecting}.
		\begin{figure}[h]
		\caption{\footnotesize GMST, ocean temperature, and OHC anomaly series before synchronization. The light gray area corresponds to the time horizon 1955 -- 2020 in the empirical study.}
		\begin{subfigure}{0.49\textwidth}
			\subcaption{\footnotesize GMST Anomalies (${ }^{\circ} \mathrm{C}$) (1850 -- 2020)}
			\includegraphics[width=\linewidth]{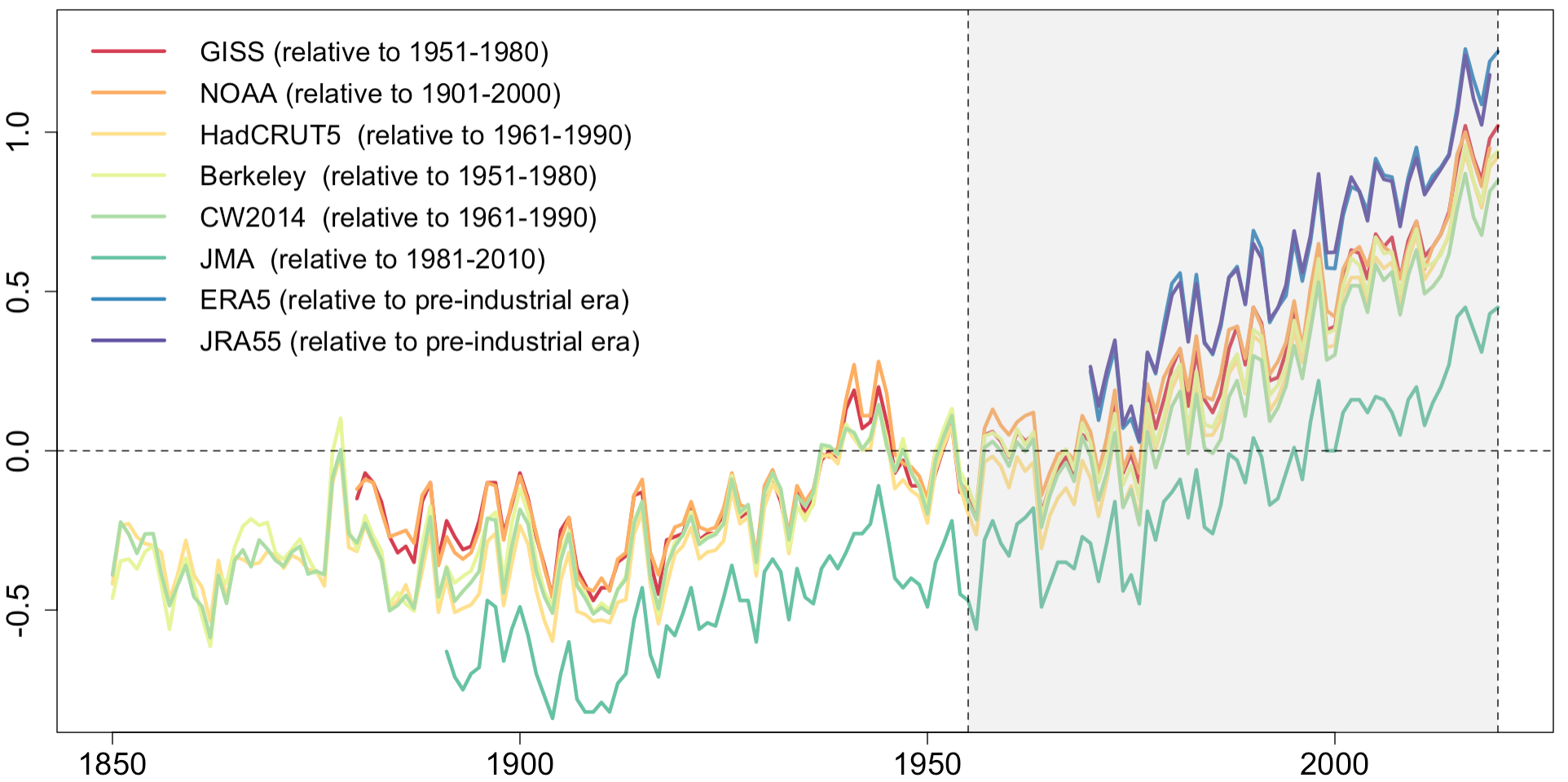}
			\label{TempAno}
		\end{subfigure}\hfill
		\begin{subfigure}{0.49\textwidth}
			\subcaption{\footnotesize ocean temperature (${ }^{\circ} \mathrm{C}$) (1940 -- 2020)}
			\includegraphics[width=\linewidth]{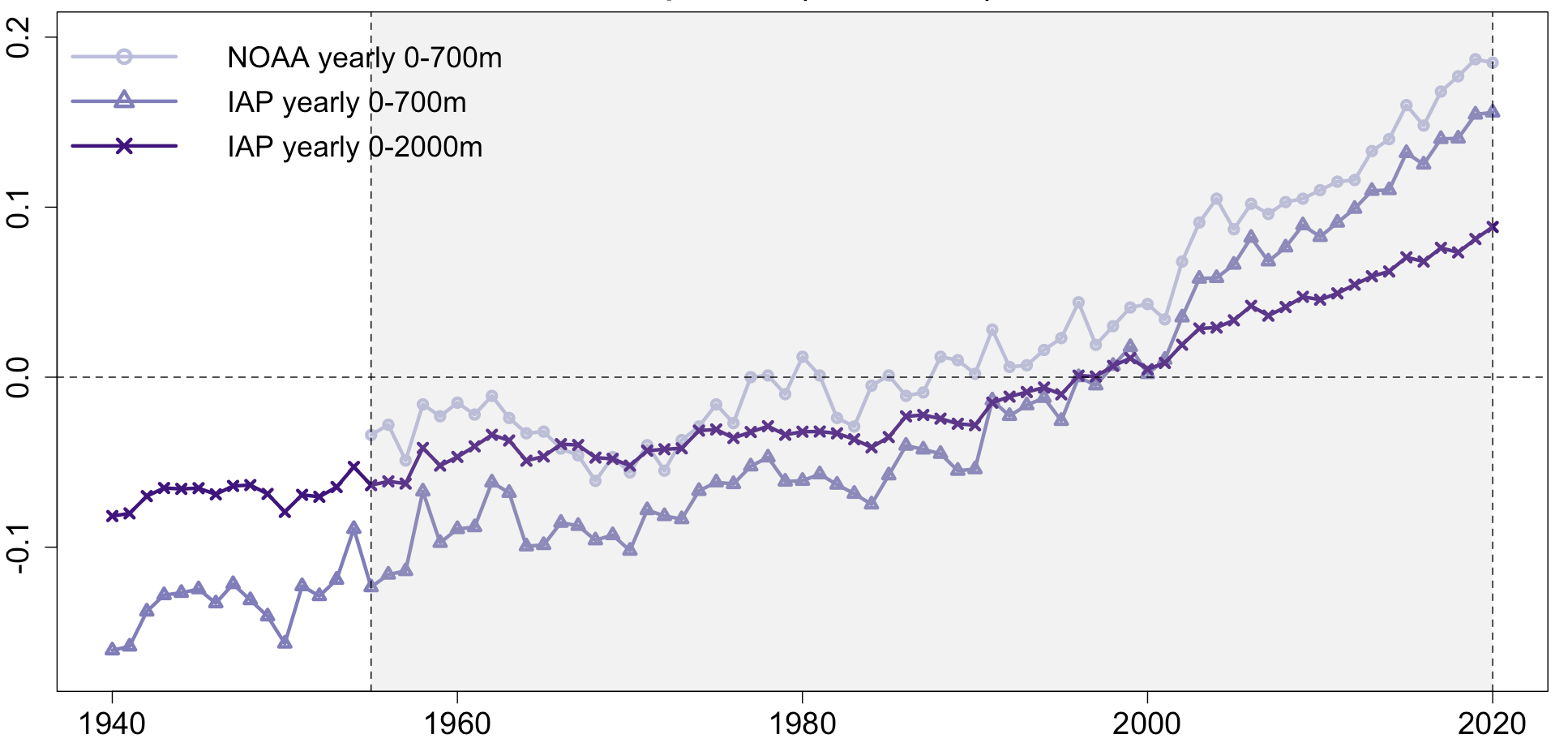}
			\label{OtempAno}
		\end{subfigure}\\
	\centering
		\begin{subfigure}{0.49\textwidth}
		\subcaption{\footnotesize OHC Anomalies (J $\text{m}^{-2}$) (1940 -- 2020)}
		\includegraphics[width=\linewidth]{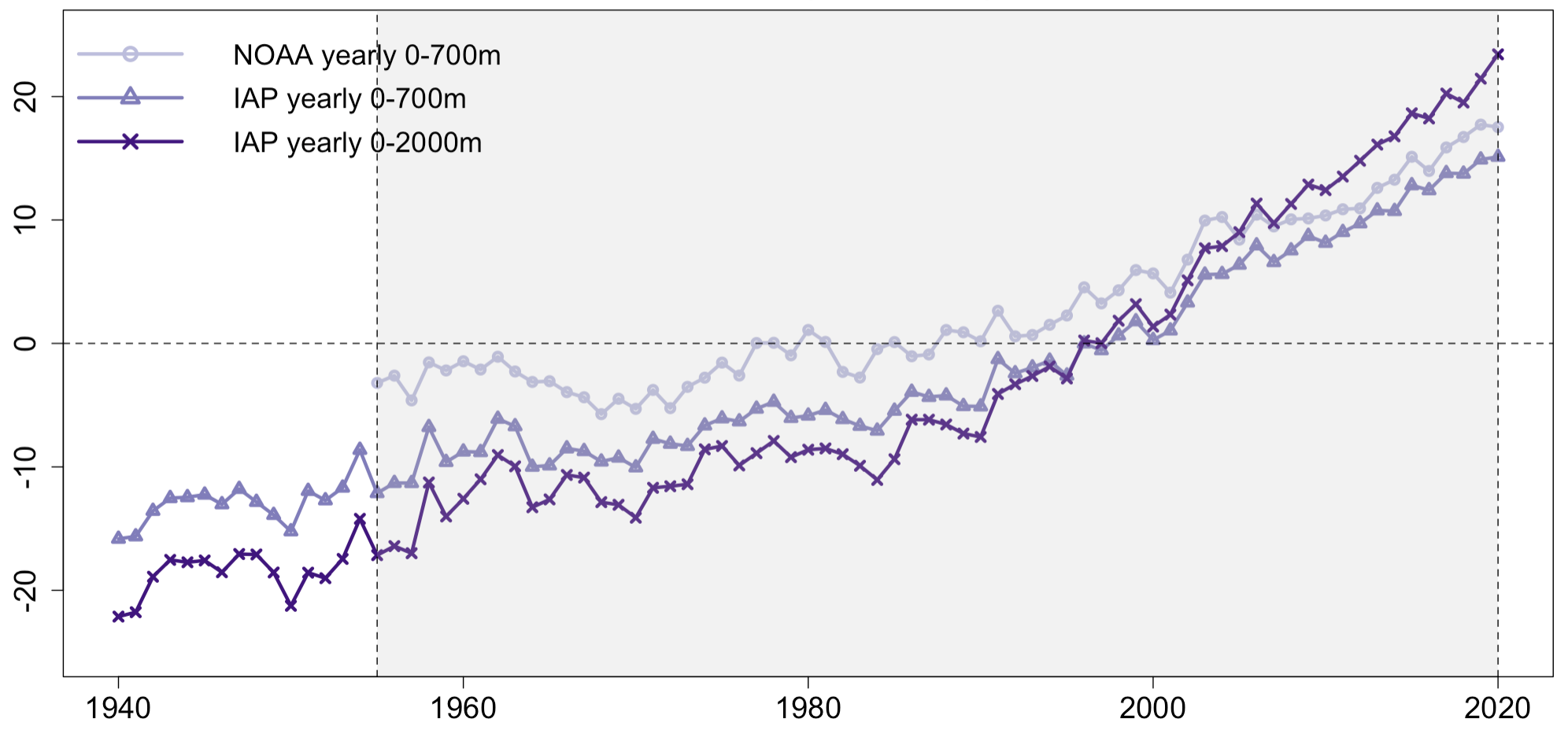}
		\label{OHCAno}
	\end{subfigure}
	\end{figure}\\
		\indent Figure \ref{TempAno} indicates that, despite having different lengths, these eight GMST series share similar upward trends and fluctuation trajectories, but the ranges they span vary to some extent, which is mainly due to the different reference periods.\\
	\indent  The datasets for the ocean (including both ocean temperature anomalies and OHC anomalies) are from two research bodies, NOAA National Centers for Environment Information (NOAA) and Institute of Atmospheric Physics (IAP), and cover 0-700 and 0-2,000 meter, respectively (see Table \ref{tableData} for details). The OHC anomaly series are estimated based on the in-situ subsurface ocean temperature measurements combined with salinity series \cite{levitus2012world}, and they have the same coverage and baseline as their ocean temperature counterparts. The 0-700m ocean series from  NOAA \cite{levitus2012world} is available since 1955, while the  IAP \cite{cheng2020improved} series begins in 1940. As we aim to only use the data based on direct observations, we choose 1955 -- 2020 as the time horizon for the empirical study. While the baseline periods for NOAA ocean temperature and OHC anomalies are not documented, IAP benchmarks the ocean series against the 30-year period from 1981 to 2010. \\
	\indent Figure \ref{OHCAno} shows that all of the OHC series agree on a warming trend but exhibit slightly different yearly variations. According to \citeA{von2020heat}, the global heating rate over the period 1971 -- 2018 is estimated as $0.47 \pm 0.1$ $\mathrm{ Wm}^{-2}$, 89\% of which is contributed by the global ocean (thereof 52\% from the layer 0-700m, 28\% from the 700-2,000m, and 9\%  beyond 2,000m). \\
	\indent We test the stationarity properties of the GMST, ocean temperature, and OHC anomalies, and we report the results in Table \ref{ADF_temp} in Appendix \ref{unitroot}. They indicate that all of these anomaly series are $I(1)$ processes. As shown in Table \ref{ADF}, total radiative forcing is also $I(1)$, and therefore all of the measurements we consider share the same integration order.\\
	\indent As shown in Table \ref{tableData}, the data series have different baselines. The parameter estimates can be distorted if we include several data series with different reference periods simultaneously in a system. Therefore, it is necessary to reconcile these datasets to the same baseline by either synchronizing them before fitting into the model or introducing some structures in the model specification to offset the discrepancies across various baselines. The differences in the baselines across different data sources can be eliminated by synchronizing these anomalies to a common reference period. The simple mathematical arguments for synchronization is given in Appendix \ref{irrelavance}. \\
 \indent  The pre-industrial era is a natural choice for this common baseline, as it is commonly used as a benchmark to measure and evaluate climate change. We follow the IPCC Global Warming of $1.5^{\circ} \mathrm{C}$ report \shortcite{2019teallenchnical} to specify 1850 -- 1900 as the pre-industrial base period for GMSTs. The synchronized GMST series are exhibited in Figure \ref{TempSyn1} in Appendix \ref{irrelavance}. The forcing series we employ in the paper is already benchmarked against 1850. To maintain consistency with the pre-industrial benchmark of GMSTs, we synchronize the anthropogenic forcing as anomalies relative to 1850 -- 1900. \\
	 \indent It is infeasible to synchronize the ocean series relative to the pre-industrial era, as the ocean information during this period is very sparse. We benchmark the NOAA ocean series against 1981 -- 2010 for comparability with IAP series, as shown in Figures \ref{OtempSyn} and \ref{OHCSyn} in Appendix \ref{irrelavance}. What remains to be accommodated is the baseline difference between 1981 -- 2010 and the pre-industrial era. A solution we consider here is to introduce a constant ${\upmu}_{T_d}^j$ to the $j$th  measurement equation of ocean temperature and to the $j$th measurement equation of the corresponding OHC accordingly:
\begin{equation}
	\begin{aligned}
			{Y}_{T_d,t}^j=&{\upmu}_{T_d}^j+T_{d,t}+\varepsilon^j_{T_d,t},\enspace \\
	{Y}_{O,t}^j=&C_d{\upmu}_{T_d}^j+C_dT_{d,t}+\varepsilon^j_{O,t},\enspace(j=1,2,\cdots J).
	\end{aligned}
\end{equation}
	\section{Empirical results}
	\label{secEst}
	\indent In this section, we fit the EBM-SS model defined in Section \ref{sec3} to the datasets described in Section \ref{sec4}. The sample period is 1955 -- 2020. 
	  The ERA-Interim and JRA-55 GMST anomalies have a shorter length of 51 years, from 1970 to 2020,  but the missing observations during 1955 -- 1969 are treated in the state space model using the techniques illustrated in \citeA{durbin2012time}. 
	\subsection{Parameter estimates by fitting historical observations}
	\label{secPara}
	 Table \ref{empirical} shows the parameter estimates using all of the eight GMST anomalies presented in Section \ref{sec4}. Using the delta method, we account for the uncertainty of the estimates of both $\lambda$ and $F_{2 \times \mathrm{CO}_{2}}$ when calculating the standard error for the ECS estimate, where $F_{2 \times \mathrm{CO}_{2}}$ denotes the radiative forcing in response to a doubling of the atmospheric CO$_2$ concentrations. The estimation results with only one data source for each of the latent states are presented in Table \ref{Empirical700+2000} in Appendix \ref{Appendix1GMST1Ocean}.  \\
	\begin{table}[h!]
	\centering
\caption{\footnotesize Parameter estimates from fitting the EBM-SS model to eight synchronized GMST and the radiative forcing series  from \protect \citeA{ha06510a}. Panel A includes two pairs of ocean temperature and OHC series covering 0-700m from NOAA $\left({Y_{T_d}^\text{NOAA}}, {Y_O^\text{NOAA}} \right)$ and IAP $\left({Y_{T_d}^\text{IAP}}, {Y_O^\text{IAP}} \right)$. Panel B includes the ocean temperature and OHC series covering 0-2,000m from IAP. All of the GMST anomalies and the ocean series have been synchronized relative to the pre-industrial period and 1981 -- 2010, respectively, where the constants for the ocean series offset the baseline difference. The standard errors for estimates are obtained using the delta method and presented in parentheses. $\hat{\mathbf{H}}$ and $\hat{\mathbf{Q}}$ denote the estimated variance-covariance matrices of the measurement errors and state disturbances, respectively, and $\hat{\mathcal{L}}$ denotes the maximized log-likelihood. $\boldsymbol{\rho}^\text{NOAA}_{{\varepsilon_{Y_{T_d}}},{\varepsilon_{O}}}$ and $\boldsymbol{\rho}^\text{IAP}_{{\varepsilon_{Y_{T_d}}},{\varepsilon_{O}}}$ denote the correlations between the ocean temperature and OHC series from NOAA and from IAP, respectively.}
	\scriptsize
	\begin{threeparttable}
		\setlength{\tabcolsep}{2pt} 
		\renewcommand{\arraystretch}{1.3} 
		\begin{tabular}{lcp{6mm}|cp{9.2mm}|cp{20mm}|cp{16mm}|ll}
			\hline\hline
			\multicolumn{10}{c}{\textbf{\footnotesize A. ocean temperature and OHC 0 -- 700m from both NOAA and IAP included}}\\
			\hline
		 & \multicolumn{2}{c|}{\footnotesize \textbf{phys. para.} }& \multicolumn{2}{c|}{ \footnotesize \textbf{$\hat{\boldsymbol{\mu}}_{Y_{T_d}} $ } } &  \multicolumn{2}{c|}{ \footnotesize\textbf{elements in   $\hat{\mathbf{H}}$} }&  \multicolumn{2}{c|}{\footnotesize \textbf{diagonal of   $\hat{\mathbf{Q}}$} } &\footnotesize \textbf{estimated linear relationships}& \\ 
			\hline
		  &  $\hat{\boldsymbol{\lambda}}$ & 1.08\newline (0.25)& {$\hat{\boldsymbol{\mu}}_{Y_{T_d}^\text{NOAA}} $ }  & $-0.27 \newline (0.09)$ & $\boldsymbol{\sigma}_{\varepsilon, Y_{T_m}}^{2}$& $0.0003\sim 0.012 \newline (0.0001)\enspace  (0.002)$ & $\boldsymbol{\sigma}_{{\eta}, T_m}^{2}$&$ 0.012\newline (0.002)$& \\
			&  $\hat{\boldsymbol{\gamma}}$ & 1.30 \newline (0.34) &{$\hat{\boldsymbol{\mu}}_{Y_{T_d}^\text{IAP}} $} &$-0.28 \newline (0.09)$ & $\boldsymbol{\sigma}_{{\varepsilon}, Y_{T_d}^\text{NOAA}}^{2}$ & $0.00014 \newline (0.00005)$ & $\boldsymbol{\sigma}_{{\eta}, T_d}^{2}$& $0.00004\newline (0.00002)$ & $T_{m,t}=  \underset{(0.09)}{0.75 }T_{m,t-1}+\underset{(0.06)}{0.14 }O_{t-1}+\underset{(0.03)}{0.10}F_{t-1}+\eta_{T_m, t}$  \\ 
			&  $\hat{\boldsymbol{C}}_m$ & {9.64\newline (2.86)}& & & $\boldsymbol{\sigma}_{{\varepsilon}, Y_{T_d}^\text{IAP}}^{2}$  & $0.00015 \newline (0.00005)$ & $\boldsymbol{\sigma}_{{\eta}, A}^{2}$ & 0.00005\newline (0.00001)&  $T_{d,t}= \underset{(0.003)}{0.01}T_{m,t-1}+\underset{(0.003)}{0.99}T_{d,t-1}+\eta_{T_d, t}$  \\ 
			&  $\hat{\boldsymbol{C}}_d$ & 98.49 \newline (0.26) &  &  &$\boldsymbol{\sigma}_{{\varepsilon}, Y_{O}^\text{NOAA}}^{2}$ & $1.53 \newline (0.05)$ & $\boldsymbol{\sigma}_{{\eta}, \beta}^{2}$& $0.00001\newline(0.000006)$  & &  \\ 
			&  $\hat{\text{ECS}}$ &3.63\newline (0.89)& &&$\boldsymbol{\sigma}_{{\varepsilon}, Y_{O}^\text{IAP}}^{2}$   &1.28 \newline (0.008) & &&  \\ 
			& & &&  & $\boldsymbol{\sigma}_{{\varepsilon}, Y_F}^{2}$   &$1.6\times10^{-11}$ \newline ($1.9\times10^{-17}$ )  && &\\
				& & &  & & $\boldsymbol{\rho}^\text{NOAA}_{{\varepsilon_{Y_{T_d}}},{\varepsilon_{O}}}$   &$0.909$ \newline ($0.034$ )  && &\\
					\cline{4-5}
					& & & $\hat{\mathcal{L}}$  & 1240.30 & $\boldsymbol{\rho}^\text{IAP}_{{\varepsilon_{Y_{T_d}}},{\varepsilon_{O}}}$  &$0.994$ \newline ($0.002$ )  && &\\
			\hline\hline
			\multicolumn{10}{c}{ \textbf{\footnotesize B. ocean temperature and OHC 0 -- 2,000m from IAP included}}
			\\
				\hline
			& \multicolumn{2}{c|}{ \footnotesize\textbf{phys. para.} }& \multicolumn{2}{c|}{\footnotesize \textbf{$\hat{\boldsymbol{\mu}}_{Y_{T_d}} $ } } &  \multicolumn{2}{c|}{ \footnotesize\textbf{elements in  $\hat{\mathbf{H}}$} }&  \multicolumn{2}{c|}{\footnotesize \textbf{diagonal of   $\hat{\mathbf{Q}}$} } &\footnotesize \textbf{estimated linear relationships}& \\ 
			\hline
			&  $\hat{\boldsymbol{\lambda}}$ & 0.66\newline (0.31)& {$\hat{\boldsymbol{\mu}}_{Y_{T_d}^\text{NOAA}} $ }  & $-0.24 \newline (0.09)$ & $\boldsymbol{\sigma}_{\varepsilon, Y_{T_m}}^{2}$& $0.0003\sim 0.012 \newline (0.0001)\enspace  (0.002)$ & $\boldsymbol{\sigma}_{{\eta}, T_m}^{2}$& {0.012  \newline(0.002)} & \\
			&  $\hat{\boldsymbol{\gamma}}$ & 1.82 \newline (0.45) & & & $\boldsymbol{\sigma}_{{\varepsilon}, Y_{T_d}^\text{IAP}}^{2}$ & $0.00001\newline (0.000006)$ & $\boldsymbol{\sigma}_{{\eta}, T_d}^{2}$& $0.00001\newline (0.000008)$ & $T_{m,t}=  \underset{(0.10)}{0.73 }T_{m,t-1}+\underset{(0.08)}{0.19 }O_{t-1}+\underset{(0.03)}{0.11}F_{t-1}+\eta_{T_m, t}$  \\ 
			&  $\hat{\boldsymbol{C}}_m$ & {9.35\newline (2.61}& & & $\boldsymbol{\sigma}_{\varepsilon, Y_O^\text{IAP}}^{2}$ & $0.70 \newline (0.002))$ & $\boldsymbol{\sigma}_{{\eta}, A}^{2}$& 0.00005\newline(0.00001) &  $T_{d,t}= \underset{(0.002)}{0.01}T_{m,t-1}+\underset{(0.002)}{0.99 }T_{d,t-1}+\eta_{T_d, t}$  \\ 
			
			&  $\hat{\boldsymbol{C}}_d$ & 269.30 \newline (0.42) &  &  &$\boldsymbol{\sigma}_{\varepsilon, Y_F}^{2}$& $3.8\times10^{-9} \newline (7.8\times10^{-9})$ & $\boldsymbol{\sigma}_{{\eta}, \beta}^{2}$& 0.00001\newline(0.000006)  &  &  \\ 
			\cline{4-5}
			&  $\hat{\text{ECS}}$ &5.91\newline (2.77)& $\hat{\mathcal{L}}$ & 1153.51 &  $\boldsymbol{\rho}^\text{IAP}_{{\varepsilon_{Y_{T_d}}},{\varepsilon_{O}}}$  &$0.985$ \newline ($0.009$ )&  & &  \\ 
			\hline\hline
		\end{tabular}
	\end{threeparttable}
	\label{empirical}
\end{table}
\indent Of the four physical parameters ($\lambda$, $\gamma$, $C_m$, and $C_d$) in the two-component EBM, the estimates of the climate feedback parameter $\lambda$, of the coefficient of heat transfer $\gamma$, and of the heat capacity of the deep ocean layer $C_d$ show pronounced increases when 0-2,000m ocean data series are employed, which account for more ocean heat uptake. The last column in Table \ref{empirical} reports the estimated two linear relationships specified by the EBM-SS model, i.e., between the state $T_{m,t}$ and the lag terms of the three latent states $T_{m,t-1}$, $T_{d,t-1}$, and $F_{t-1}$,  and between the state $T_{d,t}$ and the lag terms $T_{m,t-1}$ and $T_{d,t-1}$.  The estimated coefficients are similar regardless of whether ocean 0-700m or 0-2,000m data are employed.\\
\indent Table \ref{ECScomparison} compares the estimates of the physical parameters from the EBM-SS model with those from other studies. Comparing panel A and panel B reveals that, when fitting the eight GMST series and 0-700m ocean datasets from NOAA and IAP, our estimates are comparable to those obtained by the CMIP $4\times$CO$_2$ experiment data \shortcite[Chapter 7 supplementary material]{cummins2020optimal,ipcc2021supp7}. Our estimate of the ECS is 3.63${ }^{\circ} \mathrm{C}$, which is close to the upper bound of the estimated range of $2.5{ }^{\circ} \mathrm{C} - 3.5{ }^{\circ} \mathrm{C}$ using instrumental records in the IPCC AR6 \cite[Chapter 7]{ipcc2021c7}. It is also close to the emergent constrained ECS mean estimates from CMIP5 and CMIP6 \shortcite[Chapter 7 supplementary material]{schlund2020emergent,ipcc2021supp7}. 
\begin{table}[h!]
	\centering
	\caption{\footnotesize Comparison of estimates for the physical parameters between EBM-SS full model and other studies. The standard errors of the estimates are reported in parentheses, while some standard errors are unavailable. ``Chapter 7 SM'' represents the supplementary material to Chapter 7 in IPCC AR6.}
	\setlength{\tabcolsep}{1 pt} 
	\renewcommand{\arraystretch}{1.7} 
	\begingroup\scriptsize
	\begin{tabular}{l|cccccc}
	\hline\hline
	\multicolumn{7}{c}{\footnotesize \textbf{A. evaluation of the two-component EBM using historical data}}
	\\
		\textbf{\footnotesize model}	& 	\footnotesize$\boldsymbol{\hat{\lambda}}$ &\footnotesize $\boldsymbol{\hat{\gamma}}$& \footnotesize $\boldsymbol{\hat{C}}_m$ &\footnotesize $\boldsymbol{\hat{C}}_d$ &\footnotesize $\boldsymbol{\hat{\text{ECS}}}$ & \\ 
	\hline 
\footnotesize	{EBM-SS full, 0-700m ocean data (NOAA \& IAP)} &1.08 (0.25) & 1.30 (0.34) & 9.64 (2.86) & 98.49 (0.26) & 3.63 (0.89) &  \\
\footnotesize	EBM-SS full, 0-2,000m ocean data (IAP)&0.66 (0.31) & 1.82 (0.45) & 9.35 (2.61) & 269.30 (0.42) & 5.91 (2.77) &  \\
	\hline \hline
		\multicolumn{7}{c}{\footnotesize \textbf{B. evaluation of the two-component EBM using $4\times$CO$_2$ experiment data}}
	\\
		\textbf{\footnotesize model}	& 	\footnotesize$\boldsymbol{\hat{\lambda}}$ &\footnotesize $\boldsymbol{\hat{\gamma}}$& \footnotesize $\boldsymbol{\hat{C}}_m$ &\footnotesize $\boldsymbol{\hat{C}}_d$ &\footnotesize $\boldsymbol{\hat{\text{ECS}}}$ &\footnotesize \\ 
		\hline
\footnotesize	{CMIP6 means} \shortcite[Chapter 7 SM]{ipcc2021supp7} & 0.84 (0.38) & 0.64 (0.13) & 8.1 (1.0) & 110 (63) & 3.0 &\\
\footnotesize {CMIP5 means} \cite{cummins2020optimal} & 1.21 & 0.77 & 6.88& 97.18 & 3.41 & \\
	\hline\hline
		\multicolumn{7}{c}{\footnotesize \textbf{C. estimates of ECS using different datasets and methods}}
	\\
	\textbf{\footnotesize model and data}	& 	\footnotesize$\boldsymbol{\hat{\lambda}}$ &\footnotesize $\boldsymbol{\hat{\gamma}}$& \footnotesize $\boldsymbol{\hat{C}}_m$ &\footnotesize $\boldsymbol{\hat{C}}_d$ &\footnotesize $\boldsymbol{\hat{\text{ECS}}}$ &\footnotesize  \\ 
	\hline
		\footnotesize  {Instrumental records}  \shortcite[Chapter 7]{ipcc2021c7} &- &- &- &- & 2.5 -- 3.5& \\ 
	\footnotesize	{CMIP6 means} \shortcite{schlund2020emergent,ipcc2021supp7} & - & - & -  & - & 3.78 (1.08) &\\
	\footnotesize {CMIP5 means}\shortcite{schlund2020emergent,ipcc2021supp7} & - & - & -& - & 3.28 (0.74) &\\
\hline \hline
		\multicolumn{7}{c}{\footnotesize \textbf{D. heat capacity by physical relationship}}
\\
\footnotesize	\textbf{literature}	& \multicolumn{2}{c}{\footnotesize $\boldsymbol{C}_m$} & 	\multicolumn{2}{c}{\textbf{\footnotesize $C$ at 700m}}&	\multicolumn{1}{c}{\textbf{\footnotesize $C$ at 2000m}} \\ 
	\hline
\footnotesize	\citeA{hartmann2015global} & 	\multicolumn{2}{c}{9.32} &	\multicolumn{2}{c}{93.17} & 	\multicolumn{1}{c}{266.2} \\
\footnotesize	\citeA{gregory2000vertical} & 	\multicolumn{2}{c}{14.33} &	\multicolumn{2}{c}{66.85} & 	\multicolumn{1}{c}{191} \\
	\hline
	\hline
\end{tabular}
	\endgroup
	\label{ECScomparison}
\end{table}\\
\indent  Panel D in Table \ref{ECScomparison} reports heat capacity values indicated by physical relations. As we employ ocean data covering 0-700m and 0-2,000m in this paper, we also examine the physics-implied heat capacities at these two depths to evaluate the estimation accuracy. Here we have two benchmarks that define different depths for the mixed layer. The first benchmark is by \citeA{hartmann2015global}, who declares the average depth of the mixed ocean layer that interacts with the atmosphere on a scale of one year is 70 m, and the corresponding heat capacity is $9.32$ $\mathrm{W}$ year $\mathrm{m}^{-2} \mathrm{~K}^{-1}$\footnote{The ocean heat capacity at a specific depth $d$ is: $C=d \times 0.1331 \mathrm{WK}^{-1} \mathrm{~m}^{-2} \mathrm{~m}^{-1}$ \cite{hartmann2015global}.}.  Another benchmark considers 150 m as the mean depth \cite{gregory2000vertical} and the heat capacity value for this benchmark is $14.33$ $\mathrm{W}$ year $\mathrm{m}^{-2} \mathrm{~K}^{-1}$\footnote{According to \shortciteA{geoffroy2013transient}, $C=d \times 0.0955 \mathrm{WK}^{-1} \mathrm{~m}^{-2} \mathrm{~m}^{-1}$.}. Our estimates of the heat capacities for the mixed layer and deep ocean layer $C_m$ and $C_d$ are noticeably close to Hartmann's benchmarks. \\
\indent Figure \ref{fit700} and Figure \ref{fit2000} in Appendix \ref{2000m} give graphic summaries of the model fit for the EBM-SS full model to two pairs of 0-700m ocean datasets from NOAA and IAP, and to 0-2,000m ocean series from IAP, respectively. These two figures indicate that the smoothed states of the latent states, which are the estimated states given the entire observational trajectory, closely catch the data. \\
\indent As described in Section \ref{EBM-SS}, we assume the state disturbances and measurement errors to be serially uncorrelated and normality distributed. Then, ideally,  the standardized one-step ahead prediction errors  are also serially uncorrelated and follow a standard normal distribution \cite{durbin2012time}. In Figure \ref{fit700} and Figure \ref{fit2000}, the residuals after the fit, the standardized one-step ahead prediction errors, appear centered around zero. Diagnostic statistics of the residuals are reported in Table \ref{8Temp2OHC} in Appendix \ref{diag}. The residual series have means close to zero and standard deviations close to one. There is no violation of Gaussianity except for the residuals of ocean temperature 0-2,000m and OHC 0-2,000m from IAP due to the outliers in 1958. There are a few standardized prediction error series showing autocorrelation. This can be attributed to using a single state to fit variations from multiple data series. Overall, the EBM-SS model provides a good fit for the data. 
\subsection{Empirical evidence for estimation uncertainty reduction by using multiple data sources}
\label{evidenceEmpirical}
In Section \ref{secsim}, we have shown in simulations that the multiple-data-source structure in the EBM-SS model is effective in reducing parameter estimation uncertainty. In this section, we examine if the same conclusion can be drawn in the empirical exercise. \\
\indent In the simulation study, we can directly compare the simulation performances of the EBM-SS base model (single data source) and of the EBM-SS full model (multiple data source). This is because the simulated data for these two models are generated from the same simulations and thus have the same parametrization and randomness. However, the standard errors using different empirical datasets are incomparable due to the varying magnitudes of the mean estimates of the parameters. Therefore, we use the coefficient of variation (CV), defined as the standard error divided by the estimate of the mean, to measure the relative estimation uncertainty. \\
\indent In Table \ref{CV}, we compare CVs in the EBM-SS base model with those in the EBM-SS full model. Under the EBM-SS base setting, there are sixteen and eight combinations of different series when ocean data 0-700m and ocean data 0-2,000m are employed, respectively. As shown in Table \ref{Empirical700+2000} in Appendix \ref{Appendix1GMST1Ocean}, both the physical parameter estimates and CVs are diverse across the different combinations of series, and thus we use the median of the CVs for the EBM-SS base model. 
\begin{table}[ht!]
	\centering
	\caption{\footnotesize Comparing coefficients of variation (CVs)  of the physical parameters for the EBM-SS full model and the medians of CVs for the EBM-SS base model. CV is calculated as the standard error of the estimate divided by the estimate of the mean. ``{\% CV $\geq$} *'' reports the percentage where the individual CV of EBM-SS base model is not smaller than the CV of the EBM-SS-full model (denoted as $*$). }
	\setlength{\tabcolsep}{6pt} 
	\renewcommand{\arraystretch}{1.6} 
	\begingroup\scriptsize
	\begin{tabular}{l|l|lllll|lllll}
		\hline 
		\hline
	\multirow{2}{*}{\footnotesize\textbf{Setting}}	&	&  \multicolumn{5}{c|}{\footnotesize \textbf{NOAA / IAP 0-700m}} &  \multicolumn{5}{c}{\footnotesize \textbf{IAP 0-2,000m}} 
		\\
		\cline{3-12}
		&	&\footnotesize	$\hat{\boldsymbol{\lambda}}$&\footnotesize $\hat{\boldsymbol{\gamma}}$& \footnotesize	$\hat{\boldsymbol{C}}_m$ &\footnotesize $\hat{\boldsymbol{C}}_d$ & \footnotesize  \textbf{ECS}  & \footnotesize  $\hat{\boldsymbol{\lambda}}$& \footnotesize $\hat{\boldsymbol{\gamma}}$& 	\footnotesize $\hat{\boldsymbol{C}}_m$ & \footnotesize $\hat{\boldsymbol{C}}_d$ &\footnotesize  \textbf{ECS}  \\
		\hline
		\footnotesize 	\multirow{1}{*}{\textbf{EBM-SS-full}}	&\footnotesize\textbf{CV} (*) & 0.23 & 0.26 & 0.30 & 0.003 & 0.25 & 0.46 & 0.25 & 0.28 & 0.002 & 0.47 \\ 
		\cline{1-12}
\footnotesize	\multirow{2}{*}{\textbf{EBM-SS-base}}	&\footnotesize  \textbf{median of CVs}	& {0.39} & {0.22} & {0.37} & {0.01}&{0.40} & 1.20 & 0.19 & 0.36 & 0.002 & 1.21  \\ 
& \footnotesize \textbf{\% CV $\geq$} * & 93.8\%  &  62.5\% & 93.8\% & 50\%  & 93.8\% & 100\% & 0 & 87.5\% & 100\% & 100\%\\
		\hline\hline
	\end{tabular}
	\label{CV}
	\endgroup
\end{table} \\
\indent Table \ref{CV} shows clearly that the EBM-SS full model with 0-700m and 0-2,000m ocean series produces a lower CV than the medians of CVs for the EBM-SS base model. The last row in Table \ref{CV} indicates that the majority of individual CVs are equal to or greater than the CV of the EBM-SS full model. One exception is $\hat{\gamma}$ when IAP 0-2,000m ocean data is employed. Table \ref{CV} provides empirical evidence that including multiple data sources reduces the estimation uncertainty. 
\section{Scenario Analysis}
\label{scenario}
\begin{figure}[H]
	\centering
	\caption{\footnotesize Probabilistic projection of GMST during 2021 -- 2100 conditional on RCP 2.6, RCP 4.5, RCP 6, and RCP 8.5 forcing series from \protect \citeA{meinshausen2011rcp}. }
	\begin{subfigure}{0.5\textwidth}
		\subcaption{\footnotesize RCP 2.6 --  0-700m ocean data (NOAA \& IAP)}
		\label{TempSyn}
		\includegraphics[width=0.98\linewidth]{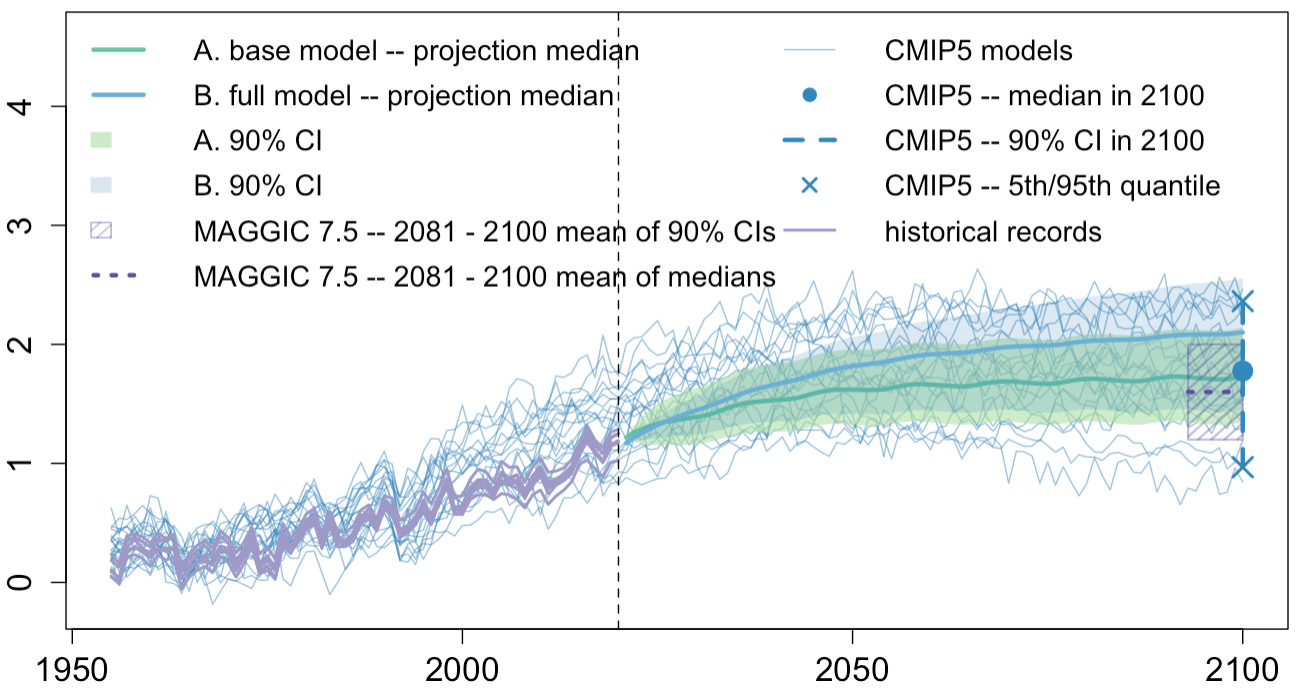}
	\end{subfigure}\hfill
	\begin{subfigure}{0.5\textwidth}
		\subcaption{\footnotesize RCP 2.6 --  0-2,000m ocean data (IAP)}
		\label{TempSyn}
		\includegraphics[width=0.98\linewidth]{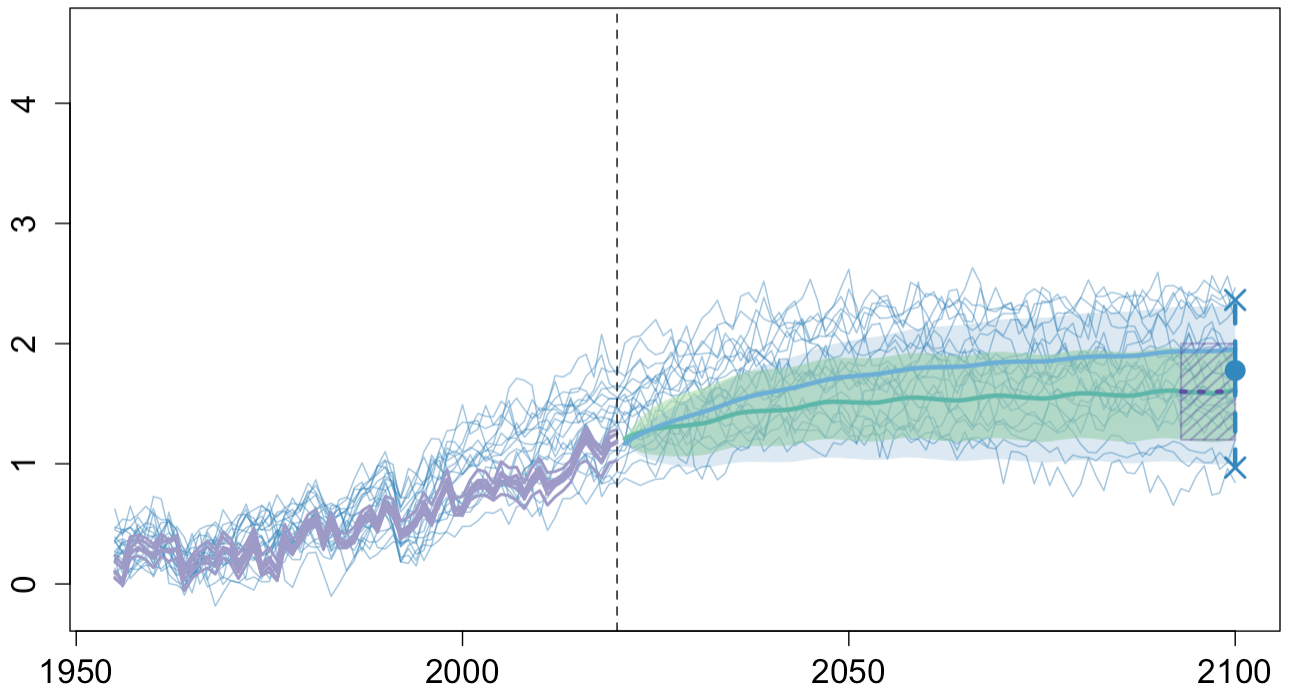}
	\end{subfigure}\\
	\begin{subfigure}{0.5\textwidth}
		\subcaption{\footnotesize RCP 4.5 --  0-700m ocean data  (NOAA \& IAP)}
		\label{TempSyn}
		\includegraphics[width=0.98\linewidth]{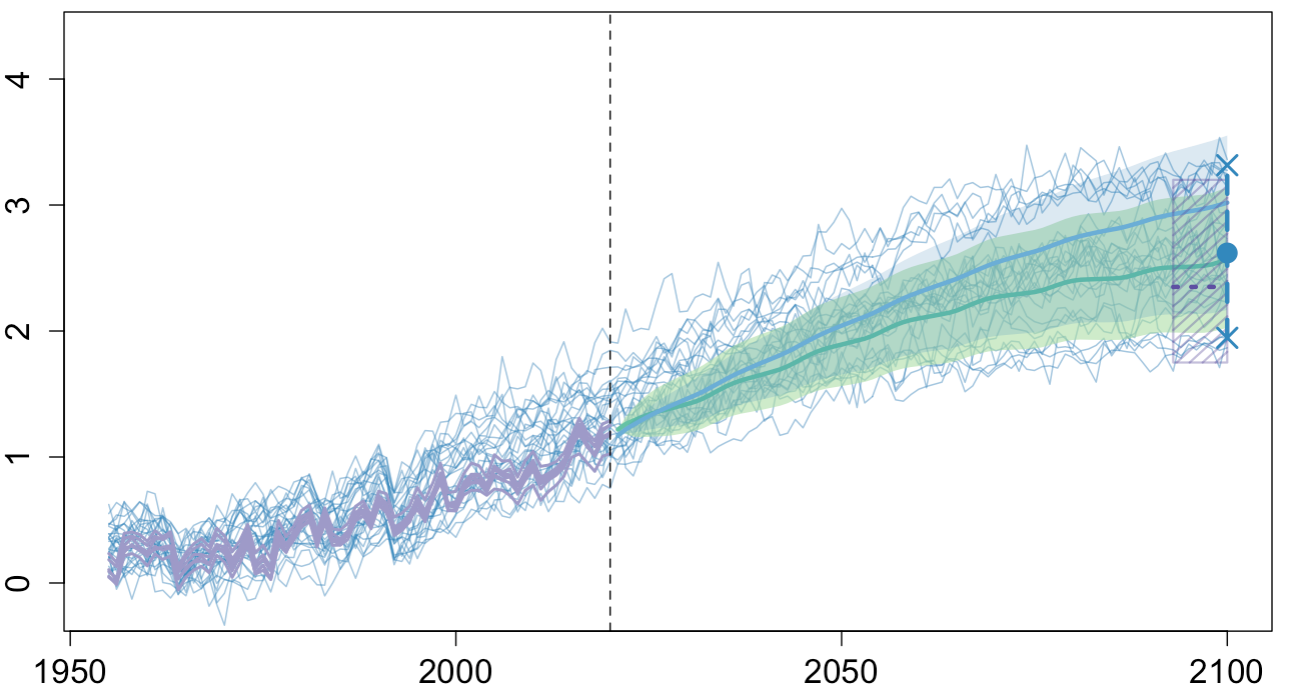}
	\end{subfigure}\hfill
	\begin{subfigure}{0.5\textwidth}
		\subcaption{\footnotesize RCP 4.5 --  0-2,000m ocean data (IAP)}
		\label{TempSyn}
		\includegraphics[width=0.98\linewidth]{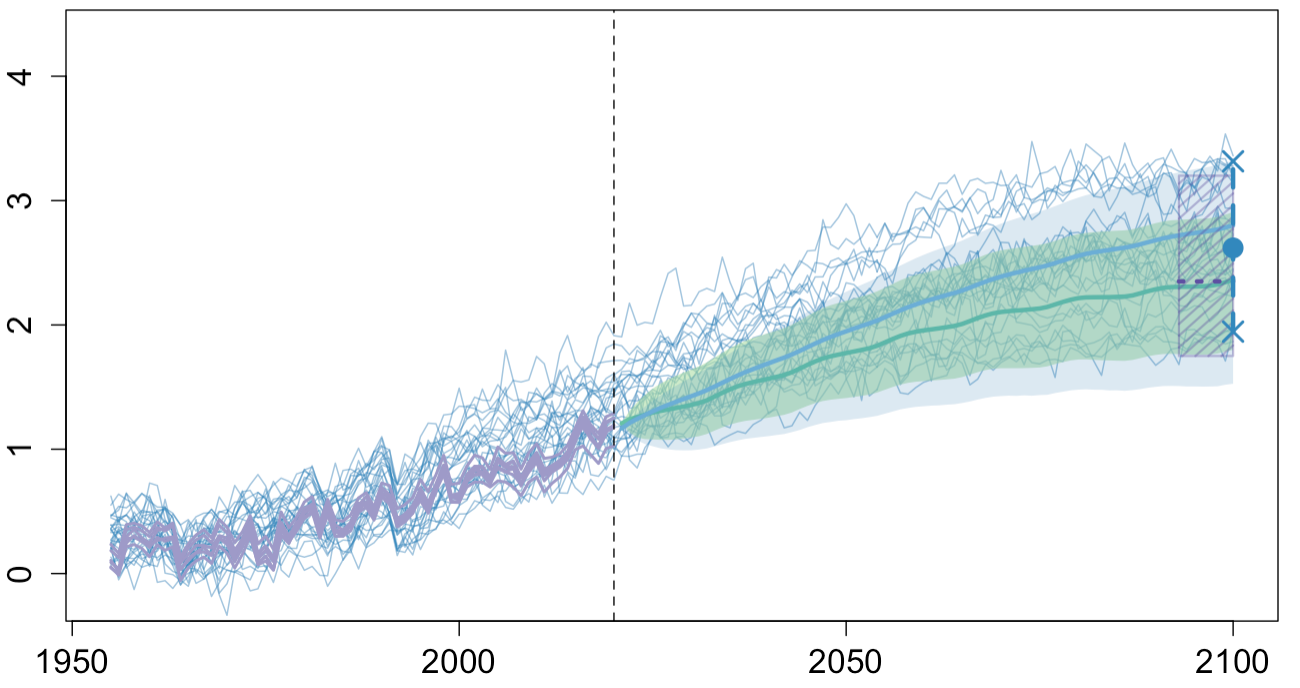}
	\end{subfigure}\\
	\begin{subfigure}{0.5\textwidth}
		\subcaption{\footnotesize RCP 6.0 --  0-700m ocean data (NOAA \& IAP)}
		\label{TempSyn}
		\includegraphics[width=0.98\linewidth]{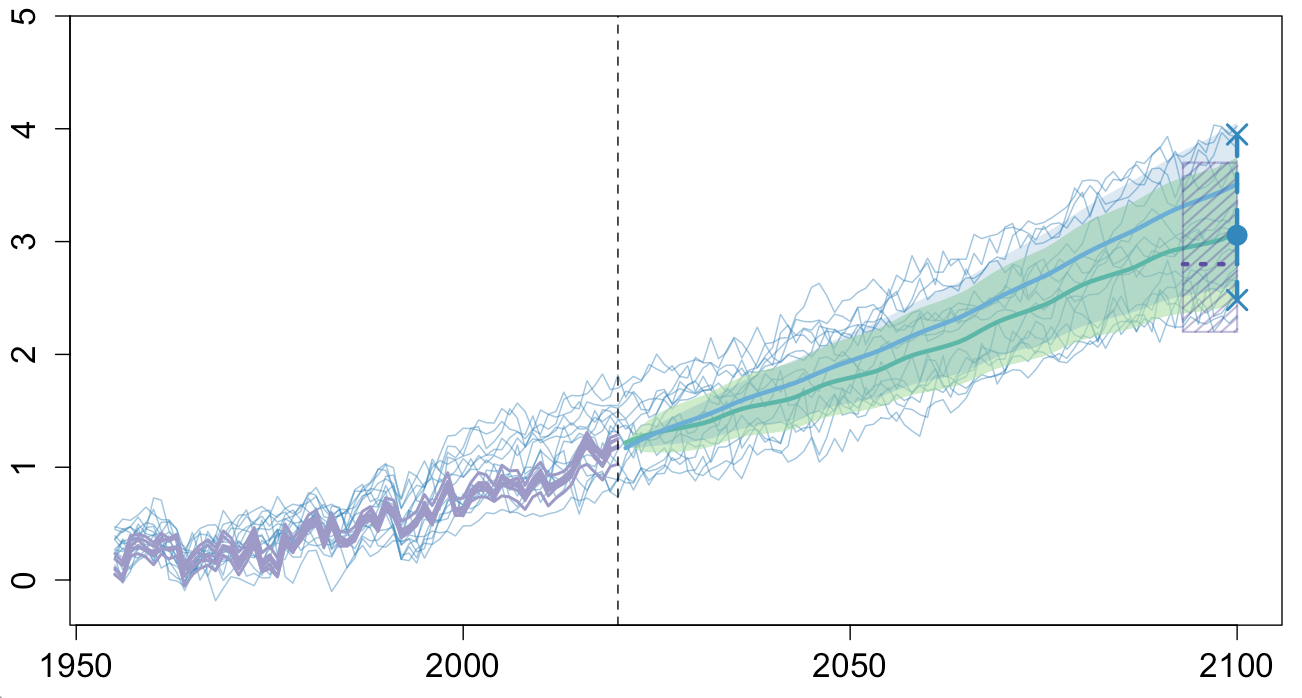}
	\end{subfigure}\hfill
	\begin{subfigure}{0.5\textwidth}
		\subcaption{\footnotesize RCP 6.0 --  0-2,000m ocean data (IAP)}
		\label{TempSyn}
		\includegraphics[width=0.98\linewidth]{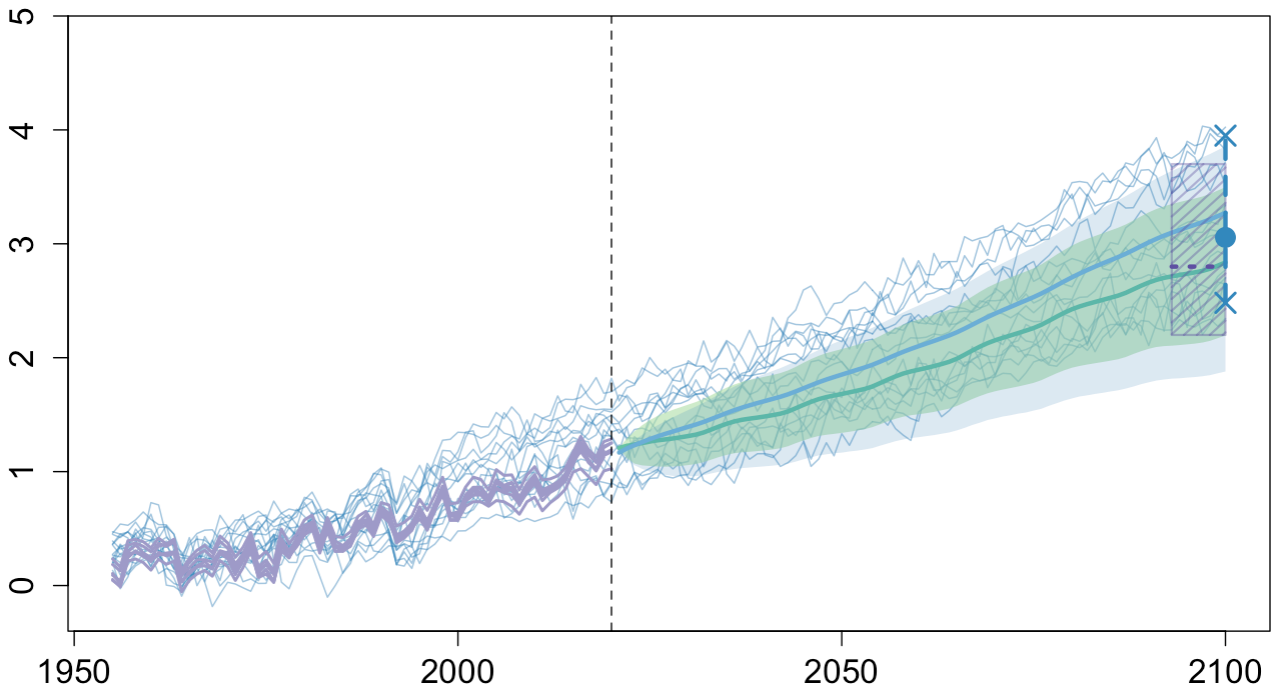}
	\end{subfigure}
	\begin{subfigure}{0.5\textwidth}
		\subcaption{\footnotesize RCP 8.5 --  0-700m ocean data (NOAA \& IAP)}
		\label{TempSyn}
		\includegraphics[width=0.98\linewidth]{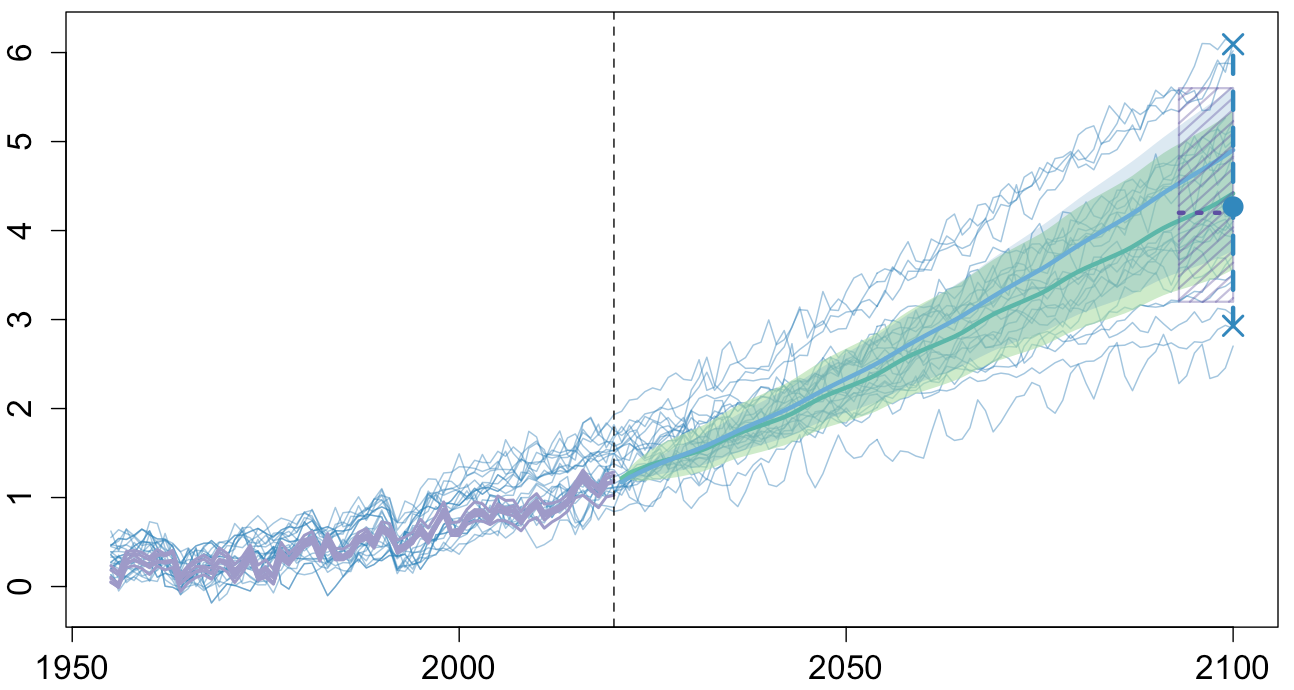}
	\end{subfigure}\hfill
	\begin{subfigure}{0.5\textwidth}
		\subcaption{\footnotesize RCP 8.5 --  0-2,000m ocean data (IAP)}
		\label{TempSyn}
		\includegraphics[width=0.98\linewidth]{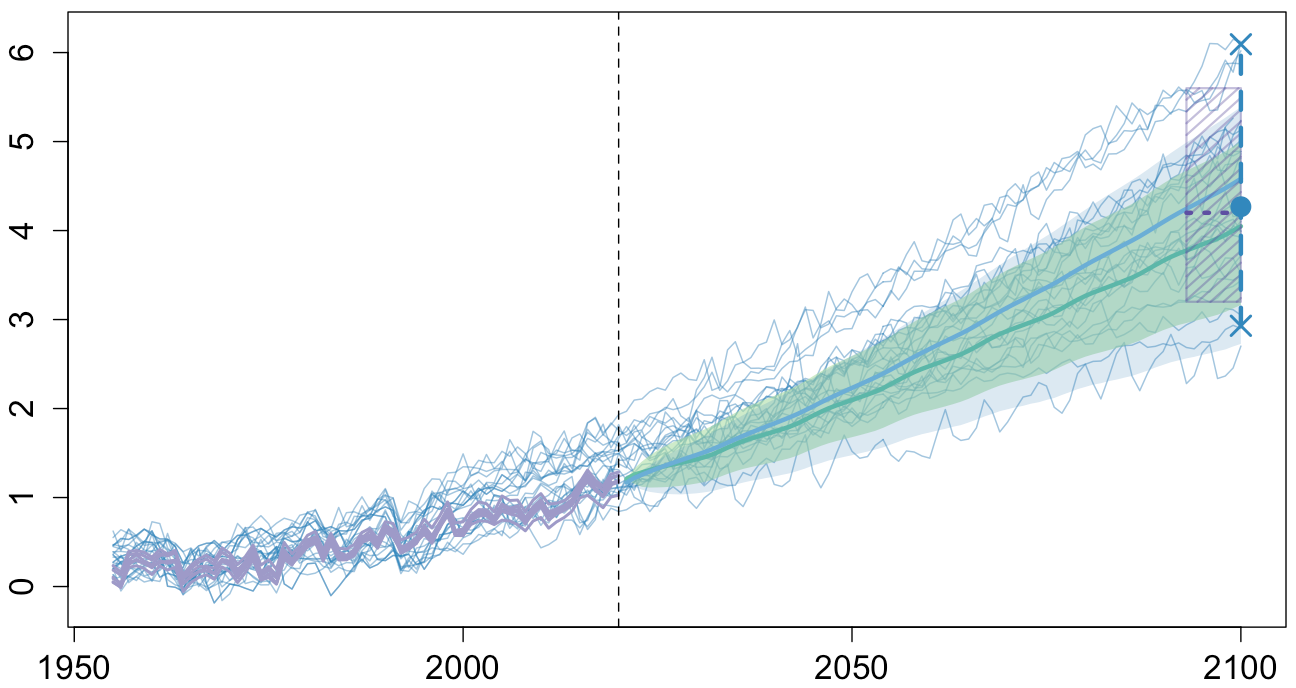}
	\end{subfigure}
	\label{projection}
\end{figure}

In this section, we study long-term projections from the EBM-SS full model using the Representative Concentration Pathway (RCP) scenarios of \citeA{meinshausen2011rcp}. \\
	\begin{figure}[h!]
	\centering
	\caption{\footnotesize Global annual mean radiative forcing ($
		\mathrm{Wm}^{-2}
		$) constructed under RCP scenarios \protect \shortcite{meinshausen2011rcp} during 2021 -- 2100. }
	\includegraphics[width=0.63\linewidth]{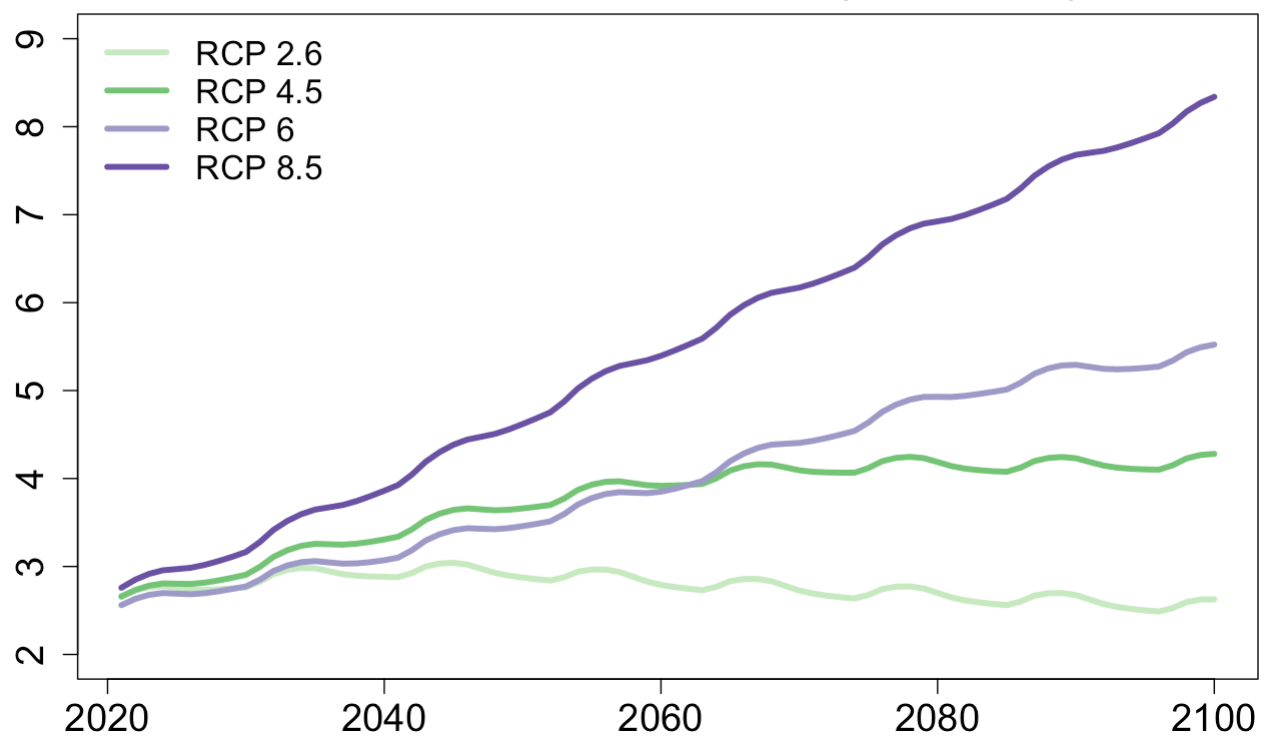}
	\label{RCPforcing}
\end{figure} 
\indent RCPs are adopted by \shortciteA{ipcc2014ipcc} to represent different possible GHG concentration trajectories in the future until 2100. The four scenarios RCP 2.6, RCP 4.5, RCP 6, and RCP 8.5 denote different levels of radiative forcing values ($
\mathrm{Wm}^{-2}
$) 2.6, 4.5, 6, and 8.5, respectively, in 2100. The reduced complexity climate model MAGICC6 \cite{meinshausen2011emulating1,meinshausen2011emulating2} was influential in constructing the RCP pathways. The RCP radiative forcing time series during 2021 -- 2100 are shown in Figure \ref{RCPforcing}. Except for RCP 2.6, which shows a peak in around 2030 and then decline, the other three pathways are rising over the period. They exhibit cyclical patterns from solar irradiance. \\
\indent We focus on the projection of GMST for 2021 -- 2100 conditional on the RCP forcing series. To take parameter uncertainty into account, we assume that the physical parameter vector follows a multivariate normal distribution $\mathcal{N}(\hat{\boldsymbol{\uptheta}}^*,\hat{\boldsymbol{\Sigma}}^*)$ and draw 10,000 sets of parameters from this distribution, where $\hat{\boldsymbol{\uptheta}}^*$ and $\hat{\boldsymbol{\Sigma}}^*$ denote the estimated mean vector and the estimated variance-covariance matrix of the physical parameter set in the empirical exercise in Table \ref{empirical}. We fix the other parameters, such as the variances of measurement errors, at their empirical estimates. We insert each of the parameter sets into the EBM-SS model, which produces point predictions of GMST at $T+h$ ($T$ = 2020, and $h$ = 1, 2, ..., 80). We show the median values and the $90\%$ projection confidence intervals in Figure \ref{projection} under each of the four RCP scenarios.\\
\indent We produce GMST projections for both the EBM-SS base and the EBM-SS full models. The EBM-SS base model is specified using HadCRUT5 GMST anomalies, together with ocean series (ocean temperature and OHC) from NOAA for 0-700m and from IAP for 0-2,000m. We choose these series because they produce parameter estimates and CVs that are closest to the medians of the values using all combinations of data series (see Table \ref{Empirical700+2000} in Appendix \ref{Appendix1GMST1Ocean}). \\
\indent  In Figure \ref{projection} and Table \ref{projec}, we compare our GMST projections with the outputs from CMIP5  models and the emulator MAGGIC 7.5, where all of the values are relative to 1850 -- 1900. The global scale averaged time series of CMIP5 projections under various RCP scenarios are aggregated by \citeA{nicholls2021regionally}. Considering data availability, we include the projection series from 21, 29, 16, and 28 CMIP5 models for RCP 2.6, RCP 4.5, RCP 6.0, and RCP 8.5 scenarios, respectively. The 2081 -- 2100 mean results from MAGGIC 7.5 runs use historical observational GHG concentration levels until 2015 and then switch to emission-driven runs \cite{ipcc2021c4}. \\
\indent Figure \ref{projection} and Table \ref{projec} show that our GMST projections produced using the EBM-SS full model and eight GMST historical datasets largely agree with those from CMIP5 models and MAGGIC 7.5. The EBM-SS base model generates both a higher mean projection and a wider confidence band. The wider confidence band is mainly due to the larger estimation uncertainty in the base case. It corroborates the crucial role of multiple data sources in producing more precise projection values. The results are robust to the depth the ocean data (ocean temperature and OHC) covers. Although employing 0-2000m ocean data generates wider confidence bands than using 0-700m, the differences are not large and the medians remain almost unchanged (Table \ref{projec}). 
	\begin{table}[h!]
	\centering
		\setlength{\tabcolsep}{6pt} 
			\renewcommand{\arraystretch}{1.5} 
		\caption{\footnotesize 5\% -- 95\% ranges and medians of GMST  ($^{\circ} \mathrm{C}$) projections in 2100 under RCP scenarios by EBM-SS full model.  }
		\label{preind}
			\begingroup\scriptsize
		\begin{tabular}{llcccccccc}
			\hline\hline
		&		& \multicolumn{4}{c}{\footnotesize \textbf{0-700m ocean data (NOAA \& IAP)}} & \multicolumn{4}{c}{\footnotesize \textbf{0-2,000m ocean data (IAP)}}\\
			\cmidrule(lr){3-6}\cmidrule(lr){7-10}
	&	& \footnotesize {RCP 2.6}&   	\footnotesize {RCP 4.5}&  	\footnotesize {RCP 6.0}&  		\footnotesize {RCP 8.5}&  \footnotesize {RCP 2.6}&   	\footnotesize {RCP 4.5}&  	\footnotesize {RCP 6.0}&  		\footnotesize {RCP 8.5}\\
			\hline
\multirow{3}{*}{\footnotesize \textbf{Base}}& \footnotesize		median & 2.10 & 3.02 & 3.51 & 4.90 & 1.96 & 2.81 & 3.28 & 4.58  \\ 
 &  \footnotesize  5\% quantile  & 1.45 & 2.17 & 2.63 & 3.77 & 0.91 & 1.39 & 1.71 & 2.49 \\ 
&  \footnotesize  95\% quantile &  2.56 & 3.55 & 4.05 & 5.62 & 2.34 & 3.34 & 3.87 & 5.41 \\ 
\hline
\multirow{3}{*}{\footnotesize \textbf{Full}}& \footnotesize		median &1.73 & 2.57 & 3.08 & 4.42 & 1.61 & 2.37 & 2.84 & 4.06 \\ 
&  \footnotesize  5\% quantile  & 1.34 & 2.02 & 2.46 & 3.56 & 1.21 & 1.80 & 2.20 & 3.17  \\ 
&  \footnotesize  95\% quantile & 2.14 & 3.14 & 3.75 & 5.35 & 1.97 & 2.91 & 3.51 & 5.02 \\ 
			\hline\hline
			&		& \multicolumn{4}{c}{\footnotesize \textbf{MAGGIC 7.5 runs 2081 -- 2100 means}} & \multicolumn{4}{c}{\footnotesize \textbf{CMIP5 outputs}}\\
			\cmidrule(lr){3-6}\cmidrule(lr){7-10}
			&	& \footnotesize {RCP 2.6}&   	\footnotesize {RCP 4.5}&  	\footnotesize {RCP 6.0}&  		\footnotesize {RCP 8.5}&  \footnotesize {RCP 2.6}&   	\footnotesize {RCP 4.5}&  	\footnotesize {RCP 6.0}&  		\footnotesize {RCP 8.5}\\
			\hline
		& \footnotesize		median & 1.6 & 2.35 & 2.8 & 4.2 & 1.78 & 2.62 & 3.06 & 4.27 \\ 
			&  \footnotesize  5\% quantile  & 1.2 & 1.75 & 2.2 & 3.2 & 0.97 & 1.95 & 2.48 & 2.93 \\ 
			&  \footnotesize  95\% quantile & 2.0 & 3.2 & 3.7 & 5.6 & 2.36 & 3.32 & 3.95 & 6.09 \\ 
			\hline\hline
			\footnotesize 
		\end{tabular}
	\label{projec}
\endgroup
\end{table}
	\section{Conclusion}
		\label{sec7}
			\indent In this paper, we present a statistical climate model (EBM-SS model), which is a multivariate linear Gaussian state space representation of the two-component EBM. The EBM-SS model provides a framework to quantify the temperature change in response to radiative forcing, while taking the thermal inertia of the ocean into account. We incorporate ocean heat content (OHC) as a second measurement of the temperature in the deep ocean, so that the heat capacity for the deep ocean, $C_d$, can be constrained using both ocean temperature and OHC historical records. \\
    \indent We incorporate multiple data sources as measurements for the latent states to reduce estimation uncertainty. To account for the different baseline periods in different anomaly series, we synchronize the eight GMST series and the anthropogenic forcing series with respect to the period 1850 -- 1900 using the information by \citeA{2019teallenchnical}. We include constants for the ocean temperature and OHC series to offset their baseline discrepancies relative to the GMST anomalies and radiative forcing series. Both the empirical and simulation exercises indicate that including eight GMST anomaly series and two pairs of ocean data series as multiple data sources reduces estimation uncertainty of the parameters compared to the models that use only one data source. We obtain physical parameter estimates that are comparable to the ones reported in \citeA{cummins2020optimal} and \citeA[Chapter 7 SM]{ipcc2021supp7}.\\
	\indent We show that fitting the EBM-SS model to a comprehensive data set of GMST anomalies, ocean temperature, and OHC over the period 1955 -- 2020 from separate research groups produces projections for GMSTs that are comparable to those of CMIP5 \cite{nicholls2021regionally} and MAGICC 7.5 \shortcite{ipcc2021c4} for RCP 2.6, RCP 4.5, and RCP 6. Our results thus corroborate earlier findings from both complex climate models and reduced-complexity models, where our statistical model exclusively uses historical data. Our model is, in contrast to earlier models, a small-scale statistical model that can be estimated using standard software packages on standard office computers. Its statistical nature allows for the assessment of parameter uncertainty. \\
	\newpage
	\bibliographystyle{apacite}
	\bibliography{EBMSS0522}
	\newpage
	\appendix
		\section{Appendix}
		\subsection{Unit root tests on the historical data }
		\label{unitroot}
		\begin{table}[ht]
			\centering
			\caption{\footnotesize Augmented Dickey-Fuller (ADF) test \protect \cite{dickey1979distribution} for unit roots on level and first-order difference of the observational series for different components of anthropogenic forcing, total anthropogenic forcing $Y_{A,t}$, and total forcing $Y_{F,t}$ during 1955 -- 2018.  The null hypothesis of the ADF test is the existence of a unit root, i.e., non-stationarity. $"\textbf{f}+\text{greenhouse gas name}"$ indicates the forcing from a specific greenhouse gas. ** and * mark significance at 1 $\%$ and 5 $\%$, respectively. We report the test statistics when lag order $k$ equals to 0, 1, 2, and 3, respectively. The values in bold indicate the optimal lag order selected by Bayesian information criterion (BIC) \protect \cite{schwarz1978estimating}. The maximum order of lags considered is 15. }
			\begingroup\scriptsize
			\setlength{\tabcolsep}{3pt} 
			\renewcommand{\arraystretch}{1.5} 
			\begin{threeparttable}
				\begin{tabular}{l|cccc|cccc|cccc}
					\hline
					\hline
					& \multicolumn{4}{c|}{\footnotesize \textbf{Level series}} &   \multicolumn{8}{c}{\footnotesize \textbf{First-order difference}} \\
					&\multicolumn{4}{c|}{\footnotesize\textbf{ with constant }}  	& \multicolumn{4}{c|}{\footnotesize\textbf{ (a). with constant }$^\text{a}$}   & \multicolumn{4}{c}{\footnotesize\textbf{ (b). with constant and trend }}     \\
					\cmidrule(lr){2-5}\cmidrule(lr){6-13}
					\setlength{\tabcolsep}{3.5pt}  \textbf{Lag order $k$}  & 0 & 1 & 2 & 3 & 0 & 1 & 2 & 3 &0 & 1 & 2 & 3 \\ 
					\hline
					\setlength{\tabcolsep}{3.5pt}   $\textbf{fCO}_2$  & \textbf{7.43} & 4.64 & 4.28 & 3.42 &$ \boldsymbol{-4.42 ^{**}}$& $-3.23^{*}$  &$-2.19 $& $-1.68 $&$\boldsymbol{-7.19^{**}}$ & $-5.93^{**}$ & $-4.41^{**}$& $-3.85^{*}$ \\ 
					\setlength{\tabcolsep}{3.5pt}   $\textbf{fCH}_4$ & $-6.90^{**}$ &$\boldsymbol{-2.54 }$& $-2.41 $& $-2.30$ & $\boldsymbol{-1.83}$ &$ -1.68 $& $-1.10$ & $-0.99$ &$\boldsymbol{-2.94}$ &$ -2.73$ &$ -2.00$ & $-1.84$ \\ 
					\setlength{\tabcolsep}{3.5pt}   $\textbf{fCFCs}$ & $-2.73$ & $\boldsymbol{-2.13}$ & $-2.24$ & $-2.18$ & $-1.19 $& $\boldsymbol{-0.93}$ & $-1.23$ & $-1.24$ & $\boldsymbol{-1.76} $& $-1.52$ & $-1.85$ & $-1.83 $\\ 
					\setlength{\tabcolsep}{3.5pt}   $\textbf{fN}_2$\textbf{O} & \textbf{4.91} & 5.11 & 3.96 & 3.18 & $\boldsymbol{-6.97^{**}} $&$ -3.72^{**}$ & $-2.66 $& $-1.92$ &$ \boldsymbol{-10.16 ^{**}}$& $-6.11^{**}$ & $-4.62^{**}$ & $-3.28$ \\ 
					\setlength{\tabcolsep}{3.5pt}   $\textbf{fO}_3$ & $\boldsymbol{-5.62^{**}}$ & $-4.25^{**}$ & $-2.47$ & $-2.09$ & $-4.58^{**}$ & $\boldsymbol{-2.13}$ & $-1.58 $& $-1.39 $& $-5.98^{**} $& $\boldsymbol{-2.67}$ & $-1.89$ & $-1.68$\\ 
					\setlength{\tabcolsep}{3.5pt}  $\textbf{f}_\text{TA+SA}$ & $-9.09^{**}$ &$\boldsymbol{-2.34}$ & $-2.04$ & $-2.18 $& $\boldsymbol{-1.60} $&$ -1.26$ & $-1.13 $& $-1.35$ & $\boldsymbol{-2.77} $& $-2.27 $&$ -2.28$ & $-2.29$ \\ 
					\setlength{\tabcolsep}{3.5pt}   \textbf{fWMGHGs} & 2.16 & \textbf{0.79} & 0.43 & 0.09 & $\boldsymbol{-4.28^{**}}$ & $-3.27^{*}$ & $-2.59 $& $-2.09 $& $\boldsymbol{-4.36^{**}}$& $-3.27^{**}$ &$ -2.53 $&$ -2.03 $\\ 
					\setlength{\tabcolsep}{3.5pt}   $\boldsymbol{Y}_{A,t}$ & 3.67 & \textbf{1.69} & 1.18 & 0.76 & $\boldsymbol{-4.41^{**}}$ & $-3.25 ^{*}$& $-2.61$ & $-2.19$ & $\boldsymbol{-4.89^{**}} $& $-3.53^{*} $& $-2.73$ & $-2.25$ \\ 
					\setlength{\tabcolsep}{3.5pt}  $\boldsymbol{Y}_{F,t}$& $-2.29$ & $-2.91$ &$\boldsymbol{-1.72}$ & $-1.56$ & $-6.01^{**}$ & $\boldsymbol{-6.92^{**}}$ & $-5.16^{**}$ & $-4.64^{**}$ & $-5.96^{**}$ & $\boldsymbol{-6.90^{**}}$ & $-5.17^{**}$ &$ -4.69^{**}$ \\
					\hline
					\hline
				\end{tabular}
				\begin{tablenotes}
					\scriptsize
					\item [a] ADF test regression with constant: $\Delta y_{t}=\alpha+\pi y_{t-1}+\sum_{j=1}^{k} \gamma_{j} \Delta y_{t-j}+\varepsilon_{t}$. Under the null hypothesis $\pi=0$, the regression equation is reduced to a random walk process with drift. 
					\item [b] ADF test regression with constant and trend: $\Delta y_{t}=\alpha+\beta t+\pi y_{t-1}+\sum_{j=1}^{k} \gamma_{j} \Delta y_{t-j}+\varepsilon_{t}$.
				\end{tablenotes}
			\end{threeparttable}
			\endgroup
			\label{ADF}
		\end{table} 
			\begin{table}[h!]
			\centering
			\caption{\footnotesize ADF test for unit roots on level and first-order difference series of different GMST, ocean temperature, and OHC anomalies during 1955 -- 2020. ** and * denote significance at $1\%$ and $5\%$. As in Table \protect \ref{ADF}, we conduct the tests with constant alone or both constant and trend included (the explicit expressions are reported in the footnote for Table \protect \ref{ADF}). Subtable (a) reports the test statistics for each of these specifications examined under lag order 0, 1, 2, and 3. The values in bold indicate the optimal lag orders selected by Bayesian information criterion (BIC). The maximum order of lags considered is 15. Subtable (b) reports the optimal lag orders and test statistics for NOAA OHC 0-700m, IAP OHC 0-700m, and IAP OHC 0-2,000m, where the optimal lag orders exceed 3. }
			\begingroup\scriptsize
			\renewcommand{\arraystretch}{1.5} 
			\setlength{\tabcolsep}{2.5pt}
		\begin{subtable}[h]{\textwidth}
			\caption{}
				\begin{tabular}{l|cccc|cccc|cccc}
					\hline
					\hline
					& \multicolumn{4}{c|}{\footnotesize \textbf{Level series}} &   \multicolumn{8}{c}{\footnotesize \textbf{First-order difference}} \\
					&\multicolumn{4}{c|}{\footnotesize\textbf{(1). with constant }}  	& \multicolumn{4}{c|}{\footnotesize\textbf{ (2). with constant }$^\text{a}$}   & \multicolumn{4}{c}{\footnotesize\textbf{ (3). with constant and trend }}     \\
					\cmidrule(lr){2-5}\cmidrule(lr){6-13}
					\footnotesize\setlength{\tabcolsep}{4pt}  \textbf{Lag order $k$}  & 0 & 1 & 2 & 3 & 0 & 1 & 2 & 3 &0 & 1 & 2 & 3 \\ 
					\hline 
					GISTEMP& $-0.99$ & $-0.53$ & \textbf{0.26} & 0.83 &$ -11.05^{**}$ & $\boldsymbol{-8.80^{**}}$& $-7.64^{**} $&$ -4.68^{**} $& $-10.99^{**}$ & $\boldsymbol{-8.85^{**}}$ & $-7.84^{**}$ & $-4.94^{**}$ \\
					NOAAGlobalTemp & $-1.27$ & $-0.95$ & $\boldsymbol{-0.07}$ &$ 0.59$ & $-10.59^{**} $& $\boldsymbol{-8.86 ^{**}}$&$ -7.50^{**} $&$ -4.70^{**} $& $-10.51^{**} $& $\boldsymbol{-8.85 ^{**}}$& $-7.64 ^{**}$&$ -4.91^{**} $\\ 
					HadCRUT5 & $-1.07$ & $-0.57$ & \textbf{0.15} & 0.70 &$ -11.36^{**}$ & $\boldsymbol{-8.89^{**}}$ & $-7.70^{**}$ &$ -4.78^{**}$ & $-11.30 ^{**}$& $\boldsymbol{-8.93^{**}}$ & $-7.88^{**}$ &$ -5.00^{**}$ \\ 
					BEST & $-1.17$ &$ -0.61$ &\textbf{ 0.16} & 0.76 & $-11.53^{**}$ &$\boldsymbol{-9.02^{**}}$ & $-7.88^{**} $& $-4.83^{**}$ & $-11.47^{**}$ &${-9.06^{**}}$ &$ \boldsymbol{-8.09^{**}}$ & $-5.08^{**}$ \\ 
					CW2014 & $-1.22$ & $-0.66$ & \textbf{0.12} & 0.69 & $-11.32 ^{**}$& $\boldsymbol{-9.07^{**}}$ & $-7.82^{**}$ & $-4.88^{**}$ & $-11.26^{**}$ & $-9.11^{**}$ & $\boldsymbol{-8.02^{**}}$ & $-5.15 ^{**}$\\ 
					JMA  & $-1.59 $& $-1.30 $&$ \boldsymbol{-0.37}$ & 0.24 & $-10.45^{**}$ & $\boldsymbol{-8.77^{**}}$ & $-7.76^{**}$ & $-4.73^{**}$ & $-10.37^{**}$ & ${-8.73^{**}}$ & $\boldsymbol{-7.83^{**}}$ & $-4.84^{**}$ \\ 
					ERA-Interim & $\boldsymbol{-1.12}$ & $-0.85$ & $-0.29 $& 0.02 & $-9.39^{**}$ & $\boldsymbol{-7.69^{**}}$ & $-5.75^{**}$ & $-4.18^{**}$ & $-9.30^{**}$ & $\boldsymbol{-7.64^{**}}$ & $-5.77^{**}$ & $-4.12^{*}$ \\ 
					JRA-55  & $\boldsymbol{-1.22}$ & $-0.94$ & $-0.36$ & 0.08 & $-9.47^{**}$ & $\boldsymbol{-7.82^{**}}$ & $-6.01^{**}$ & $-4.26^{**}$ & $-9.37^{**}$ & $\boldsymbol{-7.76^{**}}$ & $-6.04^{**}$ & $-4.19^{**}$ \\
					\hline  				
NOAA Ocean Temp$_\text{0-700m}$	&\textbf{0.49} & 1.39 & 1.52 & 1.99 & $\boldsymbol{-10.97^{**}}$ & $-6.98^{**}$ & $-5.35^{**}$ & $-3.05^{*}$ & $\boldsymbol{-11.50^{**}}$ & $-7.47^{**}$ & $-6.29^{**}$ & $-3.81^{*}$ \\ 
  	IAP Ocean Temp$_\text{0-700m}$ & \textbf{0.24} & 0.81 & 1.19 & 2.35 & $\boldsymbol{-10.46^{**}}$ &$ -7.51^{**}$ & $-6.52^{**}$ & $-4.66^{**}$ & $\boldsymbol{-10.60^{**}}$ & $-7.77^{**}$ &$ -7.48^{**}$ & $-5.52^{**}$ \\ 
  	IAP Ocean Temp$_\text{0-2,000m}$ & \textbf{0.91} & 1.27 & 1.58 & 2.62 & $\boldsymbol{-9.28^{**}}$ & $-7.05^{**}$ & $-5.78^{**}$ & $-4.24^{**}$ & $\boldsymbol{-9.56^{**}}$ & $-7.43^{**}$ &$ -6.85^{**}$ & $-5.16^{**}$ \\ 
  	\hline
  	NOAA OHC$_\text{0-700m}$ & 0.40 & 1.19 & 1.56 & 2.02 & $-10.75^{**}$ & $-7.61^{**}$ & $-5.61^{**}$ & $\boldsymbol{-3.11^{*}}$ & $-11.19^{**}$ & $-8.11^{**}$ & $-6.56^{**}$ & $-3.86^{*}$ \\ 
  	IAP OHC$_\text{0-700m}$  & 0.24 & 0.87 & 1.26 & 2.38 & $-10.80^{**}$ & $-7.69^{**}$ & $-6.45^{**}$ & $-4.60^{**}$ & $-10.96^{**}$ & $-7.97^{**}$ &$ -7.44^{**}$ & $-5.51^{**}$ \\ 
  	IAP OHC$_\text{0-2,000m}$ & 0.84 & 1.24 & 1.52 & 2.53 & $-9.45^{**}$ & $-7.14^{**}$ & $-5.76^{**}$ & $-4.24^{**}$ & $-9.72^{**}$ & $-7.50^{**}$ & $-6.78^{**}$ & $-5.12^{**}$ \\ 
					\hline
					\hline
				\end{tabular}
			\end{subtable}\\
		\vspace{10pt}
				\begin{subtable}[h]{\textwidth}
					\caption{}
					\centering
			\begin{tabular}{l|cccc|cccc|cccc}
				\hline
				\hline
				& \multicolumn{4}{c|}{\footnotesize \textbf{Level series}} &   \multicolumn{8}{c}{\footnotesize \textbf{First-order difference}} \\
				&\multicolumn{4}{c|}{\footnotesize\textbf{(1). with constant }}  	& \multicolumn{4}{c|}{\footnotesize\textbf{ (2). with constant }}   & \multicolumn{4}{c}{\footnotesize\textbf{ (3). with constant and trend }}     \\
				\cmidrule(lr){2-5}\cmidrule(lr){6-13}
				\footnotesize\setlength{\tabcolsep}{4pt}  & \multicolumn{2}{c}{optimal lag order} & \multicolumn{2}{c|}{t-stat} & \multicolumn{2}{c}{optimal lag order} & \multicolumn{2}{c|}{t-stat} &\multicolumn{2}{c}{optimal lag order} & \multicolumn{2}{c}{t-stat} \\ 
				\hline
				NOAA OHC$_\text{0-700m}$ & \multicolumn{2}{c}{14}  &  \multicolumn{2}{c|}{2.46}  &  &  &  &  & \multicolumn{2}{c}{13}  &  \multicolumn{2}{c}{$-4.00^{**}$}  \\ 
				IAP OHC$_\text{0-700m}$  & \multicolumn{2}{c}{10}  &  \multicolumn{2}{c|}{1.87}  &  \multicolumn{2}{c}{9}  &  \multicolumn{2}{c|}{$-2.52$}  & \multicolumn{2}{c}{10}  &  \multicolumn{2}{c}{$-3.61^{*}$}  \\ 
				IAP OHC$_\text{0-2,000m}$ & \multicolumn{2}{c}{10}  &  \multicolumn{2}{c|}{1.92}  &  \multicolumn{2}{c}{13}  &  \multicolumn{2}{c|}{$-1.22$}  & \multicolumn{2}{c}{10}  &  \multicolumn{2}{c}{$-3.61^{*}$} \\ 
				\hline
				\hline
			\end{tabular}
		\end{subtable}
			\endgroup
			\label{ADF_temp}
		\end{table}
	\clearpage
		\subsection{Synchronizing anomalies to a common baseline}
	\label{irrelavance}
\begin{figure}[h!]
		\centering
		\caption{\footnotesize GMST, ocean temperature, and OHC anomaly series after synchronization. The light gray area corresponds to the time horizon 1955 -- 2020 in the empirical study.}
		\begin{subfigure}{0.49\textwidth}
			\subcaption{\footnotesize GMST Anomalies $\left({ }^{\circ} \mathrm{C}\right)$ (1850 -- 2020) with respect to pre-industrial era (1850 -- 1900)}
			\label{TempSyn1}
			\includegraphics[width=0.98\linewidth]{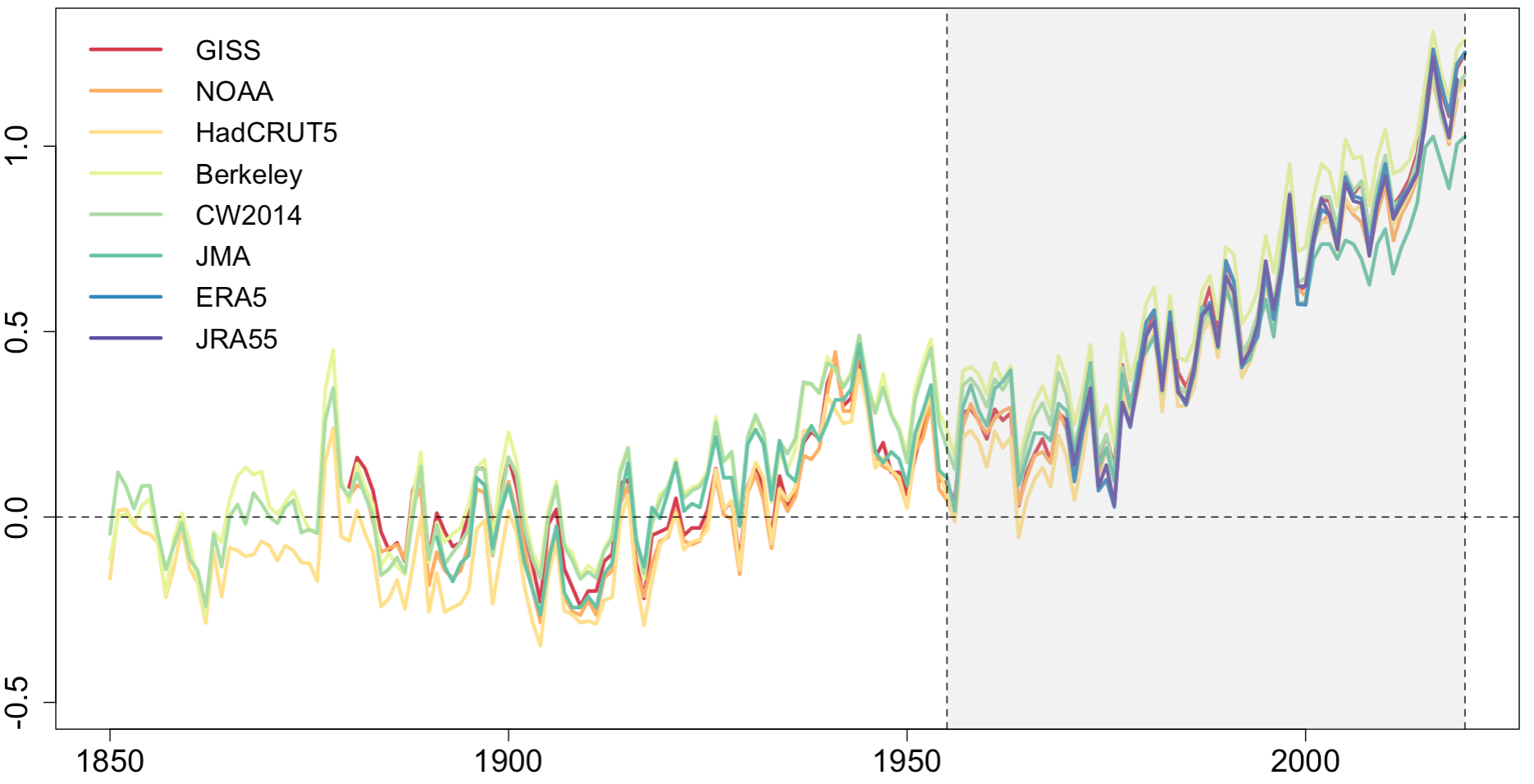}
		\end{subfigure}\hfill
		\begin{subfigure}{0.49\textwidth}
			\subcaption{\footnotesize ocean temperature $\left({ }^{\circ} \mathrm{C}\right)$ (1940 -- 2020) with respect to 1981 -- 2010 }
			\label{OtempSyn}
			\includegraphics[width=\linewidth]{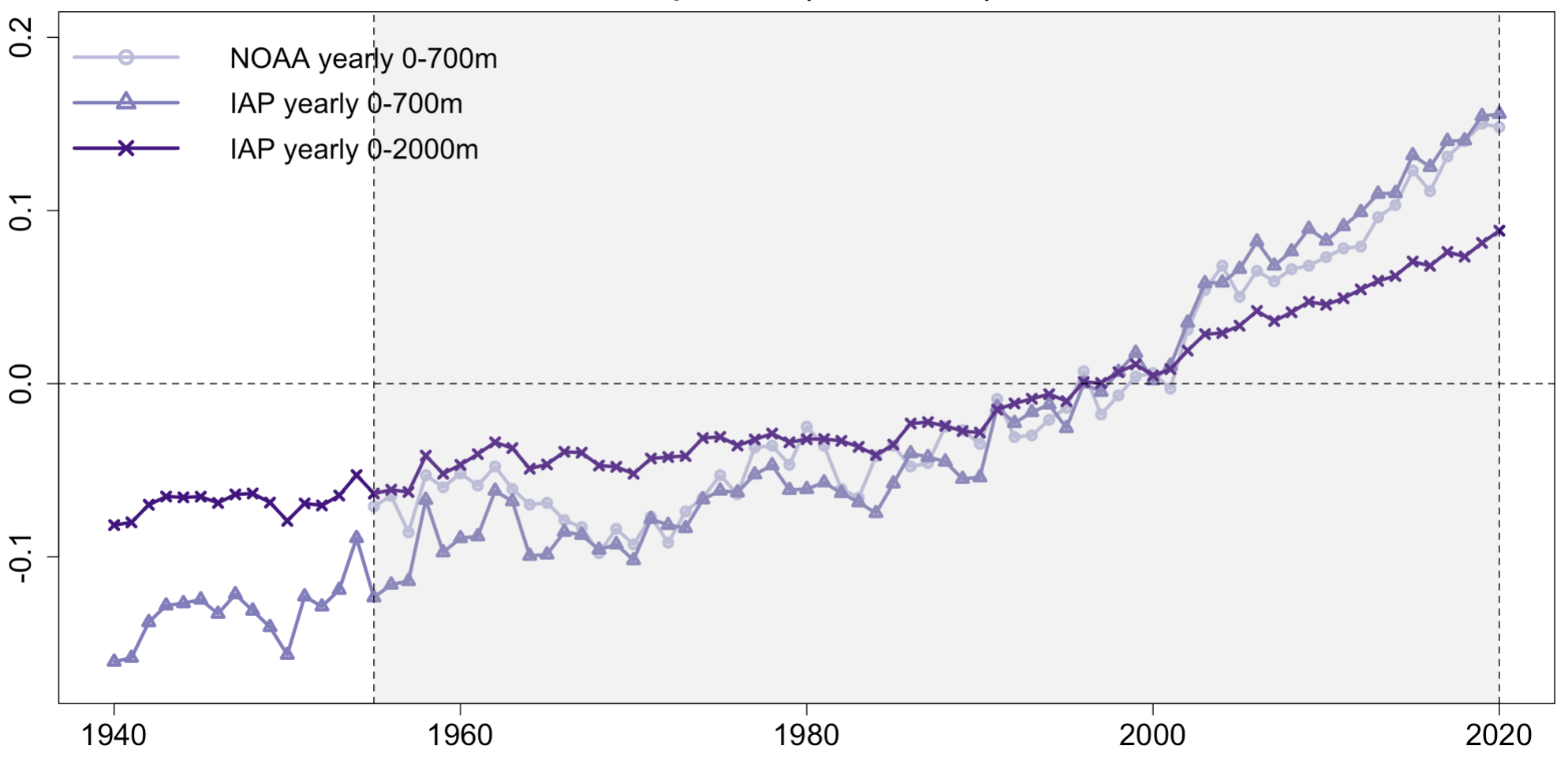}
		\end{subfigure}
		\\
		\centering
		\begin{subfigure}{0.49\textwidth}
			\subcaption{\footnotesize OHC Anomalies (J $\text{m}^{-2}$) (1940 -- 2020) with respect to 1981 -- 2010}
			\includegraphics[width=\linewidth]{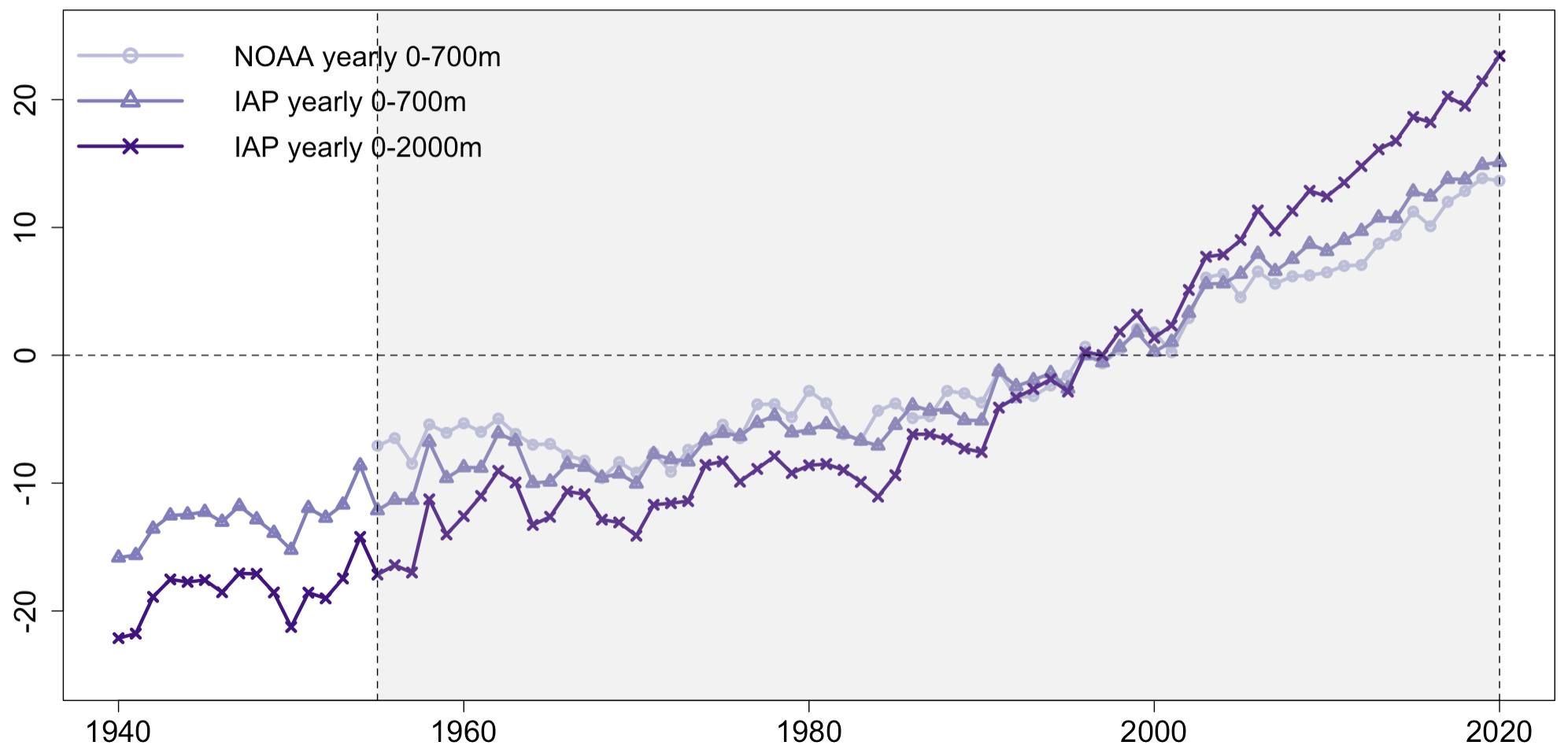}
			\label{OHCSyn}
		\end{subfigure}
	\end{figure}

	We show that we can directly synchronize anomaly series of different reference periods to a common baseline. Take a yearly global-level anomaly series $\left\{T_t^{\text{anom,1}}\right\}_{t=1}^T$ such that
	$
	T_t^{\text{anom,1}}= f\left(T_{j,\tau}^{\text{grid}} - \overline{T}_{j}^{\text{ref1}}\right)	
	$, where $T_{j,\tau}^{\text{grid}}$ is the raw gridded temperature level at time $\tau$ and location $j$, $\tau$ is the time index at higher frequency than a year, and $\overline{T}_{j}^{\text{ref1}}$ is the average of the temperatures at location $j$  over a pre-defined reference period ref1. $f(\cdot)$ is the linear operator that integrates high-resolution data into a yearly global value. Suppose we would like to get the anomalies over another reference period ref2, e.g., 1981-2010. Using linearity of $f(\cdot)$, the new anomaly $T_t^{\text{anom,2}}$ at time $t$ can be obtained by: 
	\begin{equation}
		\begin{aligned}
			T_t^{\text{anom,2}} &= f\left(T_{j,\tau}^{\text{grid}} - \overline{T}_{j}^{\text{ref2}}\right)\\
			&= f\left(T_{j,\tau}^{\text{grid}}\right)-f\left(  \overline{T}_{j}^{\text{ref2}}\right)+f\left(  \overline{T}_{j}^{\text{ref1}}\right)-f\left(  \overline{T}_{j}^{\text{ref1}}\right)\\
			&= f\left(T_{j,\tau}^{\text{grid}} - \overline{T}_{j}^{\text{ref1}}\right)-\left(f\left(\overline{T}_{j}^{\text{ref2}}\right) -f\left(\overline{T}_{j}^\text{ref1}\right)\right)\\
			& = 	T_t^{\text{anom,1}} -\left(\overline{T}^{\text{ref2}} -\overline{T}^{\text{ref1}}\right), 
			\label{Anom}
		\end{aligned}
	\end{equation}
	where $\overline{T}^{\text{ref2}} -\overline{T}^{\text{ref1}}$ is the difference between the two average global yearly temperatures over the two reference periods ref1 and ref2. Then we can synchronize the anomalies to ref2 by subtracting $\overline{T}^{\text{ref2}} -\overline{T}^{\text{ref1}}$ from the original anomalies. \\
		\indent  We use Equation \eqref{Anom} and the information in the IPCC report $1.5^{\circ} \mathrm{C}$ report \shortcite{2019teallenchnical} to get the average GMST values during the pre-industrial era,  $\overline{\boldsymbol{T}}^\text{pre-ind}$, for each of the data sources. The results are reported in Table \ref{preindGMST}. The downloaded ERA-Interim and JRA-55 datasets are already transformed relative to the pre-industrial level using the same method as in this paper, and hence we leave them as they are. We subtract $\overline{\boldsymbol{T}}^\text{pre-ind}$ from other GMST series to get the synchronized series. 
	\begin{table}[h!]
		\centering
		\begin{threeparttable}
			\setlength{\tabcolsep}{2pt} 
			\caption{\footnotesize Averages of the GMST series over 1986 -- 2005 $\boldsymbol{\overline{T}}_{i}^\text{1986-2005}$, changes of the averages over 1986 -- 2005 relative to the pre-industrial era (1850 -- 1900) $\Delta\boldsymbol{T}_\text{pre-ind}^{\text{1986-2005}}$ , and averages over 1850 -- 1900 $\overline{\boldsymbol{T}}^\text{pre-ind}_{i}$ ($^{\circ} \mathrm{C}$).  }
			\label{preindGMST}
			\begin{tabular}{lcccccccc}
				\hline\hline
				& \footnotesize \textbf{GISTEMP}&   	\footnotesize \textbf{NOAA}&  	\footnotesize \textbf{HadCRUT 5}&  		\footnotesize \textbf{BEST}&  	\footnotesize \textbf{CW2014}&  	\footnotesize \textbf{JMA}&  	\footnotesize \textbf{ERA-Interim}&	\footnotesize \textbf{JRA-55}\\
				\hline
				\footnotesize $\boldsymbol{\overline{T}}_{i}^\text{1986-2005}$ & 	\footnotesize 0.420 & 	\footnotesize 0.445 & 	\footnotesize 0.349& 	\footnotesize 0.382& 	\footnotesize 0.305& 	\footnotesize 0.014& 	\footnotesize 0.626& 	\footnotesize  0.635
				\\
				\footnotesize $\Delta\boldsymbol{T}_\text{pre-ind}^{\text{1986-2005}}$  \shortcite{2019teallenchnical} & 	\footnotesize $0.65$& 	\footnotesize $0.62$& 	\footnotesize $0.60$ $^\text{a}$& 	\footnotesize $0.73$& 	\footnotesize $0.65$& 	\footnotesize $0.59$& 	\footnotesize -& 	\footnotesize - 
				\\
				\footnotesize $\overline{\boldsymbol{T}}^\text{pre-ind}_{i}$ & 	\footnotesize $-0.230$&\footnotesize  $-0.175$ &	\footnotesize $-0.251$& 	\footnotesize $-0.210$& 	\footnotesize $-0.345$& 	\footnotesize $-0.576$& 	\footnotesize -& 	\footnotesize  	\footnotesize -
				\\
				\hline\hline
			\end{tabular}
			\begin{tablenotes}
				\footnotesize
				\item [a] The temperature change $\Delta\boldsymbol{T}_\text{pre-ind}^{\text{1986-2005}}$ for the HadCRUT 5 dataset is not avaiable in \shortciteA{2019teallenchnical}, and thus we use that for HadCRUT 4.6 instead.  
			\end{tablenotes}
		\end{threeparttable}
	\end{table}\\
\indent Figure \ref{TempSyn1} shows that these eight GMST anomalies after the synchronization have a substantial agreement, especially since the twentieth century. Figures \ref{OtempSyn} and \ref{OHCSyn} indicate that, after aligning to the same baseline, the NOAA 0-700m ocean temperature and OHC series are also comparable to their counterparts from IAP.
	\clearpage
\subsection{Estimation results using one GMST and one pair of ocean temperature and OHC series}
\label{Appendix1GMST1Ocean}
\begin{table}[ht!]
	\centering
	\caption{\footnotesize Mean estimates of the physical parameters from fitting the EBM-SS base model to different GMST and ocean  (ocean temperature + OHC) datasets. The numbers in parentheses are the  coefficients of variation.}
	\setlength{\tabcolsep}{2.5pt} 
	\renewcommand{\arraystretch}{1.4} 
	\begingroup\scriptsize
	\begin{tabular}{l|l|lllll|lllll}
		\hline \hline 
		\multicolumn{12}{c}{\footnotesize \textbf{One ocean temperature and one OHC series are included -- 0-700m}}	
		\\
		\hline
		\multirow{2}{*}{\footnotesize\textbf{No. of GMST}}	&\multirow{2}{*}{\footnotesize\textbf{GMST(s) included}}	&  \multicolumn{5}{c|}{\footnotesize \textbf{NOAA 0-700m}} &  \multicolumn{5}{c}{\footnotesize \textbf{IAP 0-700m}} 
		\\
		\cline{3-12}
		&	&\footnotesize	$\hat{\boldsymbol{\lambda}}$&\footnotesize $\hat{\boldsymbol{\gamma}}$& \footnotesize	$\hat{\boldsymbol{C}}_m$ &\footnotesize $\hat{\boldsymbol{C}}_d$ & \footnotesize  \textbf{ECS}  & \footnotesize  $\hat{\boldsymbol{\lambda}}$& \footnotesize $\hat{\boldsymbol{\gamma}}$& 	\footnotesize $\hat{\boldsymbol{C}}_m$ & \footnotesize $\hat{\boldsymbol{C}}_d$ &\footnotesize  \textbf{ECS}  \\
		\hline
	\footnotesize 	\multirow{16}{*}{\textbf{1 GMST}}	&	\footnotesize\textbf{GISTEMP} &0.63 & 1.45 & 21.44 & 96.13 & 6.24 & 0.44 & 1.21 & 25.52 & 98.23 & 8.91  \\ 
		&	&(0.47) & (0.2) & (0.38) & (0.01) & (0.47) & (1.1) & (0.36) & (0.59) & (0.002) & (1.1) \\ 
		\cline{2-12}
		&\footnotesize \textbf{NOAA} &  0.82 & 1.59 & 18.29 & 96.14 & 4.79 & 0.63 & 1.31 & 22.42 & 98.23 & 6.19  \\ 
		&	& (0.34) & (0.21) & (0.36) & (0.01) & (0.34) & (0.65) & (0.35) & (0.55) & (0.002) & (0.66) \\ 
		\cline{2-12}
		&\footnotesize	\textbf{HadCRUT5} & 0.74 & 1.3 & 20.3 & 96.15 & 5.29 & 0.63 & 1.15 & 22.12 & 98.23 & 6.29 \\ 
		&	& (0.41) & (0.22) & (0.37) & (0.01) & (0.41) & (0.61) & (0.3) & (0.48) & (0.002) & (0.62)\\ 
		\cline{2-12}
		&\footnotesize		\textbf{BEST (Berkeley)} &0.39 & 1.56 & 29.16 & 96.14 & 10.09 & 0.00017 & 1.13 & 43.18 & 98.23 & 231033.26 \\ 
		&	&(1.18) & (0.24) & (0.59) & (0.01) & (1.19) & (0.00015) & (0.29) & (0.2) & (0.002) & (0.07) \\ 
		\cline{2-12}
		&\footnotesize	\textbf{CW2014} & 0.55 & 1.61 & 26.52 & 96.13 & 7.15 & 0.07 & 1.21 & 42.28 & 98.23 & 60.06  \\ 
		&		& (0.64) & (0.21) & (0.43) & (0.01) & (0.65) & (12.56) & (0.33) & (0.72) & (0.002) & (12.56)  \\ 
		\cline{2-12}
		&	\footnotesize	\textbf{JMA} &1.01 & 2.28 & 15.81 & 96.13 & 3.88 & 0.8 & 1.83 & 21.38 & 98.22 & 4.93  \\ 
		&	& (0.26) & (0.21) & (0.34) & (0.01) & (0.27) & (0.51) & (0.38) & (0.61) & (0.002) & (0.52) \\ 
		\cline{2-12}
		&\footnotesize	\textbf{ERA5} & 0.85 & 1.2 & 15.8 & 96.14 & 4.61 & 0.87 & 1.14 & 14.11 & 98.22 & 4.52 \\ 
		&		&(0.32) & (0.26) & (0.37) & (0.01) & (0.33) & (0.26) & (0.27) & (0.32) & (0.002) & (0.27) \\ 
		\cline{2-12}
		&	\footnotesize	\textbf{JRA55} & 0.85 & 1.32 & 17.74 & 96.15 & 4.61 & 0.87 & 1.26 & 15.85 & 98.22 & 4.51  \\ 
		&		&  (0.37) & (0.27) & (0.43) & (0.01) & (0.38) & (0.3) & (0.28) & (0.38) & (0.002) & (0.31) \\ 
		\hline 
		& \footnotesize	\textbf{Median of estimate }	&	\multicolumn{2}{c}{	0.78}  & \multicolumn{2}{c}{1.51} & \multicolumn{2}{c}{19.30} & \multicolumn{2}{c}{96.14} & \multicolumn{2}{c}{5.03}  \\ 
		&\footnotesize  \textbf{Median of CV}	 & \multicolumn{2}{c}{(0.39)} & \multicolumn{2}{c}{(0.22)} & \multicolumn{2}{c}{(0.37)} & \multicolumn{2}{c}{(0.01)}& \multicolumn{2}{c}{(0.40)}  \\ 
		\hline\hline
		\multicolumn{12}{c}{\footnotesize \textbf{One ocean temperature and one OHC series are included -- 0-2,000m}}	
		\\
		\hline
		\multirow{2}{*}{\footnotesize \textbf{No. of GMST}}	&\multirow{2}{*}{\footnotesize \textbf{GMST(s) included}}	&  \multicolumn{5}{c|}{\footnotesize \textbf{IAP 0-2000m}} &  
		\\
		\cline{3-7}
		&	&\footnotesize	$\hat{\boldsymbol{\lambda}}$& \footnotesize$\hat{\boldsymbol{\gamma}}$& \footnotesize	$\hat{\boldsymbol{C}}_m$ &\footnotesize $\hat{\boldsymbol{C}}_d$ & \footnotesize  \textbf{ECS}  &  \\
		\cline{1-7}
		\multirow{16}{*}{\footnotesize \textbf{ 1 GMST}}	&	\footnotesize \textbf{GISTEMP} &   0.18 & 1.84 & 21.72 & 269.28 & 22.36 \\ 
		&	& (1.9) & (0.18) & (0.37) & (0.002) & (1.9) \\ 
			\cline{2-7}
		&\footnotesize \textbf{NOAA} &0.33 & 2.00 & 19.04 & 269.28 & 12.08 \\ 
		&	&(1.01) & (0.19) & (0.36) & (0.002) & (1.01) \\ 
			\cline{2-7}
		&\footnotesize 	\textbf{HadCRUT5} & 0.25 & 1.73 & 20.95 & 269.28 & 15.89 \\ 
		&	&  (1.40) & (0.20) & (0.36) & (0.002) & (1.40) \\ 
			\cline{2-7}
		&	\footnotesize	\textbf{BEST (Berkeley)} &0.03 & 1.93 & 27.2 & 269.28 & 148.48 \\ 
		&	&(14.61) & (0.19) & (0.48) & (0.002) & (14.61) \\ 
			\cline{2-7}
		&\footnotesize	\textbf{CW2014} & 0.10 & 1.98 & 26.16 & 269.28 & 39.12 \\ 
		&		& (3.67) & (0.19) & (0.42) & (0.002) & (3.67) \\ 
		\cline{2-7}
		&\footnotesize		\textbf{JMA} &0.47 & 2.65 & 16.9 & 269.28 & 8.44 \\ 
		&	& (0.70) & (0.19) & (0.35) & (0.002) & (0.71) \\ 
		\cline{2-7}
		&\footnotesize	\textbf{ERA5} & 0.51 & 1.55 & 13.58 & 269.29 & 7.64 \\ 
		&		& (0.49) & (0.21) & (0.27) & (0.002) & (0.49)  \\ 
			\cline{2-7}
		&	\footnotesize	\textbf{JRA55} &  0.53 & 1.7 & 14.81 & 269.3 & 7.44  \\ 
		&		&(0.49) & (0.21) & (0.3) & (0.002) & (0.5) \\ 
		\cline{2-7}
		& \footnotesize	\textbf{Median of estimate }	&	0.29 & 1.89 & 19.99 & 269.28 & 13.98  \\ 
		&\footnotesize  \textbf{Median of CV}	 & (1.20) & (0.19) & (0.36) & (0.002) & (1.21)  \\ 
		\hline \hline
	\end{tabular}
	\label{Empirical700+2000}
	\endgroup
\end{table} 
\clearpage
\subsection{Fitted estimates and standardized prediction errors using 0-2,000m ocean series}
\label{2000m}
	\begin{figure}[h!]
	\caption{\footnotesize Fit of EBM-SS full model to eight GMST series, two 0-700m ocean temperature, two OHC 0-700m, and one radiative forcing series. Panels (a), (c), (e) show  the observational series and the smoothed states from the Kalman filter, which are the estimated states conditional on the entire observational paths. Panel (g) shows the fit to OHC series using the assumption $O=C_d\mu_{T_d}$. Panels (b), (d), (f), and (g) report the standardized one-step ahead prediction errors. ``constant'' in the legends of (c) and (g) are the estimated constants for the IAP ocean temperature series. }
	\begin{subfigure}[b]{0.5\textwidth}
		\centering
		\subcaption{ \footnotesize smoothed state of $T_m$ and 8 synchronized GMST series} 
		\includegraphics[width=\linewidth]{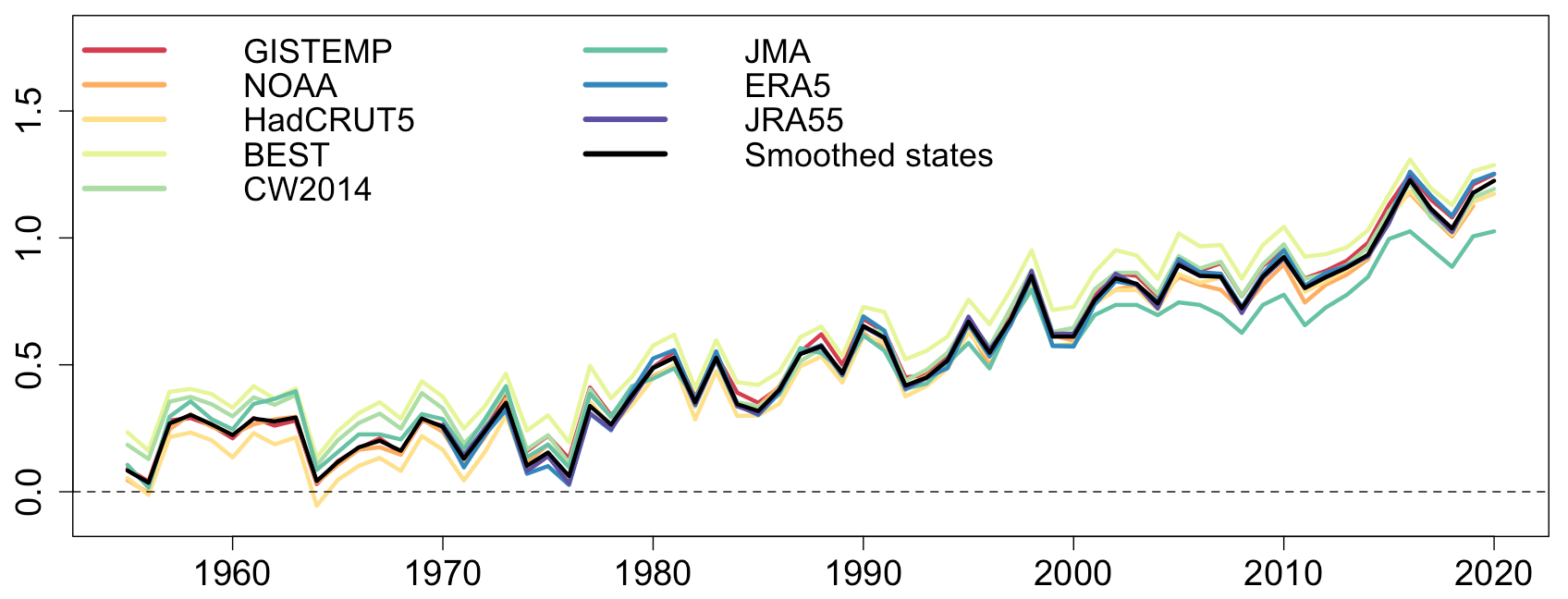}
	\end{subfigure}\hfill
	\begin{subfigure}[b]{0.49\textwidth}
		\centering
		\subcaption{\footnotesize standardized prediction errors for 8 synchronized GMST} 
		\includegraphics[width=\linewidth]{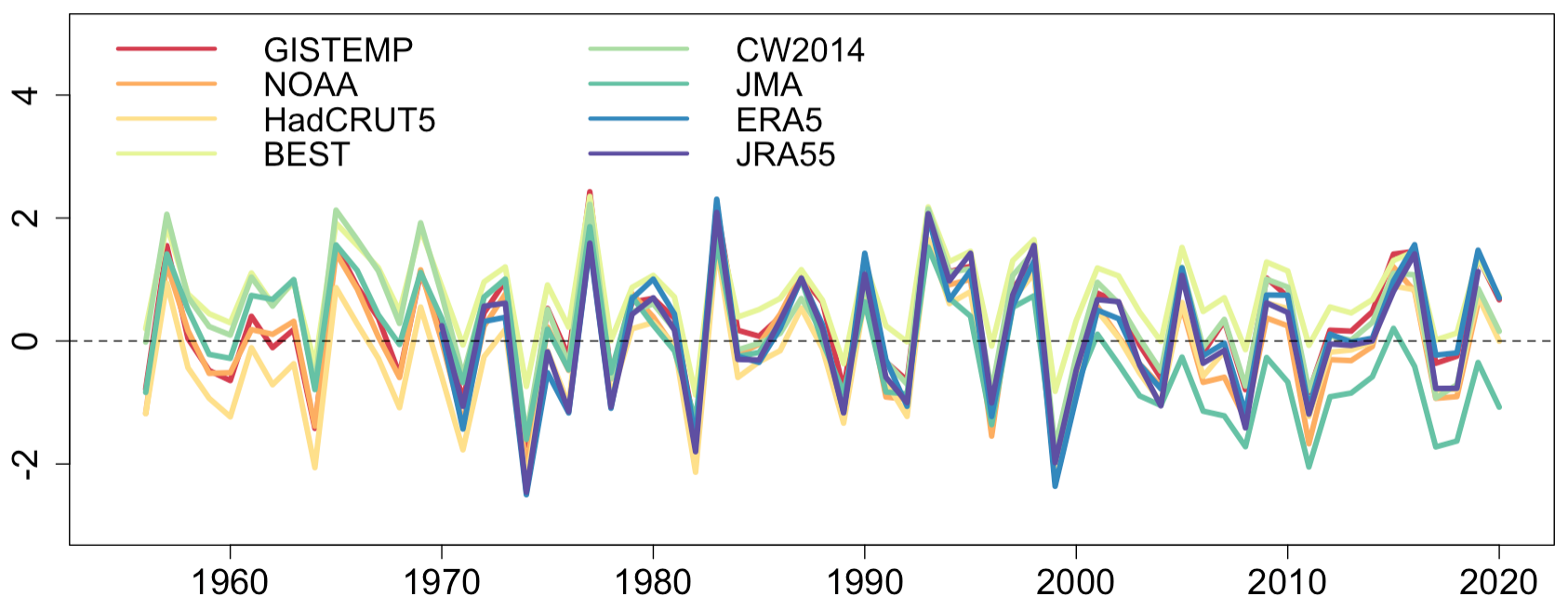}
	\end{subfigure}
	\begin{subfigure}[b]{0.5\textwidth}
		\centering
		\subcaption{\footnotesize smoothed state of $T_d$ and ocean temperature 0-700m series} 
		\includegraphics[width=\linewidth]{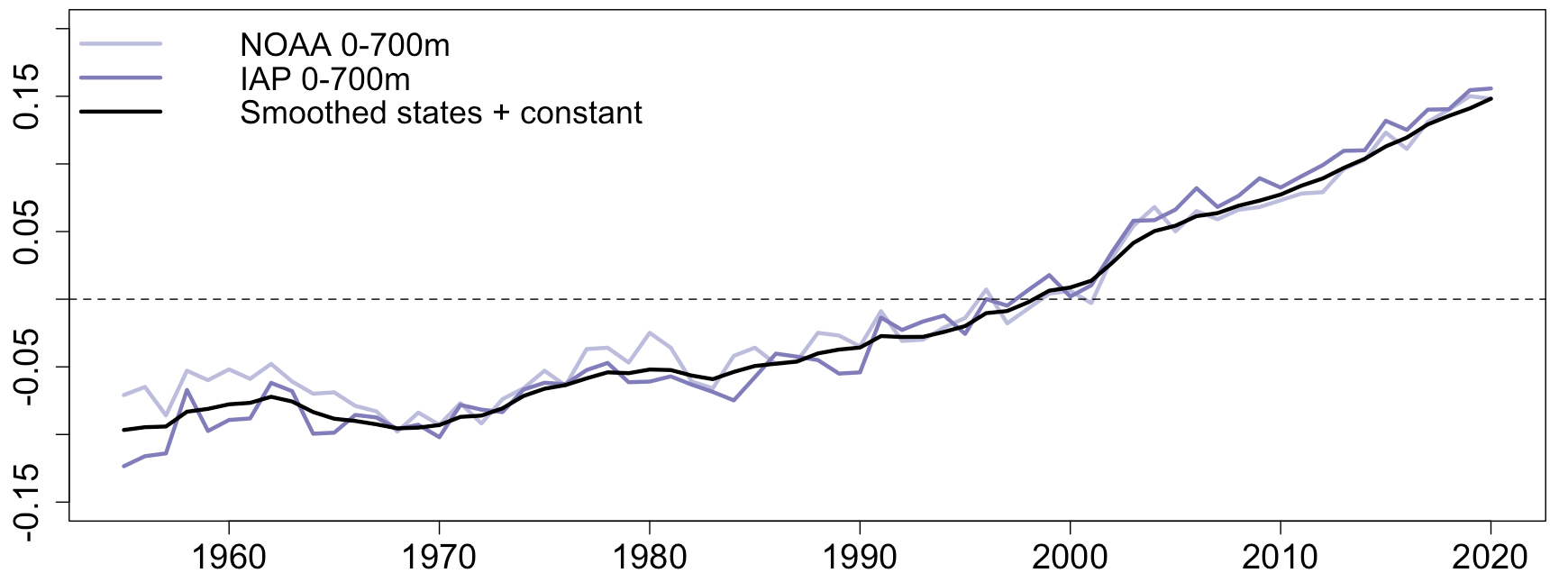}
	\end{subfigure}\hfill
	\begin{subfigure}[b]{0.5\textwidth}
		\centering
		\subcaption{\footnotesize standardized prediction errors for ocean temperature 0-700m} 
		\includegraphics[width=\linewidth]{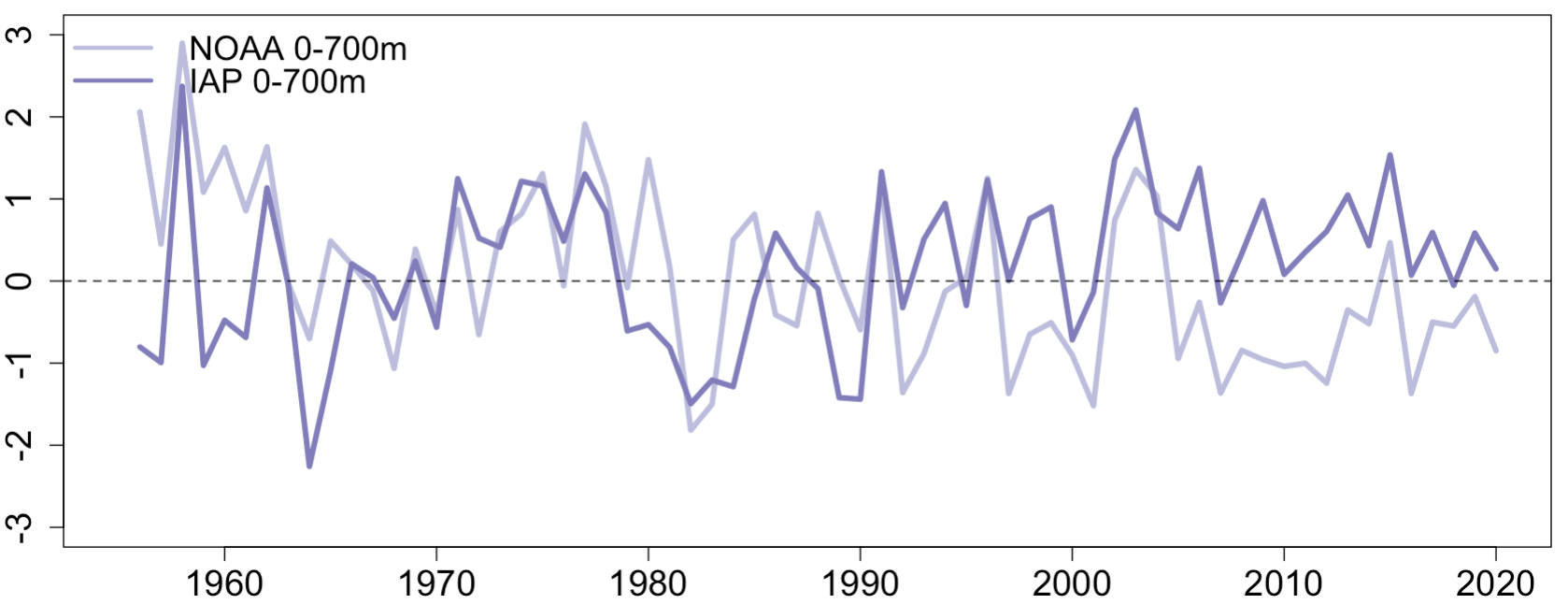}
	\end{subfigure}\\
	\begin{subfigure}[b]{0.5\textwidth}
		\centering
		\subcaption{\footnotesize smoothed state of $A$ and anthropogenic forcing series} 
		\includegraphics[width=\linewidth]{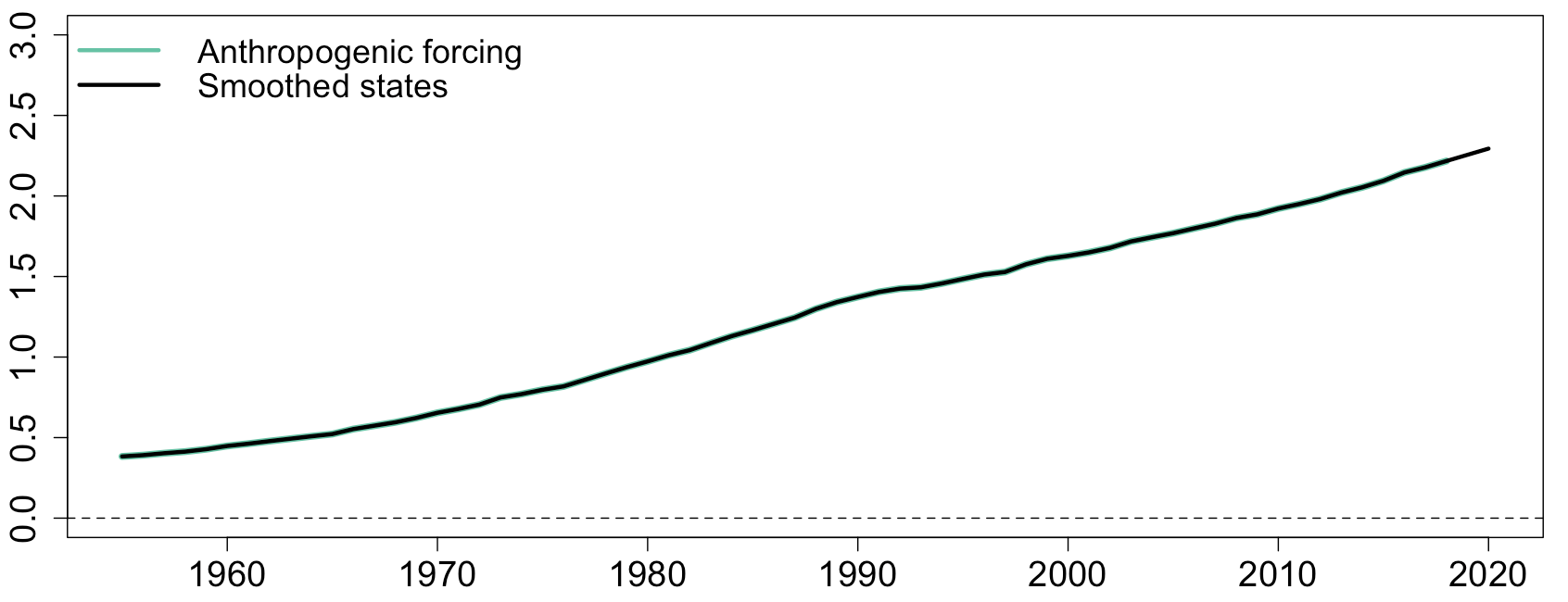}
	\end{subfigure}\hfill
	\begin{subfigure}[b]{0.49\textwidth}
		\centering
		\subcaption{\footnotesize standardized prediction errors for anthropogenic forcing} 
		\includegraphics[width=\linewidth]{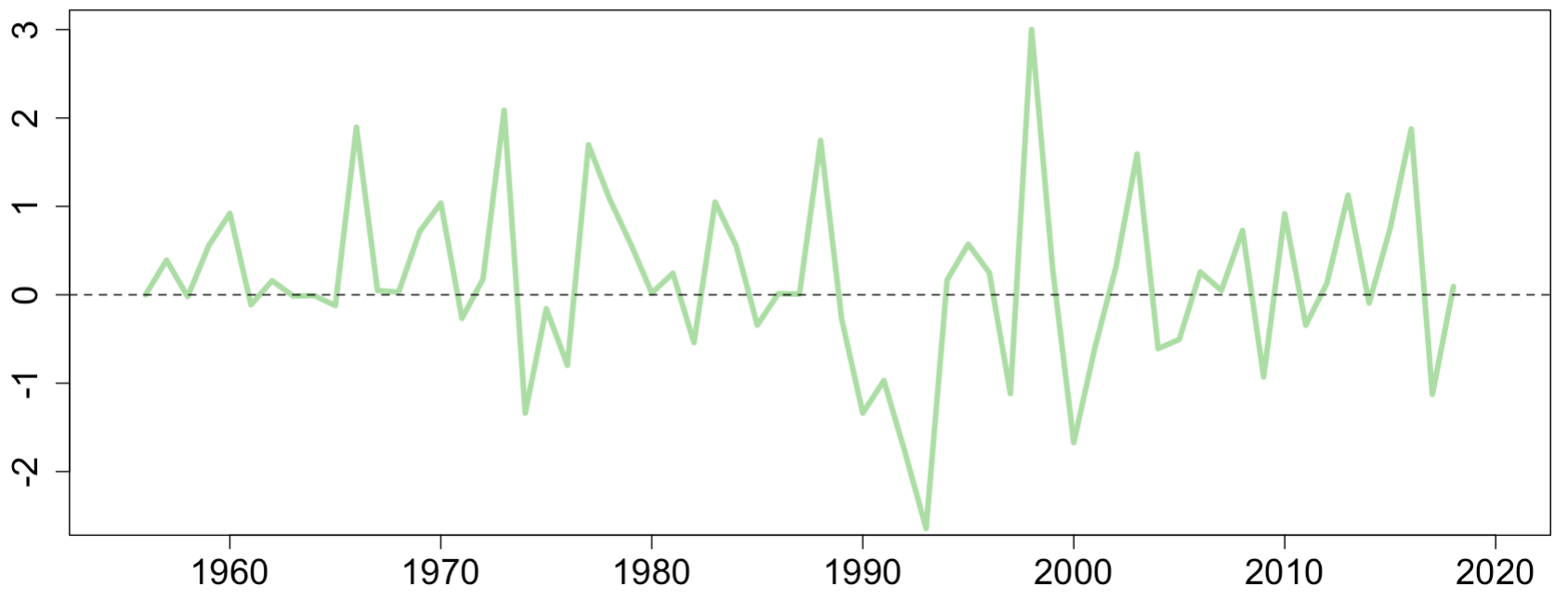}
	\end{subfigure}\\
	\begin{subfigure}[b]{0.5\textwidth}
		\centering
		\subcaption{\footnotesize Fit to OHC 0-700m series} 
		\includegraphics[width=\linewidth]{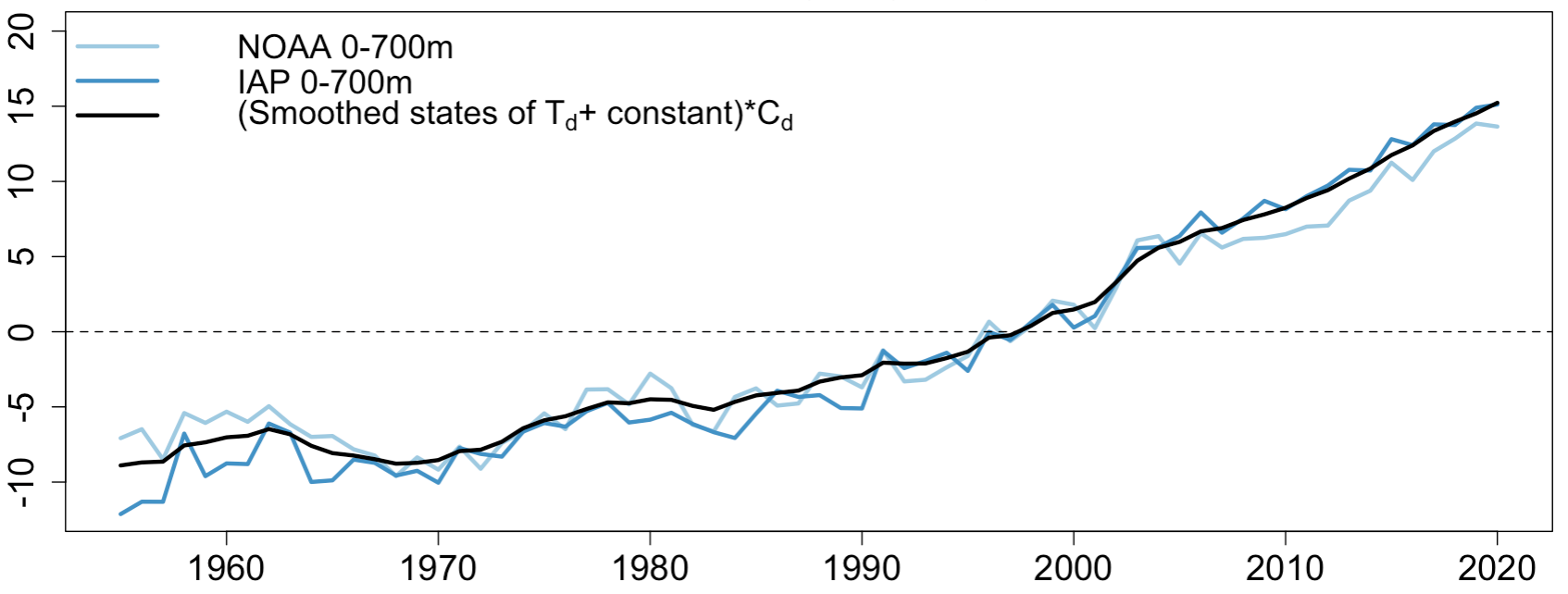}
	\end{subfigure}\hfill
	\begin{subfigure}[b]{0.5\textwidth}
		\centering
		\subcaption{\footnotesize standardized prediction errors for OHC 0-700m} 
		\includegraphics[width=\linewidth]{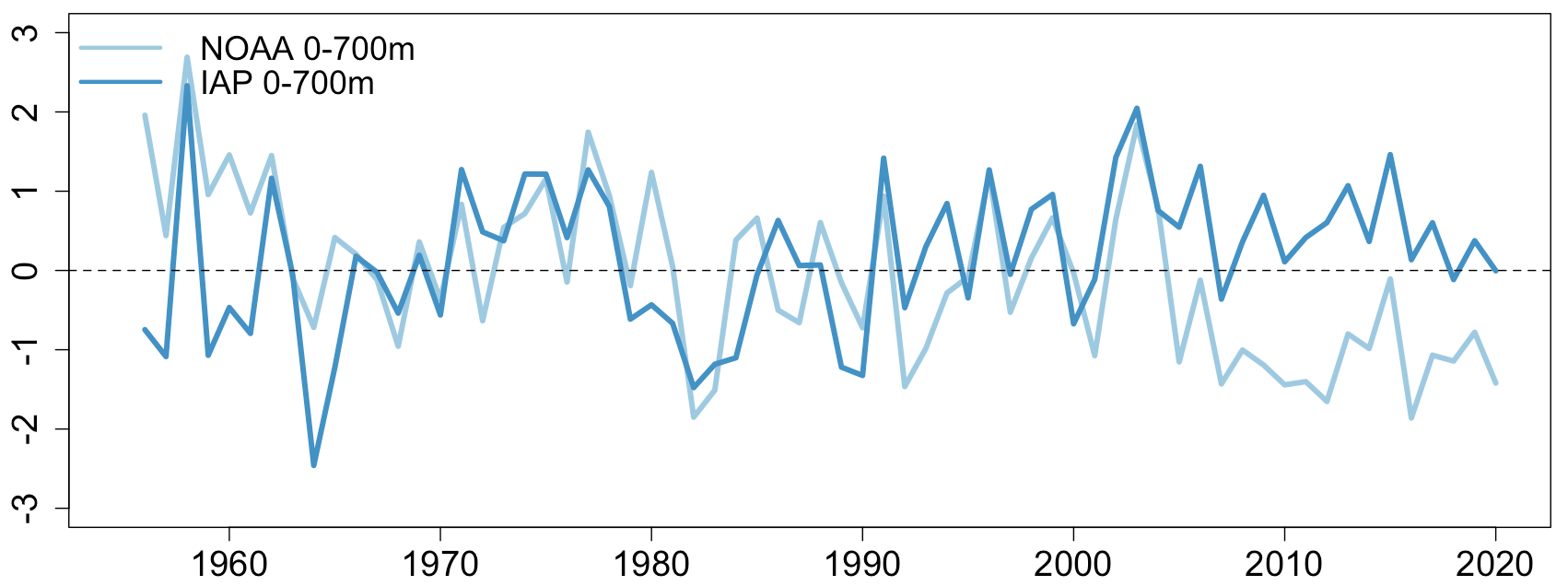}
	\end{subfigure}
	\label{fit700}
\end{figure}
	\begin{figure}[h!]
	\caption{\footnotesize \footnotesize Fit of EBM-SS full model to eight GMST series, one 0-2,000m ocean temperature, one OHC 0-2,000m, and one radiative forcing series. Panels (a), (c), (e) show  the observational series and the smoothed states from the Kalman filter, which are the estimated states conditional on the entire observational paths. Panel (g) shows the fit to OHC series using the assumption $O=C_d\mu_{T_d}$. Panels (b), (d), (f), and (g) report the standardized one-step ahead prediction errors. ``constant'' in the legends of (c) and (g) are the estimated constants for the IAP ocean temperature series. }
	\begin{subfigure}[b]{0.5\textwidth}
		\centering
		\subcaption{ \footnotesize smoothed state of $T_m$ and 8 synchronized GMST series} 
		\includegraphics[width=\linewidth]{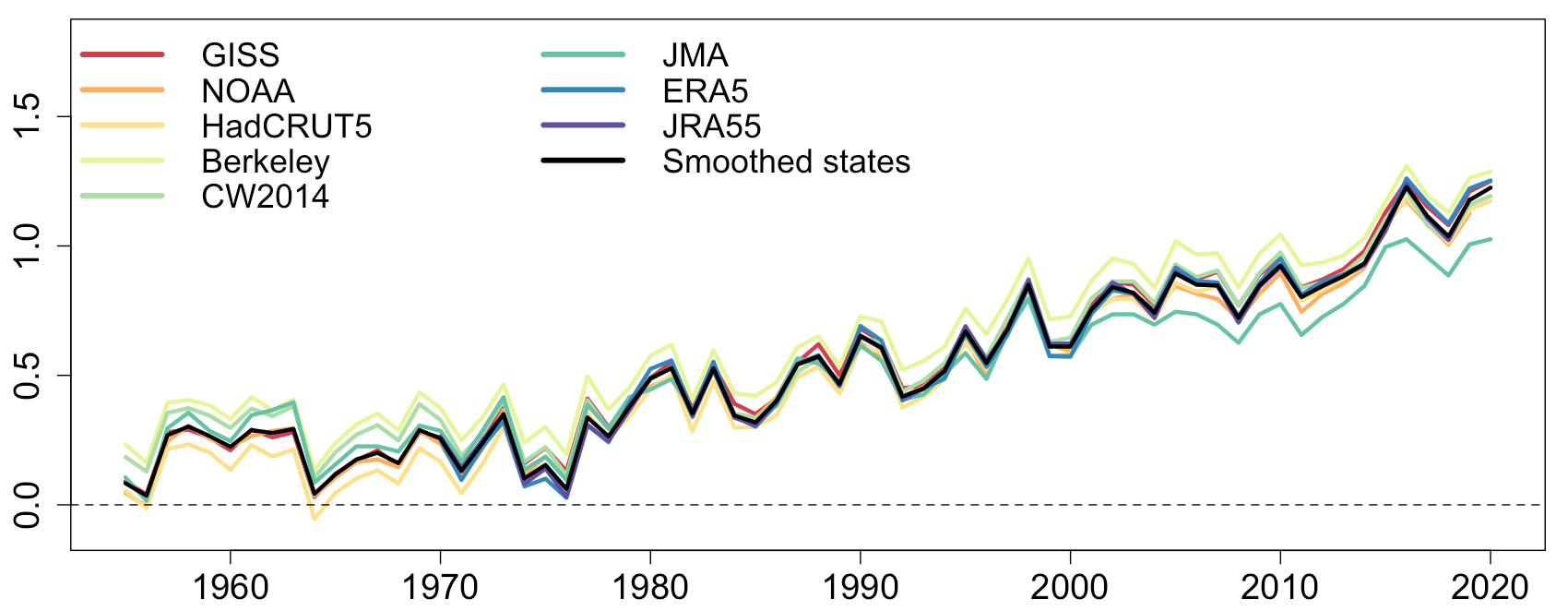}
	\end{subfigure}\hfill
	\begin{subfigure}[b]{0.49\textwidth}
		\centering
		\subcaption{\footnotesize standardized prediction errors for 8 synchronized GMST} 
		\includegraphics[width=\linewidth]{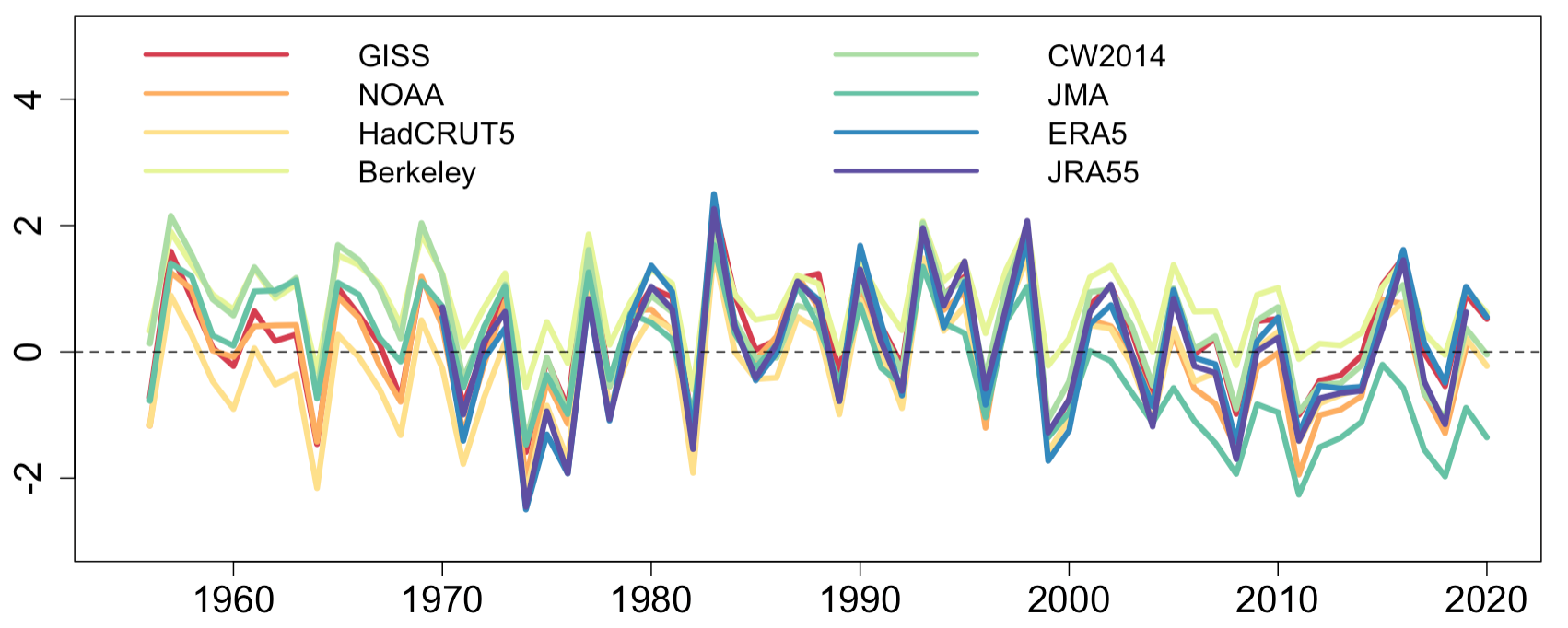}
	\end{subfigure}
	\begin{subfigure}[b]{0.5\textwidth}
		\centering
		\subcaption{\footnotesize smoothed state of $T_d$ and ocean temperature \\0-2,000m series} 
		\includegraphics[width=\linewidth]{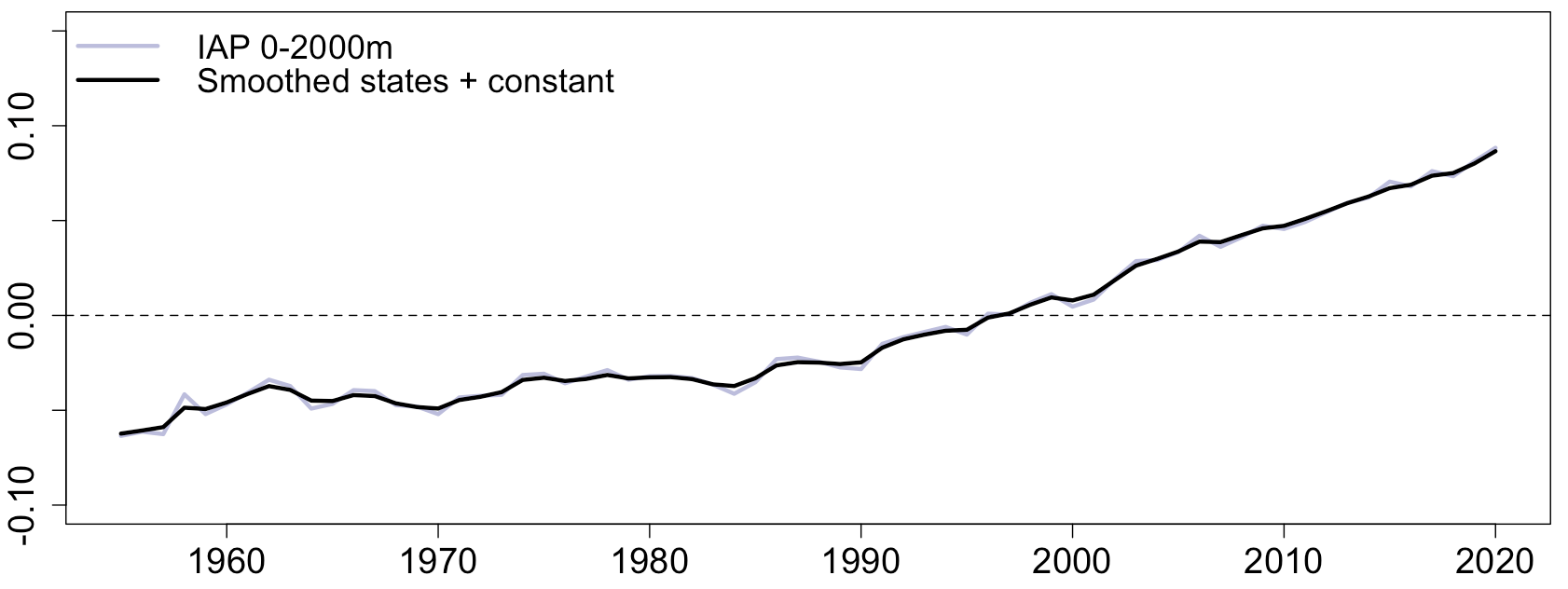}
	\end{subfigure}\hfill
	\begin{subfigure}[b]{0.5\textwidth}
		\centering
		\subcaption{\footnotesize standardized prediction errors for ocean temperature 0-2,000m} 
		\includegraphics[width=\linewidth]{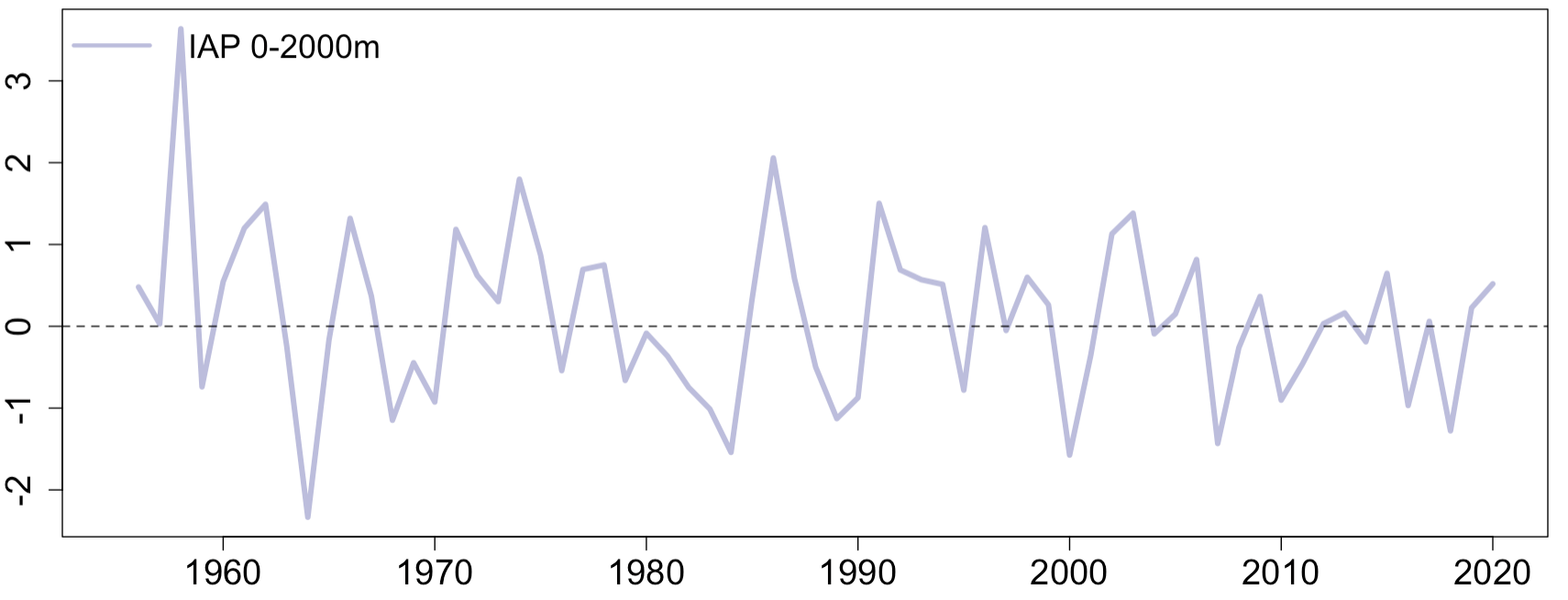}
	\end{subfigure}\\
	\begin{subfigure}[b]{0.49\textwidth}
		\centering
		\subcaption{\footnotesize smoothed state of $A$ and anthropogenic series} 
			\includegraphics[width=\linewidth]{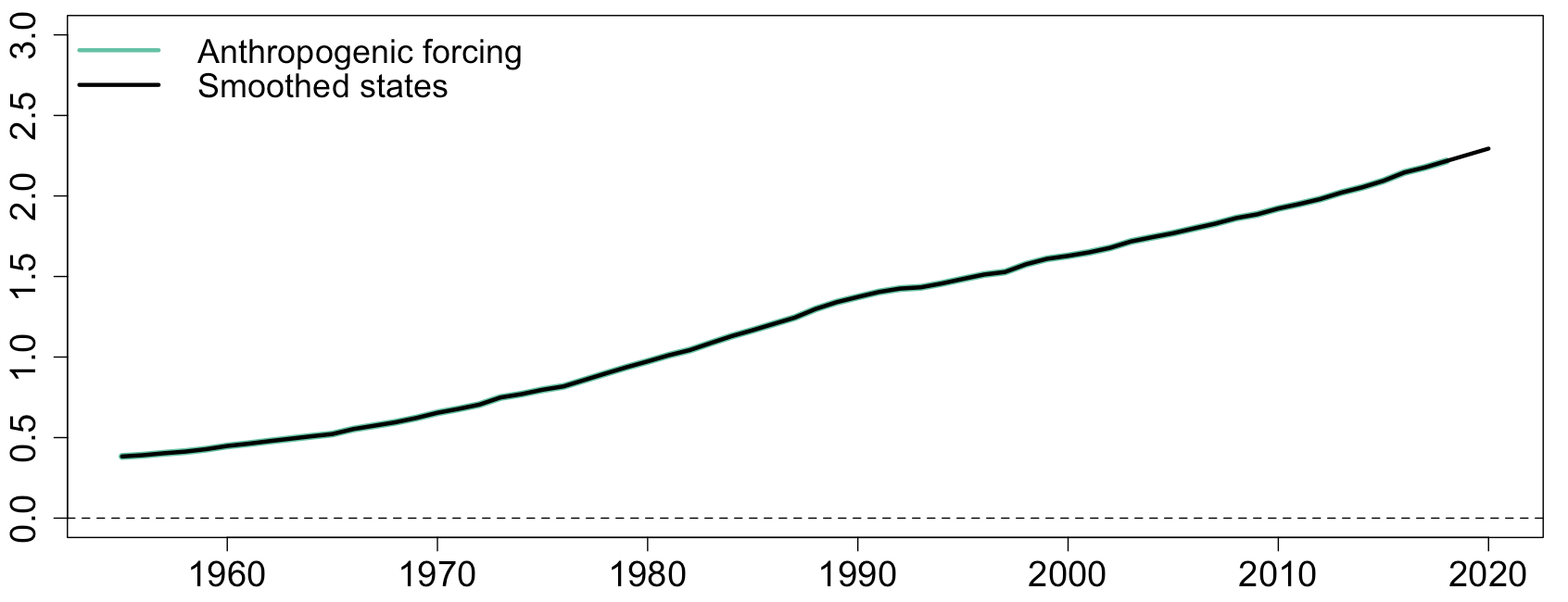}
	\end{subfigure}\hfill
	\begin{subfigure}[b]{0.5\textwidth}
		\centering
		\subcaption{\footnotesize standardized prediction errors for anthropogenic forcing} 
		\includegraphics[width=\linewidth]{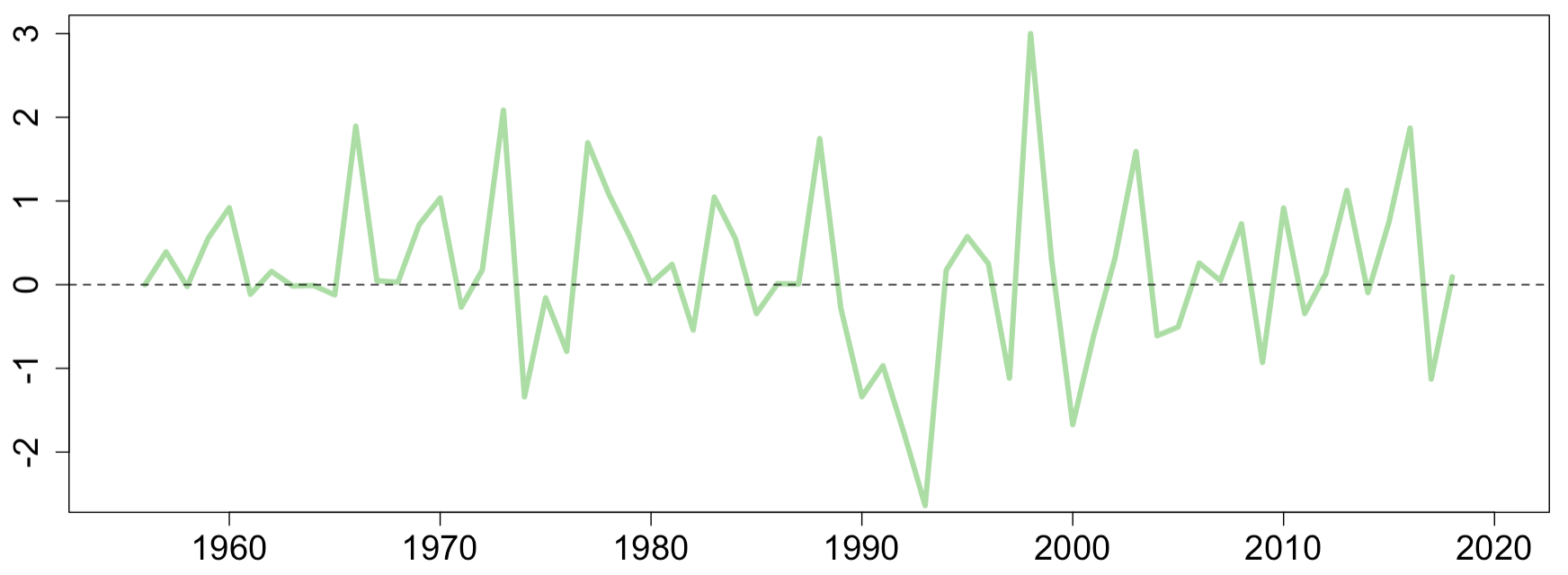}
	\end{subfigure}\\
	\begin{subfigure}[b]{0.5\textwidth}
	\centering
	\subcaption{\footnotesize Fit to OHC 0-2,000m series} 
	\includegraphics[width=\linewidth]{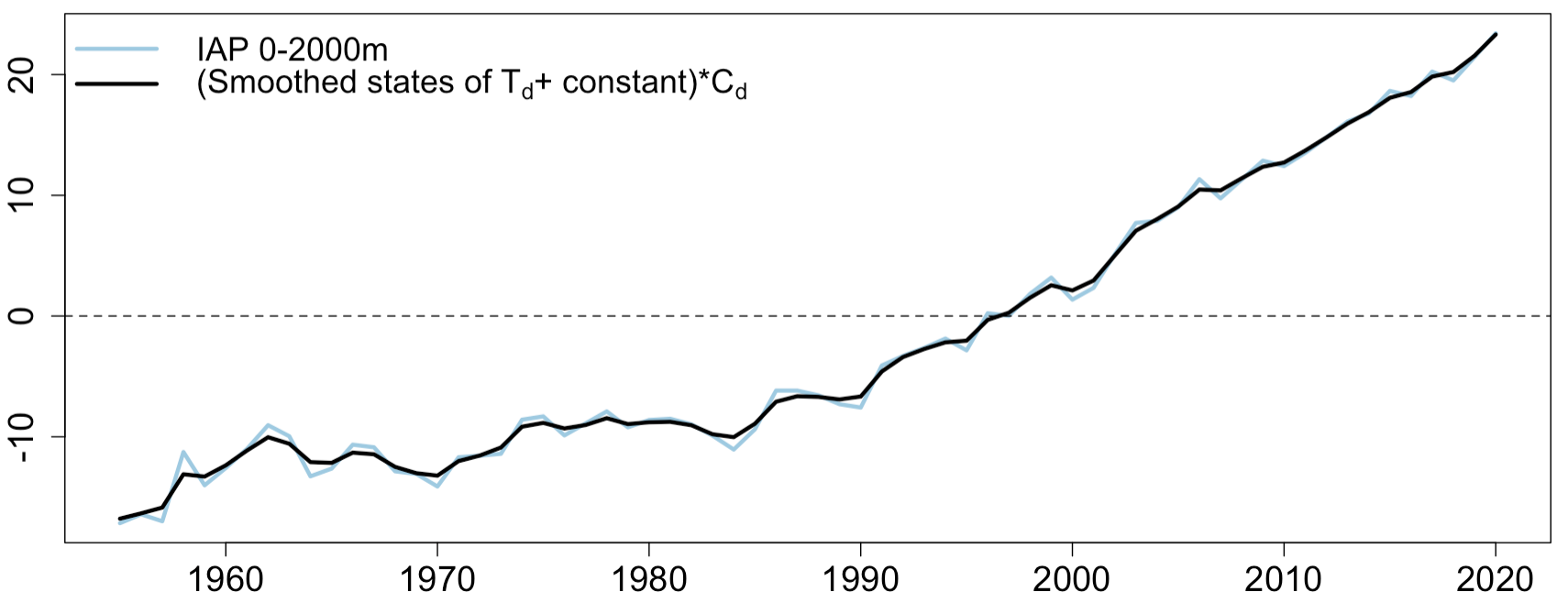}
\end{subfigure}\hfill
\begin{subfigure}[b]{0.5\textwidth}
	\centering
	\subcaption{\footnotesize standardized prediction errors for OHC 0-2000m} 
	\includegraphics[width=\linewidth]{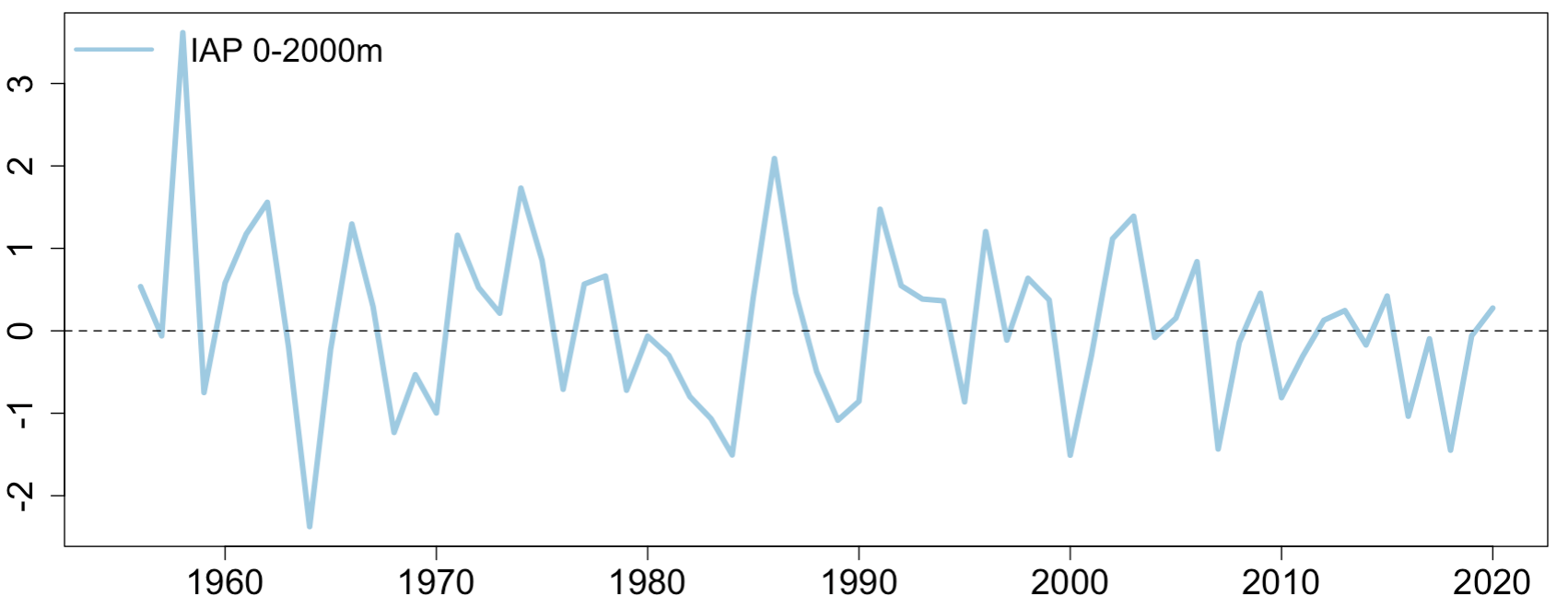}
\end{subfigure}
	\label{fit2000}
\end{figure}
\clearpage
\subsection{Diagnostic statistics}
\label{diag}
\begin{table}[h!]
	\centering
	\caption{\footnotesize Diagnostic statistics of the one-step ahead standardized prediction errors for the EBM-SS full model. The upper panel shows the results for 0-700m ocean data. The lower panel shows the results for 0-2,000m ocean data. ** and * denote significance at $1\%$ and $5\%$.}
	\setlength{\tabcolsep}{1.3pt} 
	\renewcommand{\arraystretch}{1.4} 
	\begingroup\scriptsize
	\begin{threeparttable}
		\begin{tabular}{l|ccccccccccccc}
			\hline\hline
			&  \multicolumn{13}{c}{\footnotesize \textbf{A. ocean temperature and OHC  0-700m from both NOAA and IAP are included}}
			\\
			\cline{2-14}
			& 	\textbf{GISTEMP} & \textbf{NOAA }& \textbf{HadCRUT5} & \textbf{BEST} & \textbf{CW14} & \textbf{JMA}  & \textbf{JRA55} & \textbf{ERA5} & $\boldsymbol{Y}_{T_\text{d,700 m}^\text{NOAA}}$&  $\boldsymbol{Y}_{O_\text{700 m}^\text{NOAA}}$ & $\boldsymbol{Y}_{T_\text{d,700 m}^\text{IAP}}$&  $\boldsymbol{Y}_{O_\text{700 m}^\text{IAP}}$ & \textbf{forcing} \\ 
			\hline 
			mean & 0.284 & 0.004 & $-0.225$ & 0.753 & 0.381 & $-0.138$ & 0.080 & 0.045 & 0.053 & 0.182 & $-0.074$ & 0.163 & 0.149 \\ 
			std & 0.945 & 0.989 & 0.935 & 0.724 & 0.944 & 0.962 & 1.067 & 1.052 & 1.033 & 0.930 & 1.044 & 0.928 & 0.989 \\ 
			skewness & $-0.102$ & 0.054 & $-0.115$ & $-0.079$ & $-0.106$ & 0.072 & $-0.255$ & $-0.160$ & 0.432 & $-0.167$ & 0.309 & $-0.180$& 0.093 \\ 
			kurtosis & 2.720 & 2.598 & 2.649 & 2.651 & 2.612 & 2.260 & 2.689 & 2.427 & 2.567 & 2.731 & 2.444 & 2.909 & 3.860 \\ 
			$t_\text{JB}$$^\text{a}$ & 0.325 & 0.461 & 0.476 & 0.399 & 0.528 & 1.539 & 0.759 & 0.898 & 2.529 & 0.498 & 1.873 & 0.375 & 2.034 \\ 
			Q(1)$^\text{b}$& 4.979$^*$ & 3.129 & 4.707$^*$ & 5.591$^*$ & 2.533 & 1.168 & 2.739$^*$ & 4.057$^*$ & 6.076$^{**}$ & 3.703 & 9.895$^{**}$ & 2.482 & 0.000 \\ 
			\hline \hline
			&  \multicolumn{13}{c}{\footnotesize \textbf{B. Ocean temperature and OHC  0-2,000m from IAP is included}}
			\\
			\cline{2-14}
			& 	\textbf{GISTEMP} & \textbf{NOAA }& \textbf{HadCRUT5} & \textbf{BEST} & \textbf{CW14} & \textbf{JMA}  & \textbf{JRA55} & \textbf{ERA5} & $\boldsymbol{Y}_{T_\text{d,2,000 m}^\text{IAP}}$&  $\boldsymbol{Y}_{O_\text{2,000 m}^\text{IAP}}$ & \textbf{forcing} \\ 
			\hline 
			mean & 0.259 & $-0.025$ & $-0.251$ & 0.735 & 0.358 & $-0.161$ & 0.065 & 0.028 & 0.118 & 0.089 & 0.148 \\ 
			std & 0.949 & 0.985 & 0.941 & 0.720 & 0.937 & 0.943 & 1.078 & 1.056 & 1.006 & 1.005 & 0.988 \\ 
			skewness &$-0.109$ & 0.053 & $-0.126$ &$ -0.093$ & $-0.118$ & 0.086 &$ -0.267$ & $-0.168$ & 0.456 & 0.483 & 0.095 \\ 
			kurtosis & 2.703 & 2.629 & 2.641 & 2.660 & 2.628 & 2.282 & 2.719 & 2.493 & 4.222 & 4.289 & 3.858 \\ 
			$t_\text{JB}$ & 0.368 & 0.396 & 0.520 & 0.407 & 0.527 & 1.478 & 0.772 & 0.772 & {$6.290^*$} & {$7.020^*$} & 2.026 \\ 
			Q(1) & 4.465$^*$& 3.292 & 4.012$^*$ & 5.714$^*$ & 2.880 & 0.737 & 2.144 & 3.737 & 0.234 & 0.132 & 0.000 \\ 
			\hline \hline
		\end{tabular}
		\begin{tablenotes}
			\scriptsize
			\item [a] Test statistic of the Jarque-Bera normality test \cite{jarque1980efficient}.  The null hypothesis is that the sample is normally distributed. 
			\item [b] Portmanteau test statistic for serial correlation by \citeA{ljung1978measure}: $Q(k)=n(n+2) \sum_{j=1}^{k} \frac{\rho_{j}^{2}}{n-j}$, where $k$ is the order of lag, and $\rho_{j}$ is the sample autocorrelation at lag $j$. $H_0$: the sample exhibits no serial correlation. 
		\end{tablenotes}
	\end{threeparttable}
	\label{8Temp2OHC}
	\endgroup
\end{table} 
\end{document}